\documentclass[twocolumn,twocolappendix]{aastex63}

\usepackage{graphicx}	
\usepackage{amsmath}	
\usepackage{amssymb}	
\usepackage{cleveref}
\hypersetup{breaklinks,colorlinks,urlcolor=blue,citecolor=blue,linkcolor=blue}
\usepackage{bm}
\usepackage{float}
\usepackage{booktabs}
\usepackage{listings}
\usepackage{mathtools}

\usepackage[caption=false]{subfig}

\newcommand{\code}[1]{\texttt{#1}}
\newcommand{\bs}[1]{\boldsymbol{#1}}
\newcommand\myeq{\mkern1.5mu{=}\mkern1.5mu}
\newcommand{\balrog}{\code{Balrog}}
\newcommand{\Balrog}{\code{BALROG}}
\newcommand{\galsim}{\code{GalSim}}
\newcommand{\ngmix}{\code{ngmix}}
\newcommand{\sextractor}{\code{SExtractor}}
\newcommand{\python}{Python}

\newcommand{\runtwo}{\response{\code{main}}}
\newcommand{\runtwoa}{\response{\code{aux}}}
\newcommand{\runtwomag}{\response{\code{main-mag}}}
\newcommand{\runtwoamag}{\response{\code{aux-mag}}}
\newcommand{\gridtest}{\response{\code{grid}}}
\newcommand{\noiselessgridtest}{\response{\code{noiseless-grid}}}

\newcommand{\starsample}{$\response{\delta}$\response{-\textit{stars}}}
\newcommand{\dfsample}{\response{\textit{y3-deep}}}

\newcommand{\response}{}

\newcommand{\amin}{$^{\,\prime}$}
\newcommand{\asec}{$^{\,\prime\prime}$}

\newcommand*{\rom}[1]{\expandafter\@slowromancap\romannumeral #1@}

\newcommand{\dmstar}{$\Delta\text{mag}_\delta$}
\newcommand{\dcstar}{$\Delta\text{c}_\delta$}
\newcommand{\mdmstar}{${<}\Delta\text{mag}_\delta{>}$}
\newcommand{\medstar}{$\widetilde{\Delta\text{mag}}_\delta$}
\newcommand{\medcstar}{$\widetilde{\Delta\text{c}}_\delta$}
\newcommand{\sigstar}{$\sigma_{\text{mag}_\delta}$}

\newcommand{\dmgal}{$\Delta\text{mag}_\text{DF}$}
\newcommand{\dcgal}{$\Delta\text{c}_\text{DF}$}
\newcommand{\mdmgal}{${<}\Delta\text{mag}_\text{DF}{>}$}
\newcommand{\medgal}{$\widetilde{\Delta\text{mag}}_\text{DF}$}
\newcommand{\medcgal}{$\widetilde{\Delta\text{c}}_\text{DF}$}
\newcommand{\mdcgal}{${<}\Delta\text{c}_\text{DF}{>}$}
\newcommand{\siggal}{$\sigma_{\text{mag}_\text{DF}}$}

\newcommand{\mdmgrid}{${<}\Delta\text{mag}{>}$}

\received{2021 January 14}
\revised{2021 August 16}
\accepted{2021 September 13}
\published{2022 January 12}

\reportnum{DES-2020-0538}
\reportnum{FERMILAB-PUB-20-669-E-SCD}

\submitjournal{ApJS}

\shorttitle{Measuring the DES Transfer Function with Balrog}
\shortauthors{The Dark Energy Survey Collaboration}

\usepackage[flushmargin]{footmisc}

\usepackage[mathlines]{lineno}
\relax 

\usepackage{etoolbox}
\makeatletter
\patchcmd\linenumberpar{\@LN@parpgbrk}{\penalty\@LN@parpgpen\relax}{}{}
\makeatother

\begin{document}

\title{Dark Energy Survey Year 3 Results: Measuring the Survey Transfer Function with Balrog}

\correspondingauthor{Spencer Everett}
\email{spencer.w.everett@jpl.nasa.gov}


\author{S.~Everett}
\affiliation{Santa Cruz Institute for Particle Physics, Santa Cruz, CA 95064, USA}
\affiliation{Jet Propulsion Laboratory, California Institute of Technology, 4800 Oak Grove Dr., Pasadena, CA 91109, USA}
\author{B.~Yanny}
\affiliation{Fermi National Accelerator Laboratory, P. O. Box 500, Batavia, IL 60510, USA}
\author{N.~Kuropatkin}
\affiliation{Fermi National Accelerator Laboratory, P. O. Box 500, Batavia, IL 60510, USA}
\author{E.~M.~Huff}
\affiliation{Jet Propulsion Laboratory, California Institute of Technology, 4800 Oak Grove Dr., Pasadena, CA 91109, USA}
\author{Y.~Zhang}
\affiliation{Fermi National Accelerator Laboratory, P. O. Box 500, Batavia, IL 60510, USA}
\author{J.~Myles}
\affiliation{Department of Physics, Stanford University, 382 Via Pueblo Mall, Stanford, CA 94305, USA}
\affiliation{Kavli Institute for Particle Astrophysics \& Cosmology, P. O. Box 2450, Stanford University, Stanford, CA 94305, USA}
\affiliation{SLAC National Accelerator Laboratory, Menlo Park, CA 94025, USA}
\author{A.~Masegian}
\affiliation{Kavli Institute for Cosmological Physics, University of Chicago, Chicago, IL 60637, USA}
\author{J.~Elvin-Poole}
\affiliation{Center for Cosmology and Astro-Particle Physics, The Ohio State University, Columbus, OH 43210, USA}
\affiliation{Department of Physics, The Ohio State University, Columbus, OH 43210, USA}
\author{S.~Allam}
\affiliation{Fermi National Accelerator Laboratory, P. O. Box 500, Batavia, IL 60510, USA}
\author{G.~M.~Bernstein}
\affiliation{Department of Physics and Astronomy, University of Pennsylvania, Philadelphia, PA 19104, USA}
% ...

\collaboration{147}{(DES Collaboration)\vspace{-0.175in}}




\begin{abstract}

We describe an updated calibration and diagnostic framework, \balrog{}, used to directly sample the selection and photometric biases of \response{the} Dark Energy Survey (DES) Year 3 (Y3) dataset. We systematically inject onto the single-epoch images of a random 20\% subset of the DES footprint an ensemble of nearly 30 million realistic galaxy models derived from DES Deep Field observations. These augmented images are analyzed in parallel with the original data to automatically inherit measurement systematics that are often too difficult to capture with generative models. The resulting object catalog is a Monte Carlo sampling of the DES transfer function and is used as a powerful diagnostic and calibration tool for a variety of DES Y3 science, particularly for the calibration of the photometric redshifts of distant ``source'' galaxies and magnification biases of nearer ``lens'' galaxies. The recovered \balrog{} injections are shown to closely match the photometric property distributions of the Y3 GOLD catalog, particularly in color, and capture the number density fluctuations from observing conditions of the real data within 1\% for a typical galaxy sample. We find that Y3 colors are extremely well calibrated, typically within ${\sim}$1-8 millimagnitudes, but for a small subset of objects we detect significant magnitude biases correlated with large overestimates of the injected object size due to proximity effects and blending. We discuss approaches to extend the current methodology to capture more aspects of the transfer function and reach full coverage of the survey footprint for future analyses.

\end{abstract}






\section{Introduction}

Wide-field imaging surveys have revolutionized modern astronomy. Some of the primary science goals of these projects are to extract precise constraints on cosmological models and galaxy evolution using measurements made from hundreds of millions of galaxies for ongoing surveys such as the Dark Energy Survey\footnote{\url{https://www.darkenergysurvey.org/}} (DES; \citealt{des-2005}), the Kilo Degree Survey\footnote{\url{http://kids.strw.leidenuniv.nl/}} (KiDS; \citealt{kids_overview}), and the Hyper Suprime-Cam Survey\footnote{\url{https://www.naoj.org/Projects/HSC/}} (HSC; \citealt{hsc_overview}), and even billions of sources for upcoming Stage IV experiments such as Euclid \citep{euclid_overview} and the Vera C. Rubin Observatory Legacy Survey of Space and Time (LSST; \citealt{lsst_overview}). For the largest surveys, the resulting constraints have become so precise that percent-level spatial variations in the survey's depth can cause biases that dominate over the statistical errors (see for instance \citealt{huterer_2006,wiggleZselection,2012MNRAS.424..564R,2016ApJS..226...24L,mitigating-lss-contam}). Small biases -- as small as one part in $10^4$ in some cases -- in the measurements of sizes, shapes, and fluxes of sources can have similarly important impact on the science results \citep{2013MNRAS.429..661M}.

The cumulative effect of the many selection effects and measurement biases of an astronomical survey is captured by its \textit{transfer function}. This function maps how the photometric properties of astronomical sources are distorted by real physical processes such as interstellar extinction or by our imperfect measurements at every step from detector calibration to object catalog creation. As most cosmological measurements from survey data are based on the same processed images and source catalogs, this mapping is crucial for accurately estimating the true cosmic signals imprinted on the sky such as the spatial clustering of galaxies (see \citealt{clustering_1984,clustering_sdss,clustering_y1} for a few examples) and weak lensing of galaxy light profiles by the intervening matter field (similarly, see \citealt{wl_1996, wl_overview, wl_y1}).

Unfortunately, many of these effects are in practice difficult to characterize or even identify. For example, the object catalogs derived from survey images are produced by a complex process\response{; c}alibration, detection, measurement, and validation involve a number of nonlinear transformations, thresholds applied to noisy quantities, and post-facto cuts made on the basis of human judgment. Despite significant efforts to characterize some of these effects in the past (see \citealt{connolly_2010} and \citealt{chang_2015} for the LSST and DES pipelines respectively), this complexity makes each contribution to the transfer function extremely difficult to model --  and even small errors in the estimated survey completeness can substantially bias measurements such as the amplitude of galaxy clustering or important calibration efforts like the photometric redshift inference of weak lensing samples \citep{sdssDR8,2013MNRAS.429..661M,2016MNRAS.455.3943H,kidsSimulation}.

Simulating the survey data from scratch can accurately capture some, but not all, of this complexity. Spatial variations in the effective survey completeness depend not just on the observing conditions but also on the ensemble properties of the stars and galaxies being studied. Systematic errors in the sky background estimation and biases in the measurements of galaxy and stellar properties can couple to fluctuations in the galaxy density field, leading to a completeness that depends on the signal being measured. Finally, there is a wide variety of non-astrophysical features that can affect the measurement quality and completeness such as artificial satellite trails, pixel saturation, or the diffraction spikes of bright stars. Not only are these effects difficult to model or simulate at high fidelity, but attempts to do so can introduce model-misspecification bias which can underestimate the true uncertainty in the downstream fitted photometric parameters (\citealt{misspecification, misspecification-shear}).

In contrast, injecting artificial sources directly into the real images can naturally capture many of these effects. Synthetic objects added to the real data automatically inherit the background and noise in the images as well as the biases arising from measurement in proximity to their real counterparts. Injecting realistic star and galaxy populations, convolving their light profiles with an accurate model for the point-spread function (PSF), and applying accurate models for effects not directly probed (such as Galactic reddening and variable atmospheric transparency) results in a population of simulated sources that inherits the same completeness variations and measurement biases as the real data. Mock catalogs made in this way can be used to discover, diagnose, and derive corrections for systematic errors and selection biases at high precision.

\response{Injection simulations of this kind have been used for limited calibration studies of detection efficiency and photometric calibration in the presence of realistic noise and crowded fields since at least the mid-1980's (\citealt{cluster-e3,palomar-5,Stetson87}), not long after the widespread adoption of charge-coupled devices (CCDs) in astronomical imaging. There is a rich history of mixing real and synthetic data to estimate the detection efficiency of an apparatus in hybrid Monte Carlo techniques commonly used in particle physics measurements \citep{hybrid-monte-carlo}. In addition, there have been recent examples to improve blinding procedures for rare events such as embedding fake gravitational wave signals (``hardware injections'') into the Laser Interferometer Gravitational-wave Observatory (LIGO; \citealt{ligo-overview}) data and similarly ``salting'' the data taken by the Large  Underground  Xenon (LUX; \citealt{lux-overview}) experiment with artificial events to test the robustness of their detection pipelines and guard against confirmation bias (\citealt{ligo-fakes,lux-salt} respectively).}

However, generating full-scale mocks via injection is computationally demanding for a modern wide-field (WF) galaxy survey. The injection simulations described in \cite{Suchyta_2016} for the early releases of DES data did not attempt to pass the injected galaxies through every part of the measurement process, opting to inject only onto the coadd images. The SynPipe package \citep{Huang_2018} has been used to characterize measurement biases for the HSC pipeline and includes single-epoch processing, but only on a small fraction of the survey's available imaging. The \code{Obiwan} tool, currently developed to model completeness variations for the Dark Energy Spectroscopic Instrument (DESI: \citealt{desi}), also has incorporated single-epoch processing but focuses only on the emission-line galaxies that are the primary DESI targets \response{and, so far, has only injected sources within 0.2 magnitudes of their used selection cuts for increased efficiency} \citep{obiwan}. Despite injection pipelines having shown great promise, the \response{extremely high computational cost (in addition to the difficulty in distinguishing intrinsic methodological uncertainties in their sampling of the transfer function from actual measurement biases) has, until now, largely relegated them to proof-of-concept measurements rather than being used to directly calibrate the cosmological measurements from WF surveys.}

This paper describes the generation of the \balrog{}\footnote{\balrog{} is \textit{not} an acronym. The software was born out of the original authors delving ``too greedily and too deep'' \citep{fellowship} into their data, hence the name.} injection simulations for the first three years of DES data (referred to as Y3), covering a randomly selected 20\% of the total Y3 footprint. Sources drawn from DECam \citep{decam} measurements of the DES Deep Fields (DF) \citep*{y3-deepfields} are self-consistently added to the single-epoch DES images which are then coadded and processed through the full detection and measurement pipeline. This extensive simulation and reduction effort allows us to characterize, in detail, the selection and measurement biases of DES photometric and morphological measurements as well as the variation of those functions across the survey footprint. In addition, using an input catalog with measurements from the same filters as the data resolves many of the issues in capturing the same photometric distributions as real DES objects seen in \cite{Suchyta_2016} -- particularly for color. The resulting catalogs generally follow completeness and measurement bias variations in DES catalogs to high accuracy, with mean color biases of a few millimagnitudes and number density fluctuations varying with survey properties within 1\% for a typical cosmology sample.

As the measurement pipelines for the DES DF and WF data are complex and quite technical, so too are parts of this paper. However, we also motivate interesting science cases for the presented response catalogs for both calibration and direct measurement purposes including the photometric redshift calibration of weak lensing samples, magnification effects on lens samples, and the impact of undetected sources on image noise. For readers more interested in using \balrog{} for potential science applications or as a general diagnostic tool, this is discussed in detail in Sections \ref{photometric-results} and \ref{applications}.

\begin{figure*}[ht!]
\centering
\includegraphics[width=1.0\textwidth]{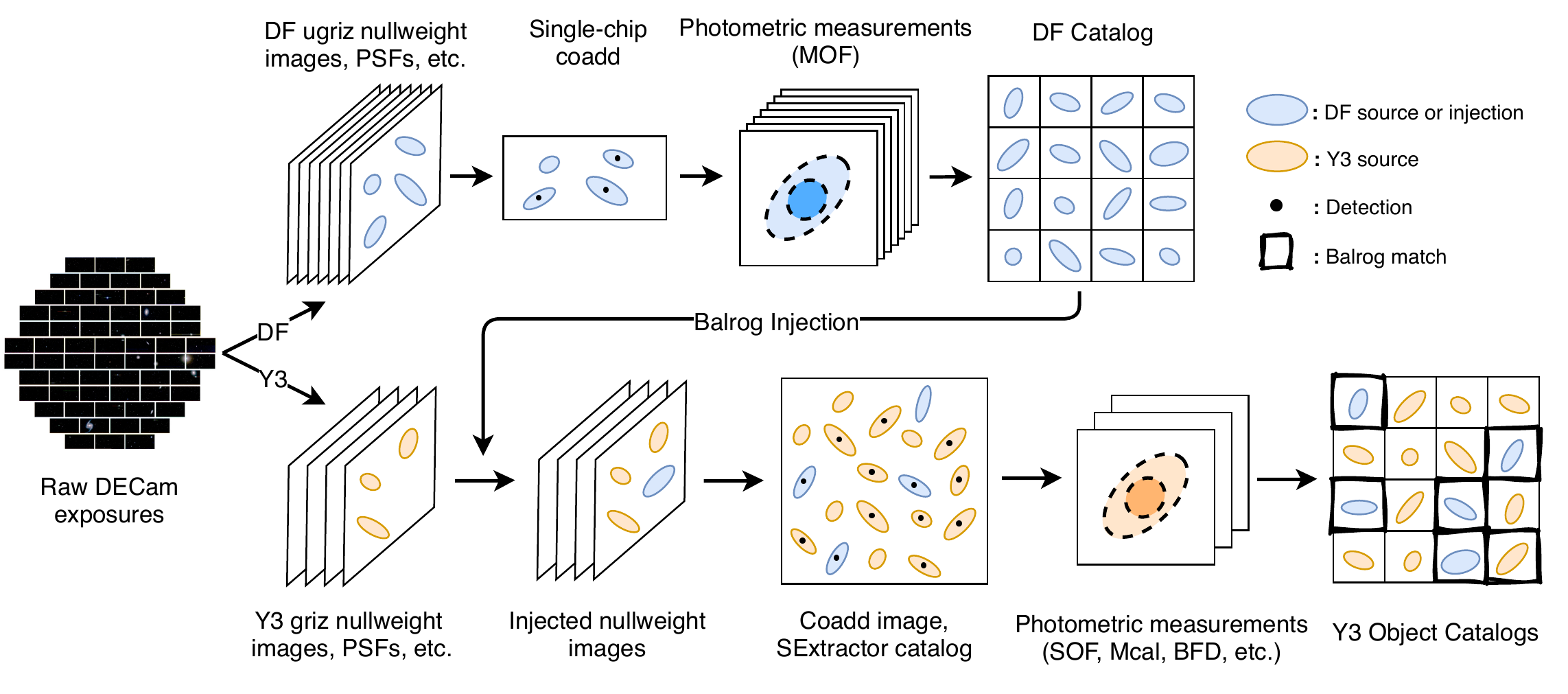}
\caption{A high-level overview of how the Deep Fields (DF) and Y3 image processing pipelines interact to create the \protect{\balrog{}} catalogs. The raw DECam exposures are used as the basis for both tracts, with the much deeper DF data being represented by the larger image stacks. The null-weight images, weight maps, PSF models, and zero point solutions are computed from the raw exposures after calibrations are applied and are the starting point of the sampled transfer function. The DF exposures are not dithered and thus single-CCD coadds are created in place of the much larger Y3 coadds. The fiducial DF catalog is created by fitting CModel profiles to detections with Multi-Object Fitting (MOF), which simultaneously models the light profiles of detected neighbors. These fitted model profiles (after a few limited cuts discussed in Section \protect{\ref{injection-strategy}}) are used as the \protect{\balrog{}} injection catalog which are added to the Y3 null-weight images directly. Afterwards, the injected null-weight images are processed in a nearly identical way to the real images including coaddition, detection, and photometric measurements. Finally, we match the output object catalog to truth tables containing the injected positions. As all sources are remeasured, there is some ambiguity in the matching; this is discussed further in Section \protect{\ref{matching}}. See \protect{\cite*{y3-deepfields}} and \protect{\cite{des-image-pipeline}} for further DF and Y3 pipeline details respectively.}
\label{fig:pipeline-diagram}
\end{figure*}

This paper is organized as follows: In Section \ref{balrog-pipeline} we introduce the significantly updated \balrog{} pipeline which now emulates more of the DES measurement stack and uses a completely new injection framework for source embedding into single-epoch images. Section \ref{balrog-in-y3} describes the injection samples and methodological choices for the Y3 \balrog{} simulations including a new scheme for handling ambiguous matches. In Section \ref{photometric-results} we compare the recovered \balrog{} samples to the fiducial Y3 object catalog (Y3 GOLD; see \cite{y3-gold} for details), as well as present the photometric response of the main star and galaxy samples. We also examine the performance of a typical Y3 GOLD star-galaxy separation estimator and investigate a set of catastrophic photometric modeling failures that enter science samples with dramatically overestimated fluxes (sometimes by multiple orders of magnitude). We then discuss novel applications of an injection catalog in cosmological analyses in Section \ref{applications} including the photometric redshift calibration of Y3 ``source'' galaxies and the effect of magnification on ``lens'' galaxy samples -- in addition to a few unexpected discoveries such as noise from undetected sources and issues with background subtraction. Finally, we close in Section \ref{discussion} with a discussion of our results, methodological limitations, and future directions before concluding remarks in Section \ref{summary}.

\section{The \Balrog{} Pipeline}\label{balrog-pipeline}

\balrog{} was introduced in \cite{Suchyta_2016} as a software package\footnote{\url{https://github.com/emhuff/Balrog}} that injects synthetic astronomical source profiles into existing DES coadd images to capture realistic selection effects and measurement biases for the Science Verification (SV) and Year 1 (Y1) analyses. However, as the precision of the subsequent DES cosmological analyses has increased, so too has the need for even more robust systematics control and more precise characterization of the survey transfer function. The main limitations of the original methodology were that (1) injections into the coadd rather than single-epoch images skip many important aspects of the measurement pipeline whose effects we want to capture, and (2) the injected objects were drawn from fitted templates to sources in the space-based Cosmological Evolution Survey (COSMOS: \citealt{cosmos}) rather than measurements consistent with DECam filters -- introducing discrepancies in the recovered colors. While the latter is solved by using the new Y3 DF catalog \citep*{y3-deepfields}, the former required significant additional complexity in the simulation framework to consistently inject objects across all exposures and bands.

To address this, we have developed a completely new software framework that is described and validated in the remainder of this section. An overview of the Y3 \balrog{} process is shown in Figure \ref{fig:pipeline-diagram}, with simplified summaries of the DF and Y3+\balrog{} measurement pipelines. Briefly, we use the significantly deeper DECam measurements of sources in the DES DF as a realistic ensemble of low-noise objects to inject into the Y3 calibrated single-epoch images. We then rerun the DES measurement pipeline on the injected images to produce new object catalogs that contain the \balrog{} injections. Finally, we match the resulting catalogs to truth tables containing the injection positions to provide a mapping of DF truth to WF measured properties.

All astronomical image injection pipelines such as \balrog{} have two distinct elements: emulation of a survey's measurement pipeline and source injection into the processed images. As our methodology for the former is intrinsically specific to DES while the latter is a fairly generic problem, development on the new Y3 \balrog{} was split into the two corresponding pieces discussed in detail in Sections \ref{desdm-emulation} and \ref{inj-framework} below.

\subsection{DESDM Pipeline Emulation}\label{desdm-emulation}

The DES survey data are processed through a set of pipelines by the DES Data Management team (DESDM) which perform basic astronomical image processing as well as applying state-of-the-art galaxy fitting, PSF estimation, and shear measurement codes. The standard processing steps applied to the DES Y3 data are described in detail in \citet{des-image-pipeline}. Ideally, to ensure that identical codes and versions were used at each stage of processing, one would implement \balrog{} as part of the standard data reduction. However, this was not an option for DES Y3 as the updated \balrog{} methodology did not exist until after the Y3 data were completely processed (this is now true for a future Year 6 (Y6) \balrog{} analysis as well). Therefore it was necessary to replicate the DESDM processing pipeline stack as closely as possible. While this usually amounted to calling the relevant codes and scripts with identical configurations and software stack components, sometimes minor changes were required due to differences in computing environments or practical considerations such as processing time. These differences will be noted whenever relevant.

A modular design for the measurement pipeline\footnote{\url{https://github.com/kuropat/DES\_Balrog\_pipeline}} was chosen both for ease of testing and for the ability to do non-standard production runs (see sections \ref{magnification} and \ref{undetected-sources} for examples). The individual \balrog{} processing stages for a single DES coadd tile (44\arcmin{}$\times$44\arcmin{}) are as follows:  

\begin{enumerate}\setcounter{enumii}{4}
  \item \textbf{Database query \& null-weighting} -- Find all single-epoch \textit{immasked} (the DES designation for flattened, sky subtracted, and masked) images in the $griz$ bands that overlap the given DES Y3 tile. Download all exposures, PSFs, photometric and astrometric solutions from the DESDM Y3 processing archive. A masking process called ``null-weighting'' is applied to these immasked images which sets weights of pixels with certain flagged features (e.g. cosmic rays) to 0. These null-weight images are the starting point of the later injection step.
    \item \textbf{Base coaddition \& detection} -- \label{step-base} Remake the tile coadds from the single-epoch exposures with no objects injected using \code{SWarp} \citep{swarp} and the detection catalogs with \sextractor{} \citep{sextractor}. Construct Multi-Epoch Data Structure (MEDS; \citealt{meds}) files with cutouts of the coadd and single-epoch images used for additional photometric measurement codes. This allows us to cross-check our measured catalogs with Y3 GOLD to ensure that we recover the same detections and base photometry, as well as easily investigate proximity effects on the injections. Can be skipped to save processing time if desired.
    \item \label{step-first} \textbf{Injection} -- Consistently add input objects in all relevant exposures and bands using the local PSF model in each exposure with corrections to the flux from the image zeropoints and local extinction -- along with any other desired modifications such as an applied shear or magnification. This is discussed in detail in Section \ref{inj-framework}.
    \item \textbf{Coaddition \& detection} -- Same as \ref{step-base} but with the injected null-weight images. The resulting photometric catalogs contain existing real objects, injections, new spurious detections, and blends between the two.
    \item \textbf{Single-Object Fitting (SOF)} -- Fit a composite bulge + disk model that is the sum of an exponential and a de Vaucouleurs profile (CModel) to every source, while masking nearby sources.
    \item \textbf{Multi-Object Fitting (MOF)} -- Fit sources with CModel, but group nearby detections into friends-of-friends (FOF) groups that have all of their properties fit iteratively to account for proximity effects. Only available for some \balrog{} runs due to its computational expense.  
    \item \textbf{Metacalibration} -- Fit a simple Gaussian profile to detections and then remeasure after applying four artificial shears \citep{metacal-practice}. This is useful for the creation of weak lensing samples where correcting for shear-dependent systematics is more important than absolute flux calibration \citep{metacal-theory}.
    \item \textbf{Gaussian APerture (GAp) fluxes} -- Fit a robust, scale-length-independent alternative to model-fitted photometry. Object flux is calculated within a Gaussian-weighted aperture with full-width at half-maximum (FWHM) of 4\arcsec{}. Described further in Section \ref{matching}.
    \item \textbf{Bayesian Fourier Domain (BFD)} -- Estimate the shear of sources without explicitly fitting a shape using the methodology described in \cite{BA2014}. Available only for a few specialized runs.
    \item \label{step-last} \textbf{Match and compute GOLD value-adds} -- Match input injections to output detections while accounting for ambiguous matches (see Section \ref{matching}). Merge truth and measured table quantities. Compute Y3 GOLD value-added quantities including flags, object classifiers, masks, and magnitude corrections (though only the dereddening component is used for \balrog{} magnitude corrections; see \S \ref{desdm-differences}).
\end{enumerate}

The resulting photometric catalogs of measured \balrog{} sources can then be used to measure the DES wide-field response of various input quantities or used directly as randoms with realistic selection effects (see \citealt{Suchyta_2016} and \citealt{obiwan} for examples). In addition, an ``injection catalog'' is created which contains information for all injected sources, detected or not, for investigations into detection and completeness properties. The emulation steps \ref*{step-first} through \ref*{step-last} can be repeated for multiple injection realizations of a given tile to obtain sufficient sampling for the needed science case. However, as discussed in Section \ref{balrog-in-y3}, for Y3 analyses we opted for a single realization with relatively high injection density due to the large computational cost of each realization.

\subsubsection{Differences from the DESDM Pipeline}\label{desdm-differences}
While \balrog{} strives to emulate the DESDM pipeline from null-weight images to science catalogs at high fidelity, there are some discrepancies due to practical limitations. The most significant are:
\begin{itemize}

    \item {\bf Reuse of existing single-epoch images, PSF models, photometric zeropoints, and WCS (astrometric solution):} \response{The injected fluxes of sources from the input catalog are modified only to account for an image's photometric zeropoint and the local extinction}. Due to this we do not recalculate the photometric and astrometric \response{calibrations or PSF estimate} for any exposures which have additional objects added to them; the Y3 DESDM solution is carried forward unchanged. This means that we cannot probe the individual systematic error contributions of steps in the DESDM pipeline before this stage, such as \response{biases in} the PSF modeling or image detrending.
    
    \item {\bf Incomplete \sextractor{} parameter list:} We chose to measure only a subset of the Y3 \sextractor{} parameters that were anticipated to be important for downstream analyses in order to save processing time. In particular, we did not compute any model-fitted magnitudes including \code{MAG\_PSF} which is needed for the WAVG quantities described in \cite{des-image-pipeline}. Ultimately, the overall time saved was small and we plan to save all \sextractor{} quantities for future runs.
    
    \item {\bf MOF is skipped for the cosmology sample:} While MOF photometry is available for the Y3 GOLD catalog, most Y3 cosmological analyses use the variant SOF which skips the multi-object deblending step in favor of masking neighbors. This approach is significantly faster, fails less often, and has negligible impact in photometric performance (E. Sheldon, private communication). As MOF is not needed for Y3 cosmology calibration and contributed roughly a quarter of all \balrog{} runtime (see Table \ref{tab:runtimes}), we elected to skip this step for the main samples.
    
    \item {\bf Zeropoint and chromatic corrections are not applied:} The Y3 photometric calibration introduces new chromatic corrections that achieve sub-percent uniformity in magnitude by accounting for differences in response arising from varying observing conditions and differences in object SEDs (see \citealt{y3-gold}). \response{However, the mean of all Y3 GOLD chromatic corrections are between 0.1 and 0.4 millimagnitudes (mmag) for all but $g$-band (0.9 mmag)}. As this is a subdominant effect that requires significant computation to correct in each injection realization, we do not account for these corrections before injecting into images. In addition, the SED-independent ``gray'' corrections that account for variations in sky transparency and instrumentation issues like shutter timing errors were not accounted for in the injection zeropoints. This was not intentional and will be included in all future \balrog{} runs. However, these corrections are also quite small, with the mean Y3 GOLD gray zeropoint correction \response{between -1 and 1 mmag for all bands}. As we do not modulate the truth fluxes with these corrections during injection, it is not necessary to apply these corrections \textit{after} measurement either.
    
\begin{figure*}[ht!]
\centering
\includegraphics[width=0.9\textwidth]{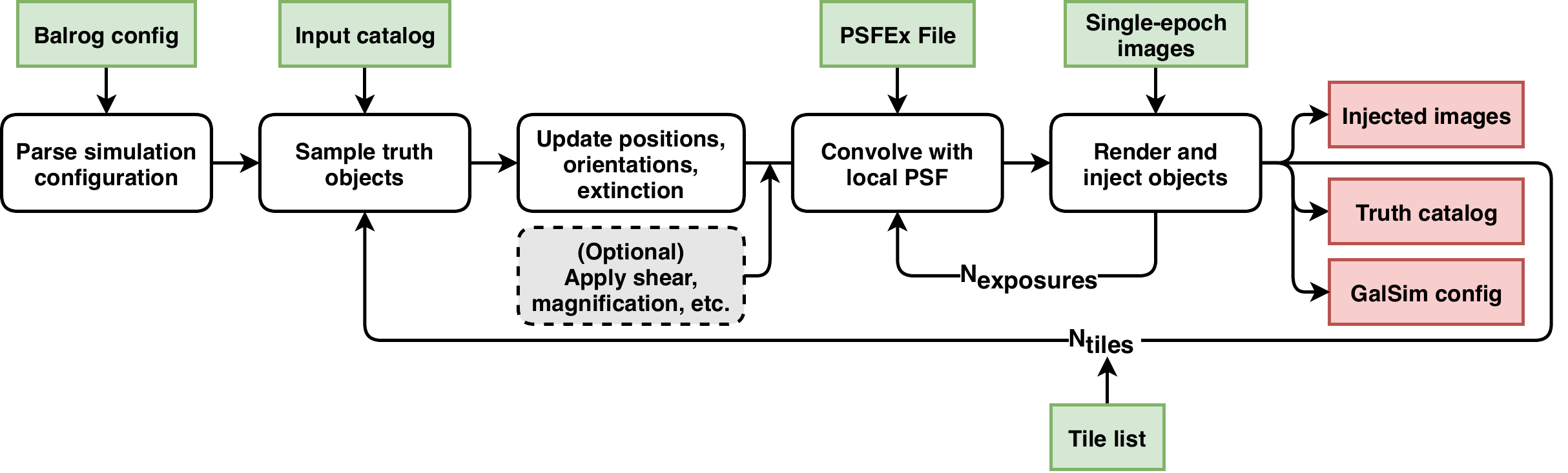}
\caption{High-level overview of the injection processing for a single realization. Green boxes are inputs to the injection framework while red boxes are outputs. The length of each loop is determined by the number of exposures and tiles considered in the full simulation. While the main runs used for Y3 cosmology calibration modify only the position, orientation, and flux normalization of the truth inputs, there are many optional transformations that can be applied such as a constant shear or magnification. The main output of our injection package is a multi-document configuration file with detailed injection specifications that is then executed by \galsim{}, with each step being executed in the physically correct order. Additional realizations replicate all steps, other than the initial configuration parsing, and produce unique outputs. }
\label{fig:inj-diagram}
\end{figure*}
    
    \item {\bf Partial GOLD Catalog Creation}: Due to the staged approach in the creation of Y3 GOLD with value-added products being incorporated as they were being developed, the exact same procedure for compiling the \balrog{} catalog could not be followed strictly as it would have produced an unnecessary and severe overhead in the production time. Scripts that approximately replicate this process were provided by DESDM, though they only reproduce the columns that were deemed to be most relevant to Y3 key science goals. Slight modifications had to be made to quantities such as \code{FLAGS\_GOLD} and the object classifier \code{EXTENDED\_CLASS\_SOF} where the required MOF columns were not available; these differences are mentioned when relevant throughout the paper.

\end{itemize}

While not technically a difference in the \textit{pipeline} emulation itself, we note here that PSF models used for injections (PSFEx; \citealt{psfex}) were found to be slightly too large in \cite{y1-wl} for bright stars in Y1 due to the brighter-fatter effect (see \citealt{brighter-fatter}). However, we still used PSFEx for our injection PSFs as the new Y3 PIFF PSF model described in \cite{y3-piff} was not yet implemented into the \galsim{} configuration structure that was required for our injection design, which is discussed below.

\subsection{Injection Framework}\label{inj-framework}

As mentioned in the beginning of this section, incorporating single-epoch injection into \balrog{} required a new software design to handle the significant increase in simulation complexity beyond what was done in \cite{Suchyta_2016} for the SV and Y1 analyses. Development on the injection framework was partitioned into its own software package\footnote{\url{https://github.com/sweverett/Balrog-GalSim}} as the injection step is fairly generic and of potential interest to other analyses outside of DES Y3 projects -- as well as upcoming Stage IV dark energy experiments such as LSST. Briefly, our injection framework maps high-level simulation choices into individual object and image-level details consistent between all single-epoch images for the simulation toolkit \galsim{} \citep{galsim} to process. With this design, \balrog{} automatically inherits much of the modularity, diverse run options, and extensive validation of \galsim{}. A schematic overview of the injection process is shown in \autoref{fig:inj-diagram}. \response{The remainder of this section will quickly summarize the most relevant aspects of each step; we leave a more detailed description of the implementation details as well as a description of the most important user options for this new software package in Appendix \ref{appendix:injection}.}

\subsubsection{Injection Configuration}\label{inj-config}

The \balrog{} configuration serves as the foundation for the final, much larger \galsim{} configuration file produced for each tile by the injection pipeline which follows the \galsim{} configuration conventions that are extensively documented\footnote{\url{https://github.com/GalSim-developers/GalSim/wiki/Config-Documentation}}. Global simulation parameters that apply to all injections are defined here such as the input object type(s) (see \S \ref{appendix:input-sample}), position sampling method, injection density, and number of injection realizations. During injection processing, the requisite simulation details needed to inject the sampled input objects consistently across the relevant survey images are appended to this file to create a multi-document \galsim{} configuration file with each document corresponding to a single CCD exposure. \response{An example configuration that was used for the two main cosmology runs is given in Appendix \ref{appendix:config}.}

\subsubsection{Input Sample and \response{Object Profiles}}\label{input-sample}

\response{While any native \galsim{} input type can be used for the simulations, most \balrog{} runs sample objects from an existing catalog with parametric properties that describe the flux and morphology of each source. The photometric measurements of the DF catalog, as well as most measurements in Y3 DES WF science catalogs, are based on Gaussian mixture model fits to various profiles by \code{ngmix}\footnote{\url{https://github.com/esheldon/ngmix}} introduced in \cite{Sheldon_2014} and most recently updated in \cite{y3-gold}. Each profile parameterization is converted to a sum of \galsim{} \code{Gaussian} objects that represent the Gaussian components used in the original fit. \balrog{} can currently inject the following \ngmix{} model types: a single Gaussian (\code{gauss}), a composite model (CModel; \code{cm}) first introduced in SDSS\footnote{\url{https://www.sdss.org/dr12/algorithms/magnitudes/\#cmodel}} which is a linear combination of an exponential disk and a central bulge described by a de Vaucouleurs’ profile \citep{devacouleurs}, and a slightly simpler CModel with fixed size ratio between the two components (\code{bdf}, for Bulge-Disk with Fixed scale ratio). In DES Y3, the DF measurements use \code{bdf} profiles while the WF uses \code{cm}.}

\response{See \S \ref{appendix:input-sample} for all provided custom input types, including the option to inject the ``postage stamp'' image cutouts of objects in MEDS files. While using the actual images of DF sources rather than parametric fits to their profiles would be a more accurate representation of the true distribution of galaxy properties and morphologies, there are significant added complexities due to adding artificial noise from stamps with larger associated PSFs than the injection image and ensuring stamp and mask fidelity of the full DF catalog; these issues are discussed in detail in Section \ref{discussion}.}

\subsubsection{Updating Truth Properties and Optional Transformations}\label{updating-truth}

\response{Measurements of the transfer function with \balrog{} require truth tables that compile the properties of injected objects. For injections that are based off of real sources, some of these object properties are modified to fit the needs of the simulation such as the positions, orientations, and fluxes. The updated source properties either replace their original columns in the output truth catalogs or are appended as new columns. Object fluxes are scaled to account for interstellar extinction and to match the photometric zeropoint of each single-epoch injection image. Additional transformations such as a constant shear or magnification factor can be applied depending on the desired science case (see Section \ref{magnification} for an example using magnification in Y3).}

\response{The position sampling of injections depends on the desired science case; uniform sampling naturally allows for \balrog{} objects to be used directly as randoms for galaxy clustering calibration, but overlapping \balrog{} injections can artificially inflate the inferred blending rate. Alternatively, a hexagonal lattice is more appropriate for a perturbative sampling of the transfer function at a given position, but this embeds an unrealistic (though correctable) clustering signal at small scales. The available options are described in \S \ref{appendix:updating-truth} and the trade-offs are discussed in more detail in Section \ref{injection-strategy}.}

\subsubsection{PSF Convolution}

\response{The PSF used for each object is determined by the local single-epoch PSFEx solution at the injection position. Simpler PSF models are also allowed for testing purposes but not recommended for science runs.}

\subsubsection{Object Rendering and Injection}\label{rendering-and-inj}

All of the previous simulation choices are ultimately encoded in a detailed configuration file that is structured to be read by \galsim{}. This design was chosen over explicit use of the software's \python{} API as the configs facilitate easily reproducible simulations and allow for runs that are identical except for minor modifications such as an added constant magnification factor. Each transformation from truth property to pixel value is automatically handled by \galsim{} processing in the physically correct order. After an object stamp is rendered \response{(including Poisson noise from the new source)}, its pixels are summed with the initial image while ignoring any part of the profile that may go off image. Rarely a profile will require an extremely large grid for the fast Fourier transform (FFT) during PSF convolution and exceed available memory. To avoid this, we set a maximum grid length of 16,384 pix$^{-1}$ (or ${\sim}63,000$ arcsec$^{-1}$ for DES) per side and skip objects that exceed this limit. While the injection framework was designed with flexibility in mind for uses outside of the Y3 cosmology science goals (and even DES itself), there are currently some assumptions made about the structure of the input data to emulate DES Y3 that we plan on generalizing in upcoming releases.

\subsection{Pipeline Validation}\label{validation}

As \balrog{} is a non-generative, or discriminative, model of the transfer function, it is difficult to disentangle any intrinsic errors in the input sample or survey pipeline emulation from actual systematic effects we are trying to characterize -- particularly since \balrog{} was run independently of DESDM processing for Y3. Therefore a series of increasingly complex test runs were completed in order to validate both the injection and emulation steps and characterize the pipeline fidelity at a detailed level. We initially ran \balrog{} with the injection step turned off to confirm that we recovered identical detection and photometry catalogs as Y3 GOLD when carefully accounting for the same random seeds in the fitters that were used in nominal Y3 processing. Once this was achieved, we verified that the injected profiles of objects drawn onto blank images matched single-object renderings made independently of the pipeline.

We then ran a series of tests where we ignored the existing survey image data during injection except for the estimated residual local sky background that is automatically subtracted from the exposures later in the pipeline. Objects were placed on a sparse grid to limit proximity effects from other injections with two types of noise depending on the run -- either only Poisson noise for the injections or Poisson in addition to low levels of zero-mean Gaussian background sky noise. These blank image runs became progressively more complex as we added the features used in the main science runs described in Section \ref{balrog-in-y3} and acted as a form of regression testing.

\begin{figure*}
\centering 
\subfloat[]{%
  \includegraphics[width=0.45\textwidth]{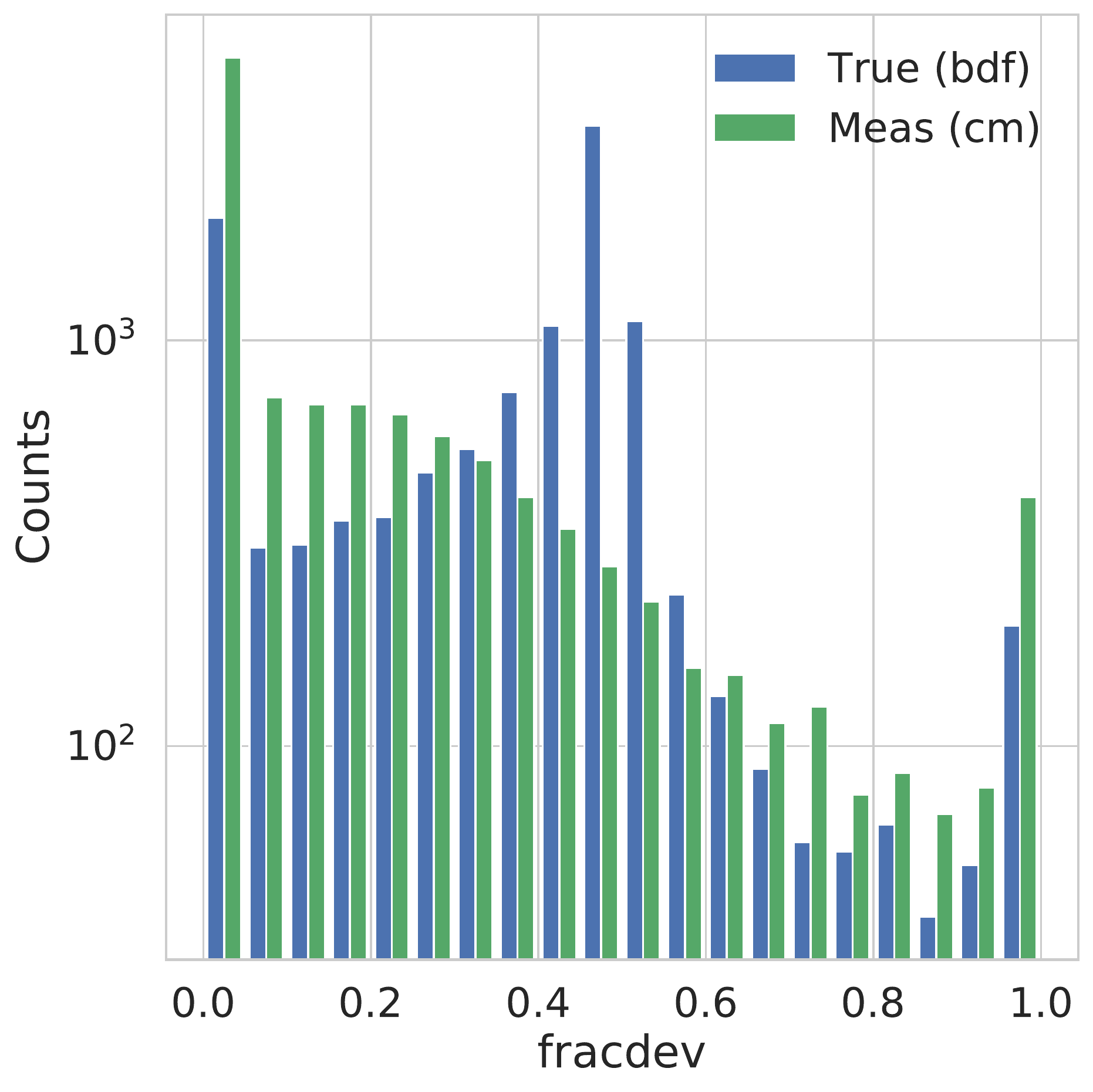}%
  \label{fig:grid1}%
}\hspace{.07\textwidth}%
\subfloat[]{%
  \includegraphics[width=0.45\textwidth]{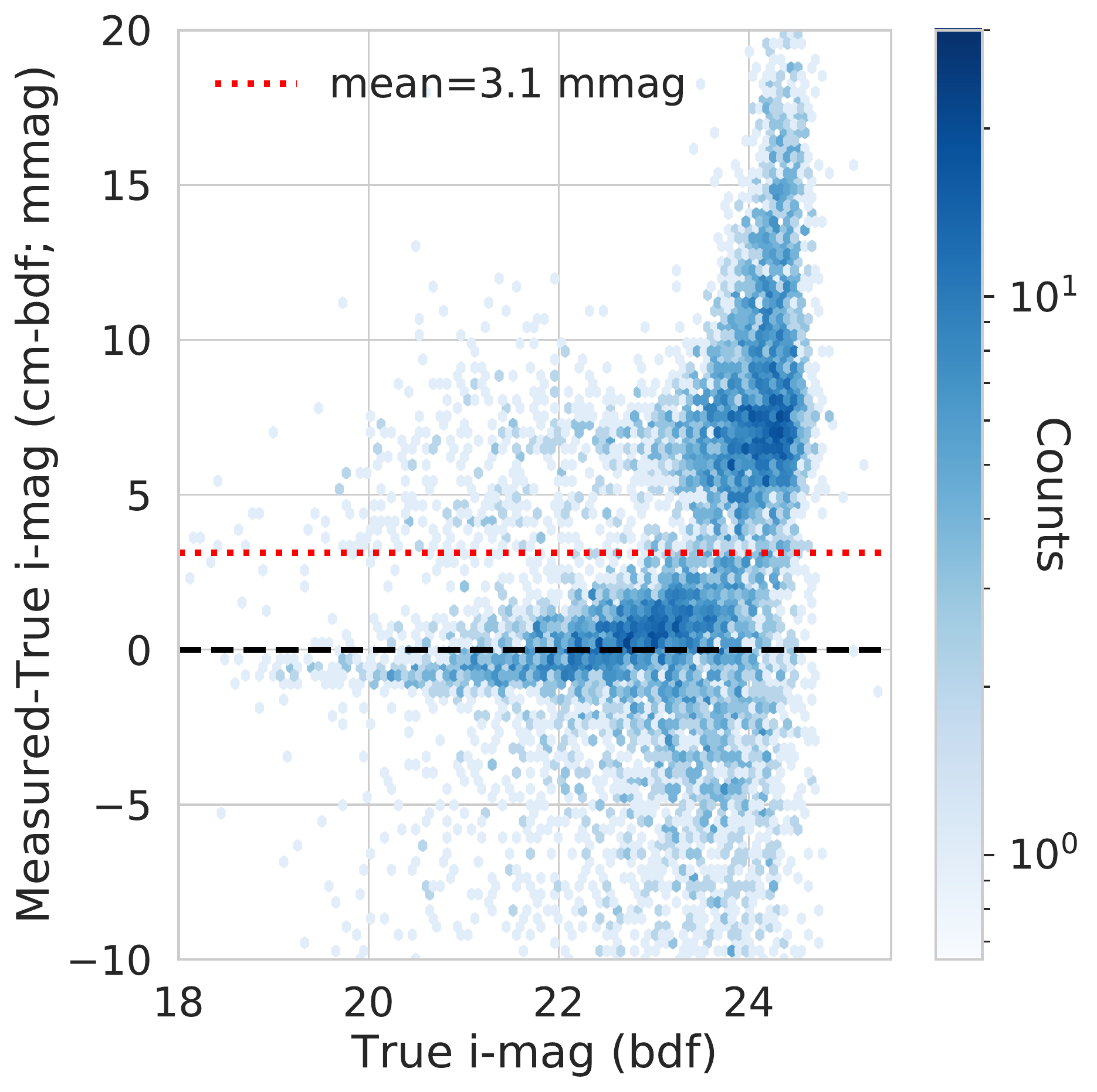}%
  \label{fig:grid2}%
}\\
\subfloat[]{%
  \includegraphics[width=0.45\textwidth]{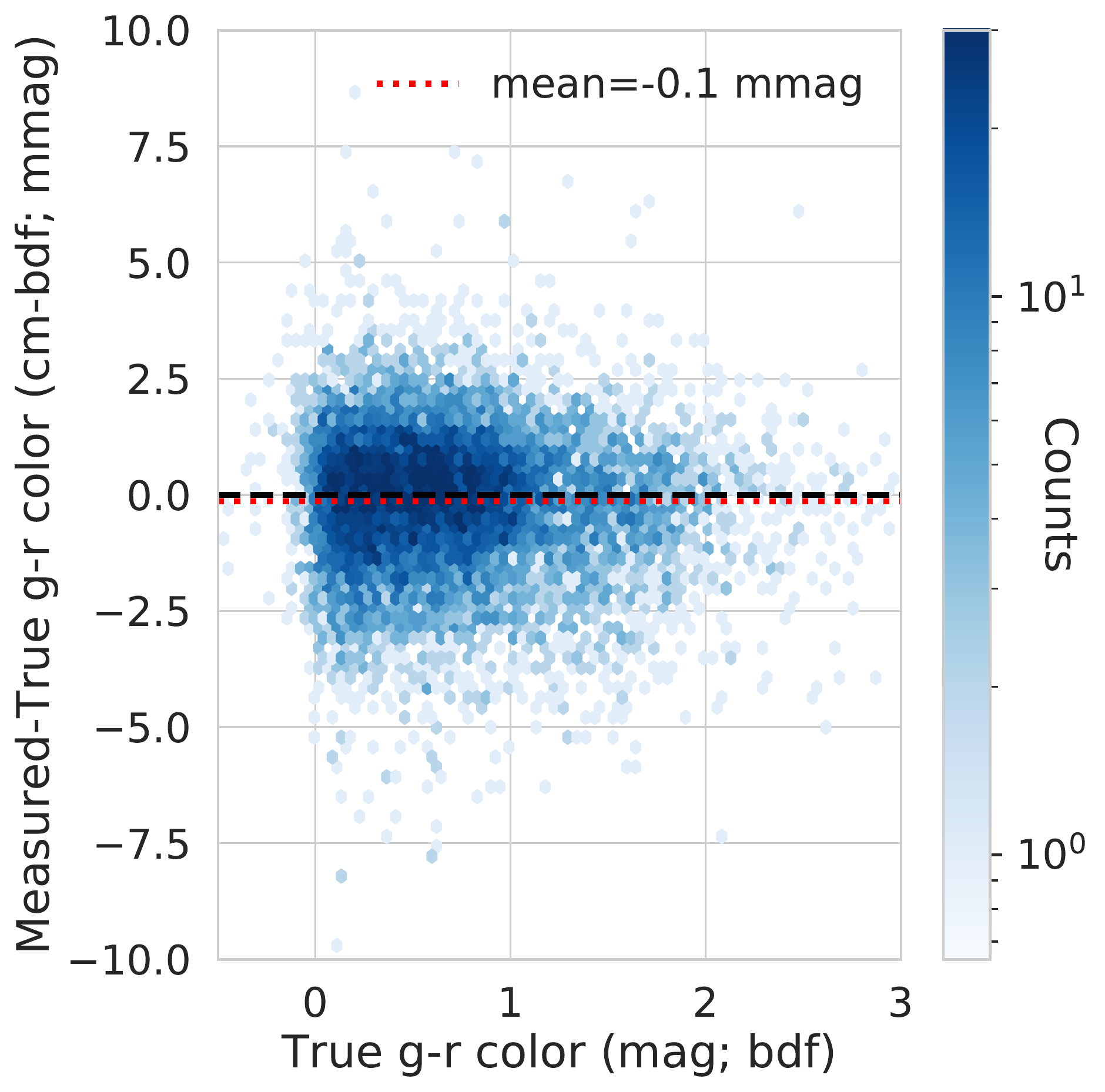}%
  \label{fig:grid3}%
}\hspace{.05\textwidth}%
\subfloat[]{%
  \includegraphics[width=0.45\textwidth]{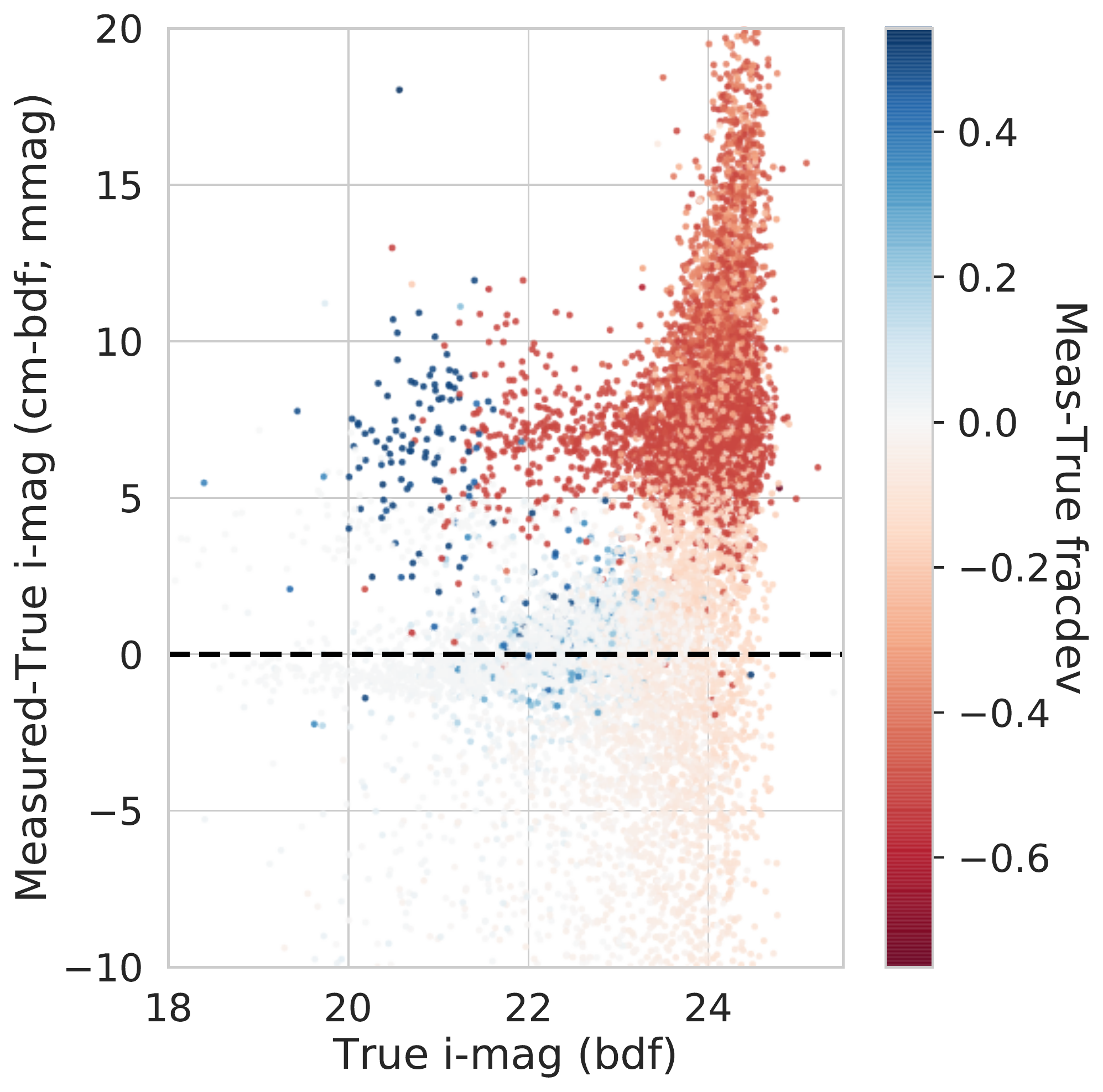}%
  \label{fig:grid4}%
}
\caption{A series of plots highlighting aspects of the noiseless blank image test described in Section \protect{\ref{validation}}. (\protect{\subref*{fig:grid1}}) The first panel shows the difference in input \code{bdf\_fracdev} vs measured \code{cm\_fracdev} for detected objects. The additional peak at 0.5 for \code{bdf\_fracdev} is a result of the slightly different model definition; for \code{bdf}, the relative size ratio between the bulge and disk components is forced to be 1. This constraint does not exist for \code{cm} and thus it has a different prior on the parameter.
(\protect{\subref*{fig:grid2}}) This panel shows the $i$-band magnitude response of these objects, where there are clearly two different populations. The first is well-calibrated with the majority of detections well within +/-2.5 mmag of truth. The second population is biased towards fainter measurements by ${\sim}$7.5 mmag on average.
(\protect{\subref*{fig:grid3}}) The $g-r$ color response for these objects. The bias in recovered magnitude is nearly identical in $griz$ and so does not translate to the recovered colors. The mean color response for $g-r$, $r-i$, and $i-z$ is 0.1, 0.3, and 0.2 mmag respectively.
(\protect{\subref*{fig:grid4}}) The final panel shows that the biased magnitude population is a result of injections with input \code{bdf\_fracdev}${\sim}0.5$ scattering to 0 or 1 to match the expected \code{cm\_fracdev} prior. As we do not believe this differential response to be of physical origin, it contributes to a lower bound on the precision in which \balrog{} can calibrate $\Delta\text{mag}$ -- though importantly this does not contribute a bias to recovered colors.}
\label{fig:grid-test-plots}
\end{figure*}
 
These tests are relevant for more than pipeline validation; effects from methodological choices can also be identified and quantified while working in a simplified environment. As an example, the runs with only Poisson noise indicated that there were two subgroups of objects with statistically significant differences in magnitude response -- one was well calibrated, the other with a mean offset of ${\sim}$7.5 mmag too faint in each of $griz$. This was ultimately discovered to be a result of different priors used for the parameter that measures the relative flux ratio between the de Vaucouleurs and exponential component, \code{fracdev}, for the \code{ngmix} profile type used to fit DF objects (\code{bdf}) and the one used to fit wide-field measurements (\code{cm}). A series of plots that show the difference in input vs. measured \code{fracdev} and examples of its downstream effect on the recovered magnitude and color responses for this test is shown in Figure \ref{fig:grid-test-plots}. 

The impact of the different \code{fracdev} fits on the magnitude response can be seen clearly in Figure \ref{fig:grid4}, where the difference in measured vs. true $i$-band magnitude as a function of injected magnitude is colored by the response in \code{fracdev} for a single tile. As the difference in profile definition between \code{cm} and \code{bdf} is largely due to fitting stability and has little to do with the true distribution of galaxy properties, this effectively puts a lower bound on the accuracy of the mean magnitude response that we are able to measure with \balrog{} when using the DF sample as inputs at around 3 mmag. Importantly, however, the effect is nearly identical in each of the $griz$ bands and has negligible impact in the recovery of colors, as seen in Figure \ref{fig:grid3}. This example highlights some of the difficulties in choosing a ``truth'' definition for injections based on model fits and the importance of carefully testing the impacts of model assumptions.

\begin{figure}
 \centering
 \includegraphics[width=.4725\textwidth]{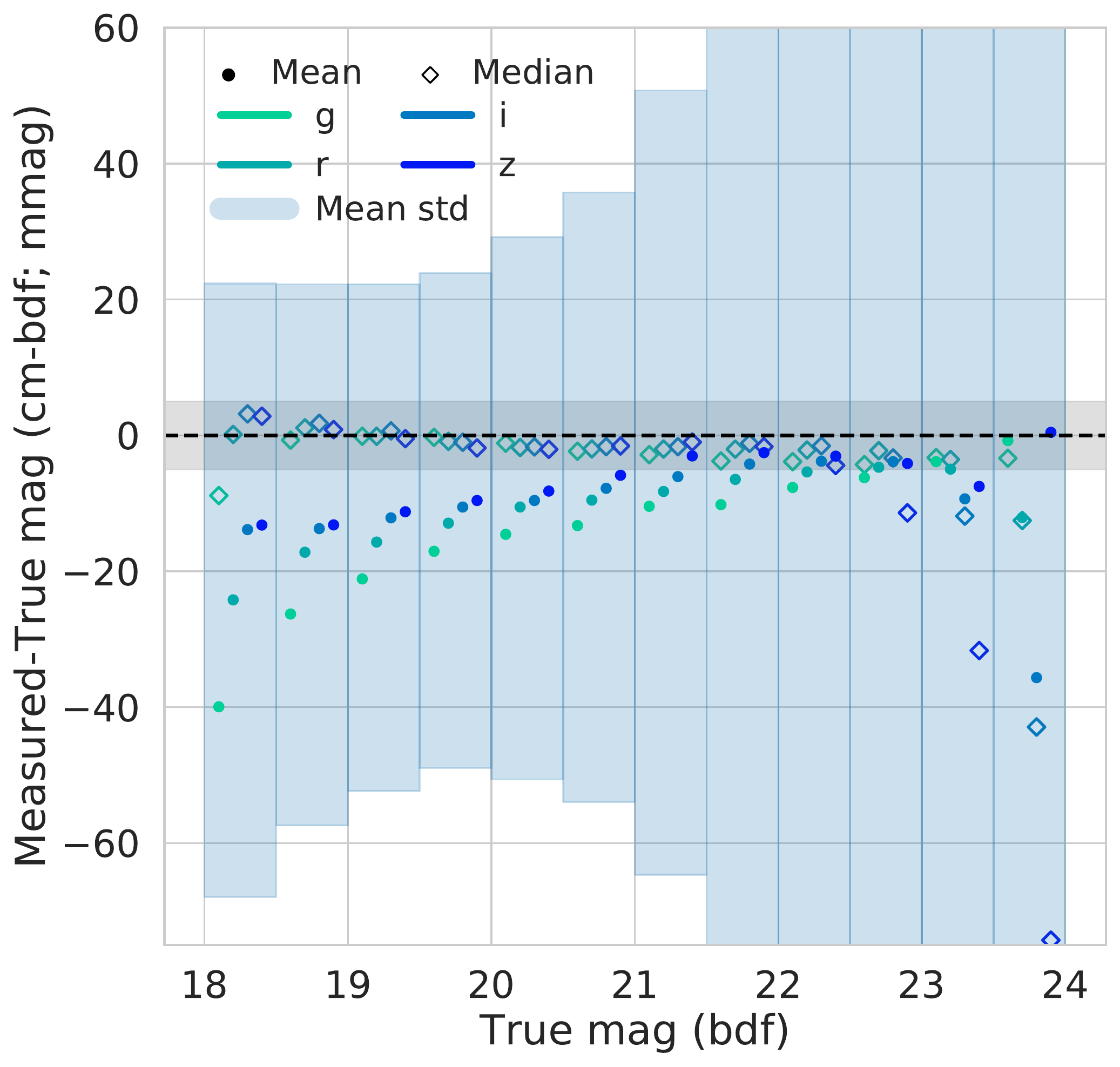}
 \caption{The mean (solid circle) and median (hollow diamond) difference in measured vs. injected magnitude (\mdmgrid{}) as a function of input magnitude for the final blank image runs with zero-mean Gaussian background noise. The vertical bars correspond to the mean of the standard deviations of $griz$ magnitude responses in each truth magnitude bin, centered at the mean magnitude response. The vertical bars represent the average of the standard deviations of $griz$ magnitude responses in each bin of size 0.5 magnitudes, centered at the mean magnitude response. The overall calibration is excellent, with the median response less than 5 mmag in all bins except for $g<18.5$ and $22.5<z<23$. We expect significant biases past magnitude 23 due to selection effects near the detection threshold. However, the mean responses show some bias -- particularly on the bright end. As discussed in the text, this is due to an asymmetric tendency for SOF to measure the fluxes of bright, extended galaxies to be too large when neighbors are contained in the object's MEDS stamps. The errors in \mdmgrid{} do not substantially decrease past input magnitudes of 20 for the same reason. This is discussed in greater detail in \S \ref{results-galaxies-mags}.}
 \label{fig:grid-test-mag-response}
\end{figure}

The final version of the blank image test was performed with identical input and configuration to that used to produce the fiducial Y3 catalogs across 200 tiles which contain over 2.3 million injections and 1.6 million detections. Zero-mean Gaussian background noise was applied to the blank images with variance set to the \response{corresponding} CCD \code{SKYVAR} value. The resulting object responses allow us to characterize the baseline performance of the photometric pipeline in ideal (though overly simplistic) conditions which in turn may provide lower limits on the intrinsic uncertainty in our sampling of the DES transfer function. The mean and median difference in recovered versus injected magnitude for $griz$ is plotted in Figure \ref{fig:grid-test-mag-response}. The vertical bars correspond to the mean of the standard deviations of $griz$ magnitude responses in each truth magnitude bin, centered at the mean magnitude response.

The medians are extremely well calibrated, with only  $g < 18.5$ and $22.5 < z < 23$ off by more than 5 mmag, or 0.45\%, through 23rd magnitude where selection effects near the detection threshold become significant. The mean responses are consistently biased towards larger recovered flux on the bright end by ${\sim}15$ mmag due to the asymmetric tendency of SOF to measure the sizes of bright, extended objects to be too large in the presence of neighbors; this is a real effect seen in the main data runs and is discussed in greater detail in \S \ref{results-galaxies-mags}. Such biases are not seen in isolated SOF measurements of similar objects (E. Sheldon, private communication) and appear in this test as it was inefficient to use a grid size large enough to keep all other grid injections outside the MEDS stamps of the largest injections. This effect also keeps the magnitude error from decreasing as the intrinsic brightness increases as one would naively expect. While the magnitude bias induced by the difference in the \code{cm} vs. \code{bdf} profile definition is present in this measurement, it is negligible compared to proximity biases for extended sources and selection effects present in the noisier images.

Importantly, there is no significant band-dependence in the median magnitude responses where the recovered sample is complete, with a typical spread in median $griz$ biases of ${\sim}3$ mmag for truth magnitudes ranging from 18.5 to 22 with no characteristic shape or distribution systematics. While there is a detectable band-dependence in the mean magnitude responses, it is nearly eliminated when binned in signal-to-noise (S/N) instead of magnitude to account for differences in sky noise.

\section{\Balrog{} in DES Year 3}\label{balrog-in-y3}

We describe here the injection samples, pipeline settings, and matching choices used to create the Y3 \balrog{} data products for the photometric performance characterization described in Section \ref{photometric-results} and downstream science calibrations described in Section \ref{applications}. For Y3, we ran \balrog{} several times with different configurations for various validation and science cases. These runs are tabulated in Table \ref{tab:runlist} which lists the following quantities: the run name, the number of simulated tiles, the total number of injected objects, the fraction of detected objects, the mean number of times a given object is injected across all tiles, the spacing between injections, and the magnitude limit used for sampling. As detection in DES is based on a composite $riz$ detection coadd, we emulate the detection magnitude by averaging the dereddened $riz$ fluxes of the injections.
 
\begin{table*}
\begin{center}
\begin{tabular}{lrrcrccl}
\toprule
\footnotesize
  Run Name & Tiles & N Det & Det-Frac & $\langle$Inj's$\rangle$ & Mag Lim & Spacing & Notes \\
  \midrule
   \runtwo{} & 1544 & 7.4 M & 0.369 & 16& 25.4 & 20\asec{} & \response{Deepest, main run used for Y3 cosmology} \\
   \runtwoa{} &  497 & 3.9 M & 0.600 & 9& 24.5 & 20\asec{} & \response{Shallower, auxiliary run used for Y3 cosmology} \\
   \runtwomag{} & 155 & 0.8 M& 0.463 & 2 & 25.4 & 20\asec{} & 2\% magnification on \runtwo{} objects \\
   \runtwoamag{} & 497 & 3.9 M & 0.607  & 9 & 24.5 & 20\asec{} & 2\% magnification on \runtwoa{} objects \\
   \gridtest{} & 200 & 1.6 M& 0.702 & 3& 24.5 & 20\asec{} & Inject \response{onto} blank images with noise \response{(see \S \ref{validation})} \\
    \noiselessgridtest{} & 196 & 2.6 M & 0.997 & 4 & 24.5 & 20\asec{} & Same as \gridtest{}, but without noise \response{(see \S \ref{validation})} \\
   \code{clusters} & 901 & 39.9 M & 0.930 & 163& 23.0 & 10\asec{} & Tiles containing rich galaxy clusters \response{(see \S \ref{clusters})} \\ 
   \code{blank-sky} & 88 & -- & --- & --- & --- & 20\asec{} & Injected zero-flux objects \response{(see \S \ref{undetected-sources})} \\
   \bottomrule
\end{tabular}
\caption{\response{A list} of Y3 \balrog{} runs and associated parameters: the number of tiles sampled, the number of total detections (N Det), the detection fraction (Det-Frac), the mean number of injections per unique DF object ($\langle$Inj's$\rangle$), the composite $riz$ detection magnitude limit, and injection lattice spacing.} 


\label{tab:runlist}
\end{center}
\end{table*}

\begin{figure}
    \includegraphics[width=0.46\textwidth]{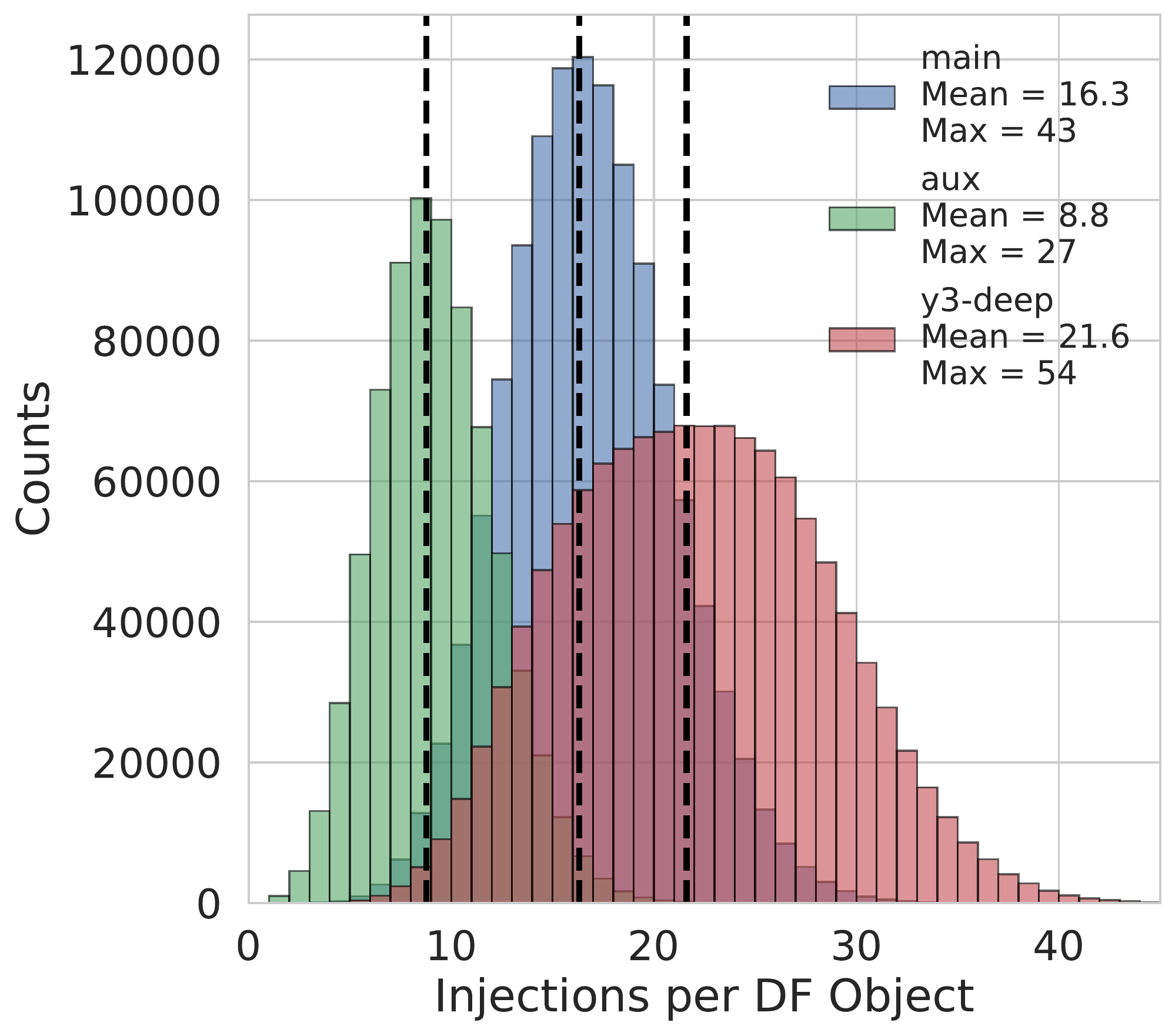}
    \caption{The number of injections per unique DF object for \runtwo{} in blue, \runtwoa{} in green, and their combination \dfsample{} in red. The mean number of injections per run is shown with dashed vertical lines and is stated along with the maximum number of injection realizations. \runtwo{} is composed of 1,544 tiles vs. only 497 for \runtwoa{}, but has a larger input catalog to sample due to the more conservative composite \protect{$riz$} detection magnitude of 25.4 vs. 24.5 for \runtwoa{}. The resulting combination is no longer a Poisson distribution, but this can be accounted for in downstream analyses \response{by using the column \code{injection\_counts} for building a weighting scheme; see \cite*{y3-sompz} for an example}. The typical \protect{\balrog{}} object in \dfsample{} has just over 20 unique injection realizations across the sampled footprint.}
    \label{fig:injection-counts}
\end{figure}

\begin{figure*}
 \centering
 \includegraphics[width=0.975\textwidth]{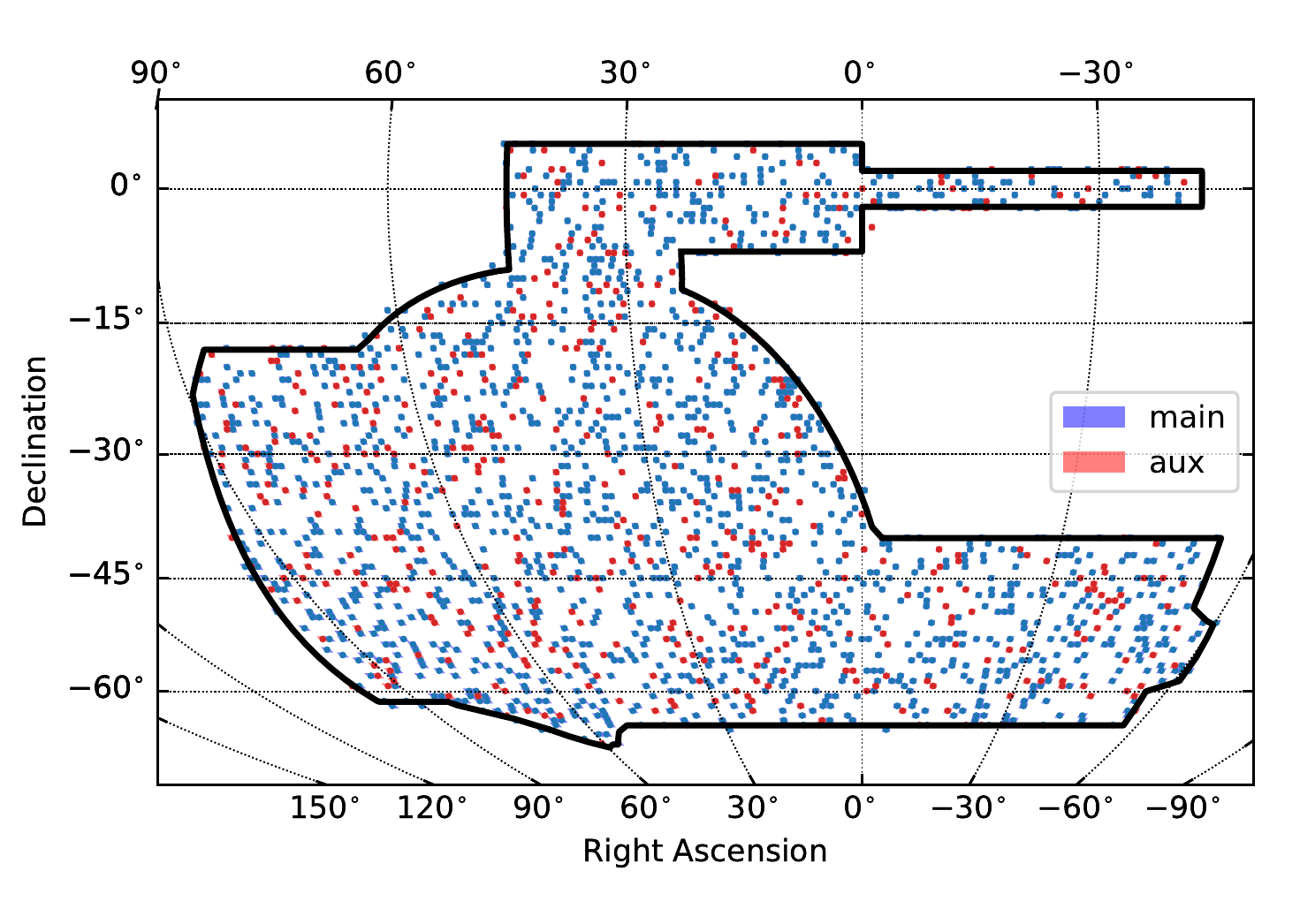}
 \caption{The spatial distribution of randomly sampled DES tiles used for \balrog{} injections. 1,544 \runtwo{} and 497 \runtwoa{} tiles are shown in blue and red respectively. The outline of the DES Y3 footprint is shown in black. Some tiles are slightly outside of the official footprint due to partial image coverage from DECam observations on the footprint edge.}
 \label{fig:footprint}
 \end{figure*}

The \response{primary} runs used for cosmological analyses are called \runtwo{} and \runtwoa{} \response{(auxiliary)}. The former samples the transfer function across 1,544 randomly chosen tiles (of the 10,338 Y3 tiles) to a detection magnitude limit of 25.4. This limit was chosen to capture DF objects that had at least a 1\% chance of being detected as measured from a 200 tile test run. \response{The latter,} \runtwoa{}, was a supplemental run at a shallower limiting magnitude of 24.5 across 497 tiles to increase the fraction of recovered injections for analyses that needed a larger total sample. These runs are combined for the fiducial \balrog{} catalogs \dfsample{} and \starsample{} which are described in upcoming sections. The distributions of the number of injection realizations per input object for these runs are shown in Figure \ref{fig:injection-counts}, and the spatial distribution of these tiles are shown compared to the full DES footprint in Figure \ref{fig:footprint}.  \runtwomag{} and \runtwoamag{} are identical to the above runs except for a constant added magnification of \response{$\mu\sim0.02$ for a limited subset of tiles}; these are described in more detail in Section \ref{magnification}. The \gridtest{} and \noiselessgridtest{} runs were used for the validation tests shown in \ref{validation}. The \code{blank-sky} and \code{clusters} runs were conducted separately from the main cosmology runs in order to facilitate two of the science cases discussed in Sections \ref{undetected-sources} and \ref{clusters} respectively.

The processing was done on a dedicated compute cluster at Fermilab, ``DEgrid'', consisting of 3000 cores with 6-8GB RAM per core available. The typical core and memory provisioning along with wall-clock running times for each stage of the pipeline is given in Table \ref{tab:runtimes}. MOF is not used for the fiducial Y3 cosmology analyses and so is excluded for \runtwo{} and \runtwoa{} -- along with their corresponding magnification runs. We include the estimated computational cost to show the difficulty in scaling this methodology to full footprint coverage and WF density; we discuss this more in Section \ref{discussion}. All output measurement catalogs were archived including the MEDS cutout images of detected objects; the injected single-epoch images and resulting coadds were only saved for validation runs.

A few additional post-processing steps were required to match changes made to the Y3 object catalogs after the fiducial GOLD catalog creation. These consisted of a correction to the metacalibration signal-to-noise (S/N) column, redefining the \code{size\_ratio} quantity from \code{mcal\_T\_r / psfrec\_T} to \code{mcal\_T\_r / mcal\_Tpsf}, and adding a shear weight to each of the metacalibration measurements \response{for the photometric redshift calibration detailed in Section \ref{photo-z-calibration}.}

\begin{table*}
\begin{center}
\hspace{-0.05\textwidth}
\begin{tabular}{lrrr}
\toprule
\footnotesize
  Stage & Cores & RAM & Clocktime \\
  \midrule
   Database Query & 1 & 64 GB & 2.0 hr \\
   Base Coaddition/Detection/MEDS  & 4 & 64 GB & 3.0 hr\\
   Injection & 16 & 64 GB & 3.0 hr \\
   Coaddition/Detection/MEDS &4 & 64 GB& 5.0 hr  \\
   MOF$^*$ &32 & 256 GB & 6.5 hr  \\
   SOF &16 & 64 GB & 1.5 hr \\
   Metacalibration & 8& 320 GB& 2.5 hr \\
   Match/Merge/Flag & 2 & 512 GB & 1 hr\\
   \midrule
   Total/tile & 16-32 & 64-512 GB & $18-24.5$ hr/tile \\
   \bottomrule
   \end{tabular}
\caption{Approximate \balrog{} stage run times and memory allocations per tile. $^*$As MOF is not used in the fiducial Y3 cosmology analysis, this step was only run for \response{preliminary tests} due to the long clocktime. The two total reported clocktimes are with MOF excluded or included in the pipeline emulation respectively.}
\label{tab:runtimes}
\end{center}
\end{table*}

\subsection{Input Deep Field Catalog for \dfsample{}}

The majority of Y3 \balrog{} analyses use injections drawn from DECam measurements of objects in the DF described in \cite*{y3-deepfields}. In brief, this catalog of nearly 3 million sources is assembled from hundreds of repeated exposures of three DES supernovae (SN) fields and the COSMOS field. The corresponding deep single-CCD coadds have S/N of ${\sim}\sqrt{10}$ times their WF counterparts and thus provide a good sample of low-noise sources to draw from for explorations of systematics in the WF measurements. There are multiple versions of the DF catalogs that provide trade-offs in the average seeing quality vs. the maximum depth. In Y3 \balrog{}, we use \code{COADD\_TRUTH} as it strikes a balance between using observations with 10 times the mean WF exposure time while ensuring that the composite DF FWHM be no worse than the median single-epoch FWHM in the WF for each of the injection bands.

We emphasize that we are not injecting the actual \textit{images} of DF galaxies but instead take the MOF \ngmix{} parameterized model fit to each detection and generate an idealized galaxy profile based on those model parameters (with added Poissonian noise). The injection framework described in Section \ref{balrog-pipeline} is capable of injecting the MEDS stamps directly which in principle would account for additional diversity in galaxy morphologies and eliminate any model bias compared to the true distribution of galaxy properties. However, this requires extensive validation of the DF stamps before injection and introduces additional complications due to image masks and added noise for injections into CCDs with better seeing than the DF composite image. We plan to revisit these issues for \balrog{} in the Y6 methodology.

The DF catalog is comprised of model fits that are very similar to the WF CModel with two major differences: the two components (bulge + disk) are fit simultaneously rather than separately, and the ratio of the size of each component, \code{TdByTe}, is fixed to be 1. While this was chosen for increased fitting stability for the fainter DF sources, fixing the relative bulge-disk size ratio reduces the total number of free parameters in the model by one and significantly changes the distribution in the relative flux fraction \code{fracdev} (recall Section \ref{validation} for how this impacts the corresponding recovered CModel photometry in idealized conditions). Ultimately, any photometry can be used for the injection truth as long as it is an unbiased estimate of the real distribution of object properties. The \code{bdf} profile will be used for all Y6 DES source fitting and for Y6 \balrog{} -- avoiding the small systematic difference in magnitudes between \code{cm} and \code{bdf}.

\subsubsection{DF Object Extinction}

The DF catalog has detailed photometric corrections to the fluxes including for extinction as described in \cite*{y3-deepfields}. However, these corrections were not yet ready when \balrog{} began the cosmology runs. Thus in order to accurately account for variations in DF extinction, as well as extinction variations among tiles in the Y3 survey footprint, we enacted the following procedure to deredden the DF input objects and then re-extinct them by an appropriate amount in the injection WF tile: For the DF objects, we sample the \cite{Schlegel_1998} extinction maps at five points (center and corners) in each input DF CCD (of size 9\amin{}$\times$18\amin{}) and record the average of the 5 E(B-V) values. We also record the five-point average of E(B-V) for the larger (size 44\amin{}$\times$44\amin{}) WF tiles. During injection, we deredden each object by the DF recorded value for its CCD of origin and apply the mean extinction value for the WF injection tile. This chip and tile-level correction is simple to implement and distorts the overall magnitude and color distribution of the DF galaxy sample from the cosmic average only slightly. However, we plan on implementing per-object extinction corrections in the Y6 methodology. The used dereddening and extinction values are preserved in the injection truth tables for later flux and magnitude corrections to enable consistent comparisons between true and measured quantities.

\subsection{Input Star Sample for \texorpdfstring{\starsample{}}{delta-Stars}}\label{star-sample}

While the majority (${\sim} 90$\%)\footnote{Most tiles were run with a 9-1 ratio between input catalogs, but the first 152 tiles of \runtwo{} were run with an 8-2 ratio.} of the injections are sources (both stars and galaxies) from the DES DF, ${\sim} 10$\% of injections are simulated stars. Other than characterizing the photometric response of stars in DES with nearly no galaxy contamination (see Section \ref{results-stars}), the \starsample{} sample is useful for quantifying the baseline performance of the DESDM pipeline for the simplest morphologies. This allows us to isolate the more complex model fitting issues for the heterogeneous \dfsample{} sample.

The morphologies are modeled as pure delta ($\delta$) functions convolved with the local PSFEx solution used during injection. The magnitude and color distributions are based on the local stellar population in each of the 10,338 tiles in the Y3 footprint. For example, areas of the survey with higher stellar density near the galactic plane received more bright stars than areas toward the south galactic pole in the center of the footprint. To represent color distributions fainter than the WF limit of $i{\sim} 24$, the color distribution near $i{\sim} 24$ was extended by two magnitudes to $i{\sim} 26$ using models of the Galactic disk and halo \citep{BLR18}. The simulated star catalog has already been corrected for extinction, so no other preprocessing is required. The measurement pipeline has no knowledge of the difference in input star/galaxy classification and returns the same CModel fits as \dfsample{}.

\subsection{Object Classification and Differences in Measurement Likelihood}\label{different-likelihoods}

While we expect \dfsample{} and \starsample{} will be used for calibration of DES galaxy and stellar systematics respectively, there are additional star injections in \dfsample{} as it draws from all sources in the DF that pass quality cuts. Sources in the DF catalog have been classified with a k-nearest neighbor algorithm\footnote{This classifier was added after the \balrog{} runs completed, and so is not included as one of the truth columns. It has to be matched to the relevant Y3 DF catalogs.} trained on a subset of objects that have near-infrared (NIR) data from the UltraVISTA survey (\citealt*{y3-deepfields}; \citealt{ultravista}). The classifier's stellar sample is not perfectly complete from magnitudes $18 < i < 24$ (an average of 93\%), but its mean weighted purity is greater than 98\% over the same range. The requirement of successful detection and measured photometry for all $ugrizJHK$ bands reduces the total number of objects with classification by 44.5\%. The cut \texttt{NearestNeighbor\_class=2} selects this star sample while \texttt{NearestNeighbor\_class=1} will select the classified galaxies. The DF stars are not used in the analysis of the Y3 stellar photometric performance in this paper but are available if a larger sample is required for a given science case.  However, we do use these classifications when estimating the galaxy contamination in Y3 stellar samples in Section \ref{star-gal-sep}.

We note that there is a subtle difference in the measurement likelihoods corresponding to each sample. The likelihood of the $\delta$-sample, $\mathcal{L_\text{star}^\delta}$, assumes perfect classification knowledge and is given by
\begin{align}\label{eq:star-likelihood-delta}
    \centering
    \mathcal{L}_\text{star}^{\delta}&=p(\bs{\theta}_\text{meas},\, c_\text{meas}|\bs{\theta}_\text{true},\, c_\text{true}\myeq\text{star}) \nonumber \\
    &=p(\bs{\theta}_\text{meas},\, c_\text{meas}|\bs{\theta}_\text{true}),
\end{align}

\noindent where $\bs{\theta}_\text{meas}$ and $\bs{\theta}_\text{true}$ are the measured and true objects' photometric parameters and $c_\text{meas}$ and $c_\text{true}$ are the corresponding object classifications. Alternatively, the likelihood of the DF star sample, $\mathcal{L}_\text{star}^\text{DF}$, accounts for the uncertainty in the truth classification:
\begin{equation}\label{eq:star-likelihood-df}
    \centering
    \mathcal{L}_\text{star}^\text{DF}=p(\bs{\theta}_\text{meas},\, c_\text{meas}|\bs{\theta}_\text{true},\, c_\text{true}).
\end{equation}

\noindent This becomes particularly relevant if one wants to combine results from Sections \ref{results-stars} and \ref{results-galaxies} for modeling errors of the composite sample. The needed conditional probabilities that capture the stellar efficiency and galaxy contamination of \dfsample{} can be derived from the results in Section \ref{star-gal-sep}.

\subsection{Sample Selection \& Injection Strategy}\label{injection-strategy}

While in principle we would randomly sample from all sources in the DF, there are some methodological and practical considerations that led to the following conservative cuts:

\begin{center}
\small
\begin{tabular}{ll}
     & \texttt{flags = 0} \\
    \texttt{AND} & \texttt{mask\_flags = 0} \\
    \texttt{AND} & \texttt{in\_VHS\_footprint} \\
    \texttt{AND} & \texttt{bdf\_T $<$ 100} \\
    \texttt{AND} & \texttt{bdf\_flux / bdf\_flux\_err > -3} \\
    \texttt{AND} & \texttt{bdf\_det\_mag < \{25.4, 24.5\}} \\
\end{tabular}
\end{center}

\noindent First, we eliminate any objects flagged with model fitting errors or in manually masked regions. We also require injections be from regions with external observations in the near-infrared (IR) as these IR bands are critical for the photometric redshift calibration (\ref{photo-z-calibration}). We restrict the characteristic size of the injections (\code{bdf\_T}) to be less than 100 arcsec$^2$ (corresponding to ${\sim}10$ arcsec) to reduce the rate of \balrog{}-\balrog{} blends and proximity effects on the injection grid -- though this selection may result in slightly over-sampling large, highly-elliptical galaxies. In addition, this choice may be in conflict with other potential science cases such as measuring the detection efficiency and photometric response of low-surface-brightness (LSB) galaxies \citep{low-sb}. Next, we remove objects with flux to error ratios of less than -3 in any band; this cut was needed after inspection of the DF catalog showed that there was an excess of objects with extremely negative flux values compared to WF measurements (though \ngmix{} fluxes are clipped below $10^{-3}$ when computing magnitudes).

Finally, we apply a detection magnitude limit of 25.4 to limit the time spent on injections that have almost no chance of being detected while still using a source catalog that is ${\sim}2$ magnitudes deeper than WF. As described in the beginning of Section \ref{balrog-in-y3}, this limit was derived from the mean dereddened $riz$ \code{bdf\_flux} of injections that had at least a 1\% chance of being detected during a 200 tile test of \runtwo{}. We do not consider the flux in $g$ in this calculation as it is not used in the detection image in DESDM processing. The \runtwoa{} limit of 24.5 was chosen based based on requirements for the lens magnification measurement detailed in \cite{y3-2x2ptmagnification} (and described further in Section \ref{magnification}). After making this selection, the DF injection catalogs used in \runtwo{} and \runtwoa{} have just over 1.23 million and 746,000 objects respectively.

The star catalog was sampled to its full depth of 27th magnitude in $g$ at a fraction of 10\% of the total objects injected into \runtwoa{} and (most) \runtwo{} tiles. No additional cuts were made. Since the relative contribution of Galactic stars to the total object count peaks at about 21st magnitude in a standard Y3 tile, these injections do not dominate the faint end of the distribution.

Choosing the injection density per realization is a trade-off between increasing the statistical power of the catalogs, reducing the rate of \balrog{}-\balrog{} blends, and reaching the desired footprint coverage given available computational resources. Ideally, we would measure the response of a single source added to DES images for a high number of realizations. As this is unfeasible we instead add objects on a hexagonal lattice with 20\asec{} spacing using a \code{MixedGrid} (see \S \ref{appendix:updating-truth}) for a single realization, corresponding to a density of ${\sim}7.8$ objects per arcmin$^2$ (or about 40\% of the total Y3 density). 

We can achieve a much higher injection density than that used in \cite{Suchyta_2016} as we do not randomly sample the positions which greatly reduces the self blending rate of injections. This is crucial as running a single \balrog{} tile realization in Y3 takes ${\sim} 40$ times longer than in SV and Y1 due to the increased complexity of the injection framework and additional photometric measurements. \response{While this does in principle limit the ability to use \balrog{} injections as randoms to measure clustering signals on scales at and below the grid size, this is currently well below the scale cuts of order 10\arcmin{} used in the Y3 analysis. In addition, we note that \balrog{} can still be used for studies of samples with intrinsic clustering by sub-sampling the full catalog of grid injections to match the desired clustering signal.}

However, this relatively high density could have significant implications for a non-local deblender like the one used in MOF. In early testing, we found that this level of injection density can sometimes lead to nearly all objects in a tile becoming a single MOF FOF group. Such non-local effects are less relevant for SOF except in cases where blends of other nearby injections with large, real sources may change how the masking of the blend is handled (or for extremely large injections that would be captured in the MEDS cutout of other injections, which is why we cut on the injection size). Dealing with non-local contributions to the measurement likelihood may be an important consideration for Y6 as the object detection threshold is lower and proximity effects are more of a concern.

\subsection{Blending and Ambiguous Matches} \label{matching}

An important caveat in using an object injection pipeline like \balrog{} is that there is often inherent ambiguity in the matching of the new object catalogs to the injections. Remeasurement on the injection images changes the number of detections and catalog ID assignments in unpredictable ways, and light profiles that were previously considered distinct detections can be blended together into single objects. While we will show that the fraction of ambiguous cases is relatively small at our injection density in DES images (${<} 1.5\%$) and can in principle be removed for our photometric tests, this ignores the increased shear noise and root mean square (RMS) of the measured ellipticity distribution for these objects which may be a dominant systematic for weak lensing measurements in deeper surveys like LSST \citep{Dawson_2015}. In addition, highly non-linear detection and photometry algorithms can often respond in unexpected ways to perturbations (particularly deblenders that are intrinsically non-local) which can lead to additional spurious detections and splitting of objects. As a rule: \textit{Any matched catalog from an injection pipeline has made assumptions about ambiguous matches and blending!} For these reasons, we save the full remeasured photometry catalogs so that different matching procedures can be applied depending on the desired science case. This is distinct from the approach in \cite{Suchyta_2016} which ran remeasurement in \sextractor{}'s association mode near injection positions.

\begin{figure}
    \centering
    \includegraphics[width=0.45\textwidth]{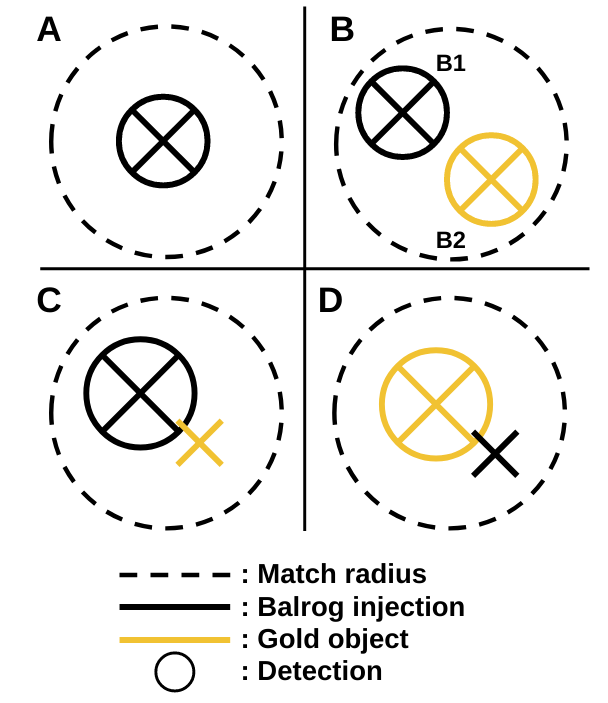}
    \caption[caption with a footnote]{An overview of how ambiguous matches can arise in the case of a two-object blend. A black cross mark denotes the position of a \balrog{} injection while a gold cross mark denotes the position of a Y3 GOLD detection. A circled cross mark indicates a detection in the \protect{\balrog{}} catalog while the dashed circle indicates the region inside of the search radius \protect{$r_2$}. Case (A) is by far the most common and is unambiguously a \protect{\balrog{}} injection. Case (B) has both the injection and the GOLD object detected within \protect{$r_2$} but is \textit{extremely} rare; in this case we select the closer detection. Cases (C) and (D) are true blends where there is ambiguity in whether to classify it as a \protect{\balrog{}} object with properties blended by the GOLD source or as a GOLD object that was blended by an injection. In this case we assign the object with the larger average \protect{$riz$} GAp flux as the antecedent. Only Case (D) is removed from the \balrog{} catalogs when applying a \protect{\code{match\_flag}} cut.}
    \label{fig:matching-diagram}
\end{figure}

However, it is useful to have a standard catalog sample with consistent matching for downstream cosmological analyses. Unless otherwise specified, Y3 analyses using \balrog{} catalogs use a catalog which applied the following matching prescription: We define the \textit{antecedent} of any blend as the ``brightest'' of the individual objects that contributes to it by some metric. Each blend thus comprises a noisy version of the antecedent as well as the non-detection of all other contributors to the blend. This approach gives a consistent and complete assignment of detection, non-detection, and antecedent to all objects of interest in the remeasured images and strikes the desired balance of including photometric scatter by blend contributors while excluding extreme outliers due to faint injections near existing bright objects. In addition, in the absence of measurement noise this scheme sets a maximum for the possible flux error of the antecedent in a two-object blend to be $|\Delta\text{mag}|{\sim}0.75$; a factor of 2. An overview of how this scheme applies to the most common case of a two-object blend is shown in Figure \ref{fig:matching-diagram}.

\begin{figure*}[ht!]
    \centering
    \includegraphics[width=\textwidth]{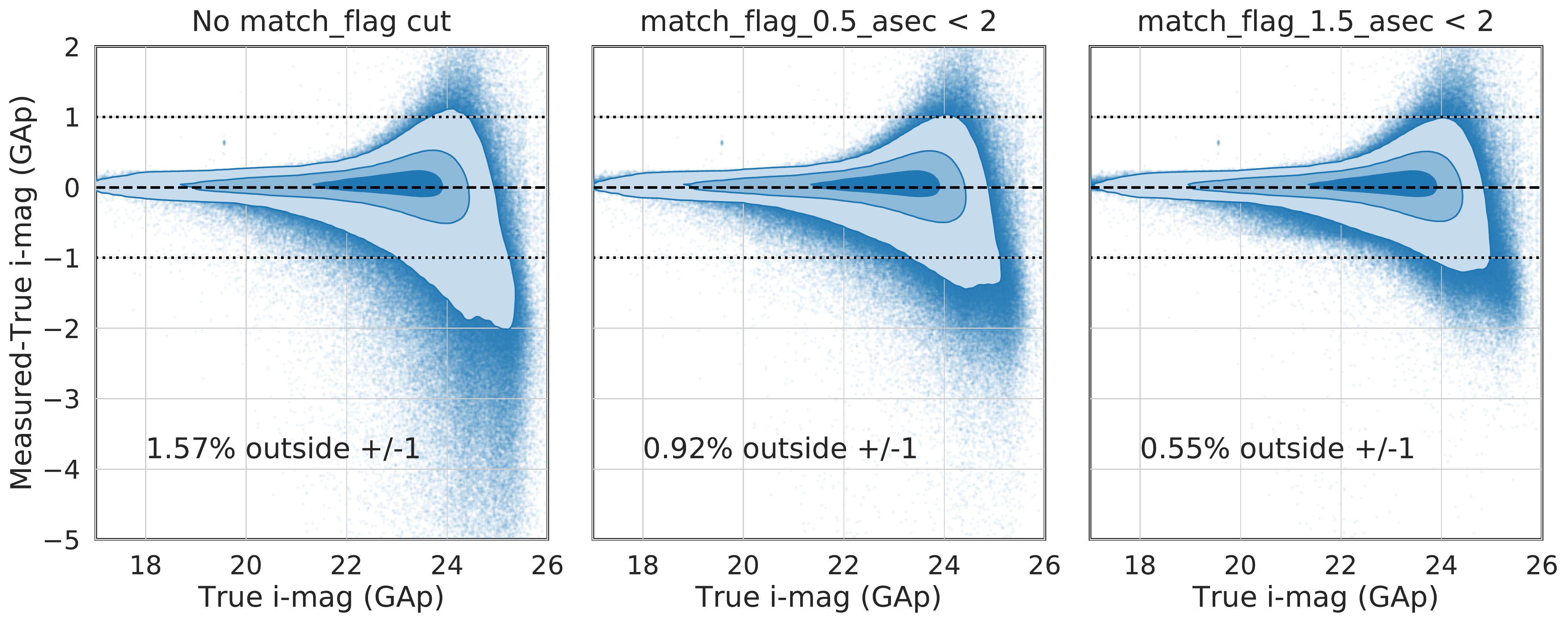}
    \caption{The effectiveness of our ambiguous matching scheme, illustrated by the difference in measured vs true $i$-band GAp magnitude ($\Delta \text{mag}_\text{gap}$) as a function of input GAp magnitude for three ambiguous matching choices. The overplotted contours contain 39.3\%, 86.5\%, and 98.9\% of the data volume, corresponding to the volume contained by the first three $\sigma$'s of a 2D Gaussian distribution respectively. The percentage of detections outside of the dashed region denoting $|\Delta \text{mag}_\text{gap}|<1$ for each choice is labeled in the bottom left of each panel. The left panel shows the $\Delta \text{mag}_\text{gap}$ response for \dfsample{} when no cut is made to handle ambiguous matches. There is an extremely long outlier tail of injections measured to be significantly brighter than the injected flux both from ambiguous blends and real effects (see \S \ref{catastrophic-fitting}, though GAp fluxes are much less sensitive to these failures). The outlier tail significantly decreases in size as more ambiguous blends are accounted for.}
    \label{fig:gap-scatter-per-radius}
\end{figure*}

The above prescription requires a brightness metric to determine the antecedent. We use the average of the dereddened Gaussian-weighted aperture (GAp) fluxes in each of the DES detection bands ($riz$). GAp fluxes are conceptually similar to GAaP fluxes described in \cite{gaap} but instead measure the aperture flux for source profiles \textit{before} convolution with the PSF. These fluxes are computed analytically from the MOF \code{bdf} fits to the DF injections and the SOF CModel fits to Y3 GOLD objects using a Gaussian weight function with FWHM of 4\asec{}. This allows us to use an estimate derived from our best guess of the flux of the PSF-deconvolved profile near the relevant object centroids while discounting variations in measured flux due to morphological differences -- particularly those arising from significant flux contributions from the wings of extended profiles. We use the average of the detection band $\delta$ fluxes for \starsample{} since an equivalent GAp flux is not well defined. This difference only becomes relevant for the brightest star injections, though in these cases they are very likely to be the antecedent.

The matching procedure is implemented in two separate steps. First, the injection positions are matched to the closest object in the remeasured photometry catalogs within a search radius of $r_1=0.5$\asec{}. All objects that have a match are saved in the output \balrog{} catalogs and undergo the aforementioned post-processing steps. Afterwards, the output catalogs are matched against the Y3 GOLD catalog to compare the relative brightness of any existing detections within a second match radius $r_2$ for a series of radii from 0.5\asec{} to 2.0\asec{} in increments of 0.25\asec{}. Over 96\% of candidate objects have no GOLD sources within the search aperture and are unambiguously a \balrog{} injection. Candidates that have an existing GOLD object within $r_2$ with mean $riz$ GAp flux below their own are considered the antecedent and given a \code{match\_flag\_\{r2\}\_asec=1} to indicate the presence of a nearby real source. Candidates that have a match within $r_2$ but have a smaller mean GAp flux than the existing object are assigned \code{match\_flag\_\{r2\}\_asec=2} and are recommended to be cut from science analyses. We encode this information as a flag instead of cuts to the fiducial catalog to allow \balrog{} users more flexibility in choosing how to handle blending and ambiguous cases as needed. In this paper, we cut on \code{match\_flag\_1.5\_asec < 2} as we found that only 0.1\% and 0.5\% of Y3 GOLD objects were separated at distances less than 1.5\asec{} at $i$ magnitudes of 21 and 22.5 respectively (or about 1.3-1.8 times the median PSF size depending on the band).

We show in Figure \ref{fig:gap-scatter-per-radius} the
the difference between the recovered and injected GAp magnitude, $\Delta \text{mag}_\text{gap}$, for all recovered \runtwo{} objects for three choices of ambiguous matching cuts. In the left panel where no cut on ambiguous matches has been made, there is a long, asymmetric tail for negative $\Delta \text{mag}_\text{gap}$ where the recovered GAp flux is up to 10 magnitudes brighter than the input. While there can be extremely large magnitude responses to model-fitted photometry in crowded fields or extreme imaging conditions (see \S \ref{catastrophic-fitting}), we expect GAp magnitudes to be less sensitive to these failure modes and most large discrepancies to be due to ambiguous matches. This is indeed the case: In the following panels where a match flag with $r_2$ of 0.5\asec{} and 1.5\asec{} are used to create the sample, the worst GAp response outliers have been removed and the fraction of detections where $|\Delta \text{mag}_\text{gap}|>1$ falls by 41\% and 65\% respectively. Some remaining scatter beyond $|\Delta \text{mag}_\text{gap}|=0.75$ is expected even for an optimal $r_2$ due to ambient light in dense fields, blends with extended sources, and image artifacts, though the number of objects below $\Delta \text{mag}_\text{gap}=-1$ for the 1.5\asec{} cut falls by over an order of magnitude for each bin of unit size.


\section{DES Y3 Photometric Performance}\label{photometric-results}

Here we present the photometric performance of the Y3 \balrog{} DF sample \dfsample{} along with the synthetic star sample \starsample{}. While there are many photometric catalogs and science samples of interest for Y3, here we largely focus on the SOF CModel photometry of a basic Y3 GOLD sample \citep{y3-gold} used as a starting point for more restrictive samples. Unless otherwise specified, the cuts for this sample are given by

\begin{center}
\begin{tabular}{ll}
\small
     & \texttt{FLAGS\_FOREGROUND = 0} \\
    \texttt{AND} & \texttt{FLAGS\_BADREGIONS < 2} \\
    \texttt{AND} & \texttt{FLAGS\_FOOTPRINT = 1} \\
    \texttt{AND} & \texttt{FLAGS\_GOLD\_SOF\_ONLY < 2} \\
    \texttt{AND} & \texttt{EXTENDED\_CLASS\_SOF >= 0} \\
    \texttt{AND} & \texttt{MATCH\_FLAG\_1.5\_ASEC < 2}, \\
\end{tabular}
\end{center}

\noindent along with any appropriate object classification cut which will be mentioned when relevant. Note that \texttt{FLAGS\_GOLD\_SOF\_ONLY} is used in place of the typical \texttt{FLAGS\_GOLD} as we are unable to compute the first bit flag without \dfsample{} MOF runs. While ${\sim}3.5\%$ of Y3 GOLD objects have \texttt{FLAGS\_GOLD=1}, no Y3 cosmology analyses currently use this flag bit due to the use of SOF or Metacalibration photometry in favor of MOF. Additional samples for a few interesting \balrog{} applications are discussed in more detail in Section \ref{applications}.

We begin by examining how representative the \balrog{} catalog properties are compared to Y3 GOLD in Section \ref{consistency}, including a detailed look at how the number density fluctuations of both samples vary with respect to survey property maps. We then show the magnitude and color responses of \starsample{} and \dfsample{} along with a discussion of interesting photometric failure modes in Sections \ref{results-stars} and \ref{results-galaxies} respectively. We then end by characterizing the performance of the \code{EXTENDED\_CLASS\_SOF} star-galaxy separator, using the extremely pure \starsample{} sample whenever possible. As it is not practical to plot the photometric responses of all quantities of interest, one-dimensional Gaussian summary statistics for many relevant parameters are provided in Appendix \ref{appendix:tabular-results}.

\subsection{Consistency with DES Data}\label{consistency}

Even without perfect emulation fidelity, we expect the measured \balrog{} property distributions to closely resemble DES catalogs if we are indeed sampling an adequately representative transfer function and input sample. We will broadly check this agreement at various steps along the measurement path: object detection, photometric properties, and correlations with survey systematics -- along with how these differences impact a typical clustering signal measurement. As we are primarily interested in the consistency in the transfer function of galaxies for cosmology, we use the \dfsample{} sample throughout and mention any classification cuts when relevant.

\subsubsection{Completeness}\label{completeness}

We begin with object detection. Of the nearly 26.5 million galaxies injected in \dfsample{}, just over 41.9\% were detected during re-measurement after accounting for ambiguous matches. However, as this catalog is the merger of two runs with different magnitude limits, it is more accurate to say that 36.3\% and 59.4\% of objects were recovered for \runtwo{} and \runtwoa{} respectively. The fraction of injections contained in the fiducial sample drops to 14.4\% and 44.2\% after considering the basic flag and mask cuts described above. To simplify the comparison on the faint end we use only \runtwo{} for the following comparison as it is about a magnitude deeper.

\begin{figure}
    \centering
    \includegraphics[width=0.475\textwidth]{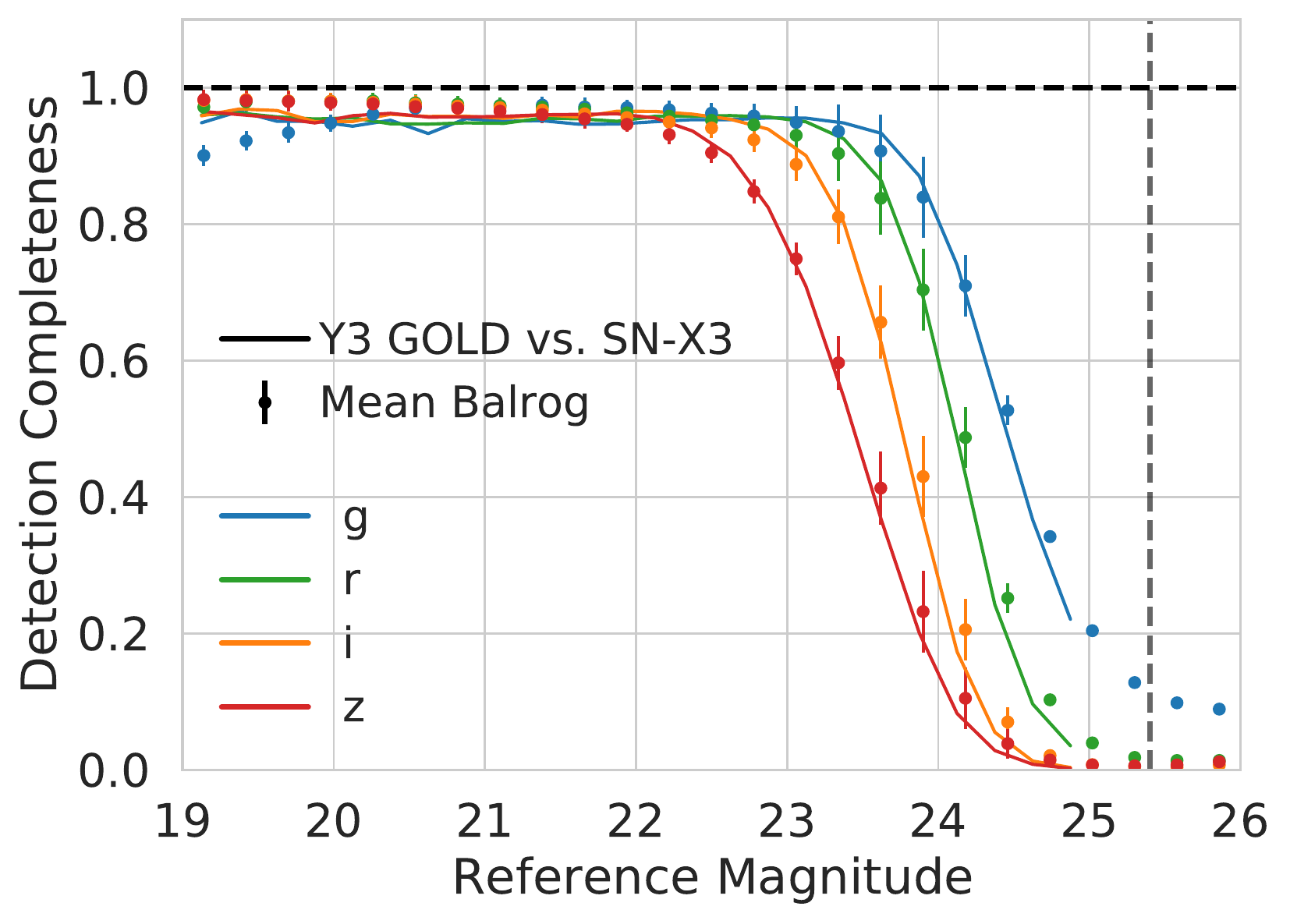}
    \caption[Caption with a citation]{The fraction of objects recovered by band and input injection magnitude. Solid lines show completeness measurements comparing the wide and deep samples on the SN-X3 field as described in Section 5.2 of \protect{\cite{y3-gold}}. Points with error bars are the \balrog{} mean completeness measurements for the full sampled \protect{\runtwo{}} footprint. Errors are the standard deviation of 50 jacknife samples of the sampled footprint, rescaled as appropriate for the area of the SN-X3 field. The dashed vertical line indicates the injection effective magnitude limit of 25.4.}
    \label{fig:balrog-detection-efficiency}
\end{figure}

The detection completeness of sources in $griz$ for \runtwo{} (points) compared to Y3 GOLD objects in the X3 supernovae field (lines) is shown in Figure \ref{fig:balrog-detection-efficiency}. The completeness is plotted as a function of reference magnitude; the injection magnitudes for \balrog{} and the DF measurements of objects in the X3 field for Y3 GOLD. As we are comparing the mean completeness of the \balrog{} sample across all \runtwo{} tiles to only a small region for Y3 GOLD, to make a fair comparison we estimate the uncertainty in the difference with 50 jacknife samples of the \runtwo{} footprint. Note that the inferred completeness is only robust until the forced magnitude limit cutoff of 25.4 indicated by the dashed vertical line; beyond this point, the sampled injection objects have inherited a selection bias that forces at least one of the other detection bands to be significantly brighter than the magnitude limit and thus is more likely to be detected.

Overall the completeness measurements are quite similar, with the only discrepancies greater than twice the estimated error occurring for the brightest $g$-band magnitudes and the faintest $i$ and $z$ bin. The \balrog{} $g$-band completeness dips on the bright end despite the very high S/N as $g$ is not included in the composite detection magnitude image limit, and thus objects bright in $g$-band but not in other bands are sometimes not detected. This is not seen as significantly in the Y3 GOLD sample which suggests that the input DF sample over represents these kinds of objects. It is more difficult to determine possible discrepancies past the detection threshold in each band without careful examination of both measurements, though their residuals are only marginally beyond 1-$\sigma$ and could simply be statistical fluctuations. While it is encouraging to see similar detection properties between \balrog{} and the data, that alone is not enough to ensure sufficient similarity for science calibrations.

\subsubsection{SOF Photometry}\label{1d-compare}

\begin{figure*}
    \centering
    \includegraphics[width=\textwidth]{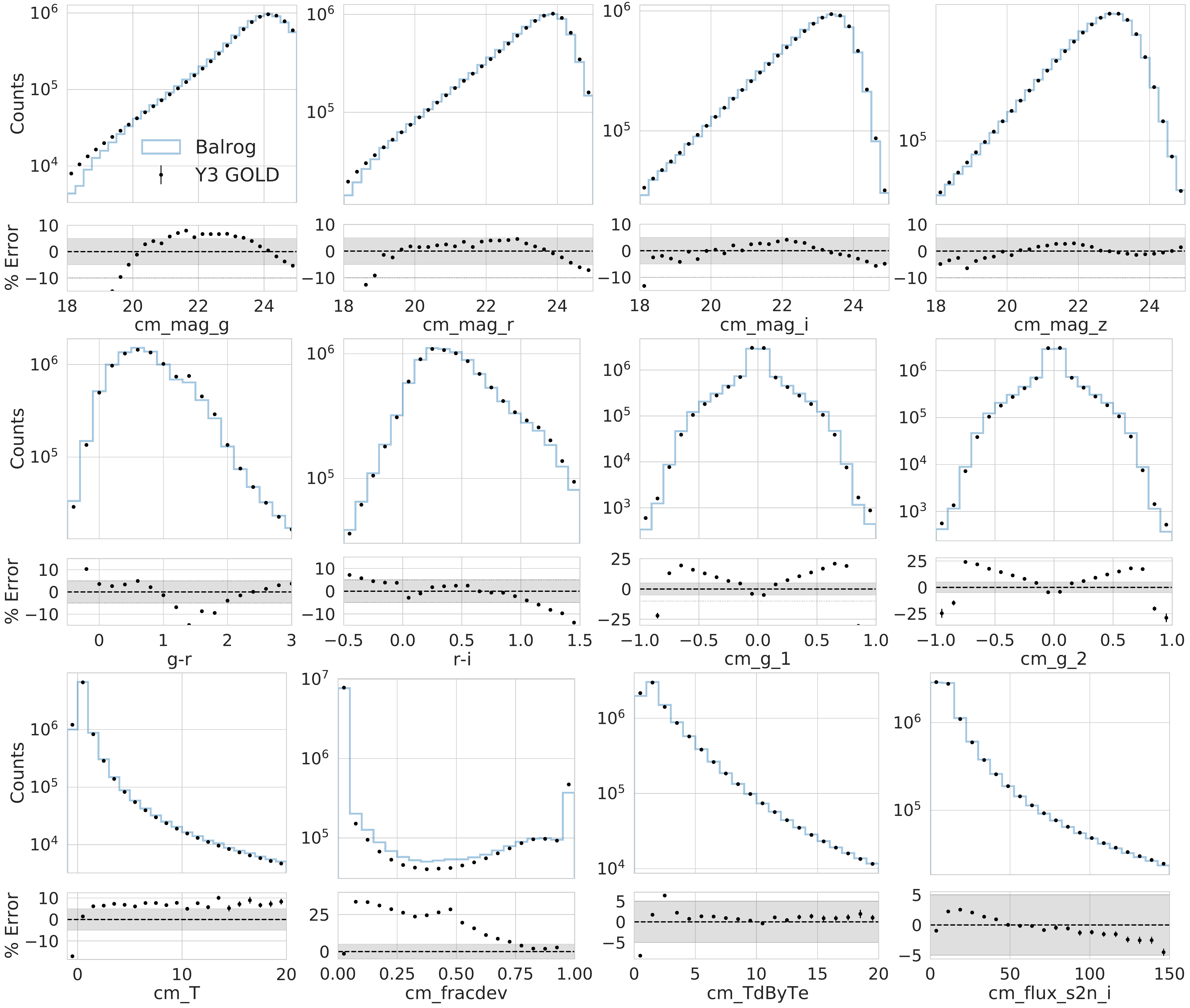}
    \caption{Comparison of the \dfsample{} sample (in blue) vs. Y3 GOLD (in black) for measured $griz$ magnitudes, $g-r$ and $r-i$ color, shape parameters \code{cm\_g\_1} and \code{cm\_g\_2}, size \code{cm\_T}, flux component ratio \code{cm\_fracdev}, size component ratio \code{cm\_TdByTe}, and $i$-band S/N. Both samples have had the basic cuts applied as described in Section \ref{photometric-results}. To compare the distributions, we resample Y3 GOLD with replacement to match the size of the \dfsample{} catalog 100 times and plot the mean and std of these bootstrap samples in black. The percent error of the binned residuals are shown below each distribution, which have been zoomed in to show the results of the most relevant regions. The region corresponding to +/-5\% has been shaded in gray. When quantities do not have hard boundaries, we include at least the 2nd-97th percentiles of the values. The residuals are very sensitive to selection cuts. For example, the discrepancies at \code{cm\_T}$<0$ and $|\code{cm\_g\_\{1/2\}}|{\sim}1$ are significantly smaller after cutting out suspected stars from the sample.}
    \label{fig:balrog-gold-1d-compare}
\end{figure*}

We can make similar comparisons of the measured photometry. Figure \ref{fig:balrog-gold-1d-compare} compares the recovered \balrog{} SOF $griz$ magnitudes, $g-r$ and $r-i$ colors, and a few morphological parameters to Y3 GOLD after both samples have applied basic cuts. The comparison is in absolute counts with \balrog{} in blue and the mean of 100 GOLD bootstrap subsamples of identical size to the \dfsample{} sample in black. The standard deviation of the subsample counts in each bin are used to estimate the uncertainty and the percent errors of the binned residuals are plotted below each distribution.

Qualitatively, the distributions are extremely similar in the most dense regions of parameter space for most quantities, with the most obvious discrepancies occurring in the low-density tails of the distributions. This is particularly noticeable for the magnitudes and colors. The relative residuals confirm this: While nearly all \balrog{} magnitude bins have fractional distribution differences below 5\% of the mean Y3 GOLD sample from 18 to 24, the region of interest for most Y3 cosmological analyses, \balrog{} counts in magnitudes below 18 underestimate GOLD by 10 to 50\% by magnitude 16. The colors are similar with the only discrepancy above 5\% in the densest regions occurring at $1.3<g-r<1.5$; values typical of M-dwarf stars \citep{m_dwarf}. A few other notable discrepancies are that \balrog{} appears to underestimate the number of objects with ellipticities \code{cm\_g\_\{1/2\}}${\sim}0$ and negative size parameter \code{cm\_T} relative to the Y3 GOLD sample - both of which are again values typical of stars.

We stress that these binned residuals are still a largely qualitative check on the agreement between property distributions as they are very sensitive to sample selection. For example, the relative error in \code{cm\_T}, \code{cm\_g\_1}, and \code{cm\_g\_2} near zero are all significantly smaller after applying the stellar cut \texttt{EXTENDED\_CLASS\_SOF > 1} which indicates that the \dfsample{} sample does not capture the transfer properties of stars as well as galaxies. Yet the shape of these residuals often indicate important real differences. The change in residual sign near the detection threshold in each band indicates potential small differences in the effective depth of the samples, and the overabundance of \balrog{} objects with \texttt{cm\_fracdev} near 0.5 reflects the effect of parameter priors not matching the true underlying distribution as discussed in section \ref{validation}.

In addition, residuals consistent with zero even under the assumption of perfect emulation fidelity requires a completely representative input sample. There are many known reasons for why our input sample fails this requirement, a few of which we discuss here: (i) The DF sample underestimates cosmic variance as it only uses objects from a tiny fraction of the sky, which is particularly a problem for the stellar population as its distribution varies across the sky much more strongly than galaxies. (ii) The photometric pipeline used to make measurements of DF objects is not identical to the one used in the WF in order to deal with non-dithered observations, an increased blending rate, the large number of exposures per detection, and instabilities in the detection of very faint sources in the presence of diffuse emission (see \citealt*{y3-deepfields}). (iii) The morphological model fits to the DF objects are subtly different (\code{bdf} vs \code{cm}) which we have shown can introduce small biases in other parameters such as the magnitude. (iv) CModel is not an appropriate photometric model for all objects in the sky. There are simple practical limitations that contribute to these discrepancies as well, such as limiting the size and magnitude distribution of objects to reduce \balrog{}-\balrog{} blends and the computational time spent on injecting near certain non-detections. We discuss these issues more in Section \ref{discussion}.

\subsubsection{Spatial Variation and Property Maps}\label{map-trends}

While the overall similarities in the photometries are encouraging, what is most critical is how well \balrog{} reproduces the measurable signals used in cosmological analyses as well as correlations with spatially varying image conditions and survey properties. These systematic trends are particularly important when measuring the galaxy clustering signal where local observing conditions can imprint fluctuations in number density that are not cosmological in origin such as variations in seeing, depth, and sky brightness \citep{y3-galaxyclustering}. We now investigate the similarity of these systematic trends in \balrog{} and Y3 GOLD for a highly incomplete sample where the variation is more apparent, before looking at their contribution to the clustering signal itself for a cosmology-like sample in \S \ref{power-spectra}.

\begin{figure*}[htp]
    \centering
    \includegraphics[width=.9636\textwidth]{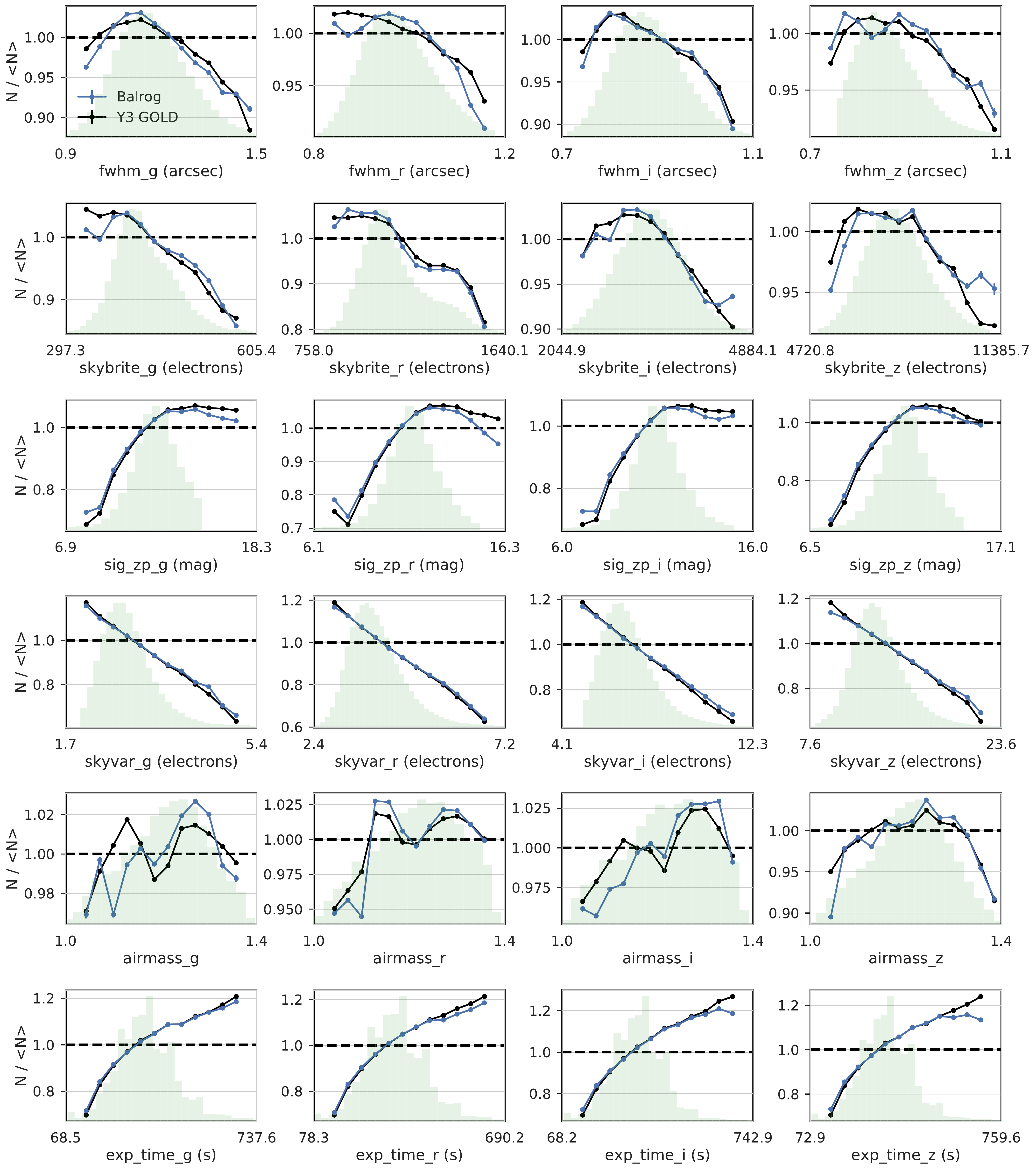}
    \caption[A caption with a citation]{
    The trend in number density fluctuations \protect{$N/{<}N{>}$} as a function of various survey observing properties for the full (and highly incomplete) \protect{\balrog{}}, in blue, and Y3 GOLD, in black, samples after basic cuts for overlapping HEALPix pixels of \protect{\code{NSIDE}=2048}. The distribution of survey condition values for the rescaled Y3 GOLD map is displayed in the background in green to highlight typical values. The errors have been estimated by resampling the pixels used in each sample with replacement for 100 realizations. 
    The property maps are described in Table \protect{E.1} in \protect{\cite{y3-gold}}, but we briefly defined them here in order from the top: the mean PSF size, the local sky brightness, the quadrature sum of the zeropoint uncertainties, the variance of the sky brightness, the airmass, and the exposure time. \balrog{} captures many of the nonlinear features in the trend lines, though there are some unexplained band-dependent discrepancies in some property maps.}
    \label{fig:balrog-gold-systematics-compare}
\end{figure*}

Figure \ref{fig:balrog-gold-systematics-compare} compares the number density of all \dfsample{} and Y3 GOLD galaxies with basic cuts as a function of survey property in overlapping HEALPix \citep{healpix} pixels of \code{NSIDE=2048}, corresponding to an area of 2.95 arcmin$^2$. The survey properties are assigned from the Y3 HEALPix maps in \cite{y3-gold} (based off the methodology in \citealt{y1-gold}) that have been rescaled\footnote{The map rescaling is done by averaging all non-empty pixels.} from a $N_\text{side}$ of 4096 to 2048 to smooth out irregularities in the pixel occupation distribution due to the regular structure and lower density of \balrog{} sources. The uncertainty in number density was estimated by resampling the pixels used in each sample of equal size with replacement for 100 bootstrap samples. The distribution of the rescaled survey properties for the Y3 GOLD sample are plotted in the background in green to highlight typical property values.

With a few notable exceptions, the number density of the two samples match closely in both amplitude and shape. It is especially encouraging to see \balrog{} capturing the high frequency structure in the dependence of a few of the more complex trends such as the local sky brightness (\code{skybrite}) and airmass. The largest differences in recovered number density occur for extremely rare values of a few properties such as the quadrature sum of zeropoint uncertainties (\code{sig\_zp}) and exposure time (\code{exp\_time}) and are not particularly concerning. However, there are still some more serious unresolved discrepancies in amplitude -- particularly in $r$-band seeing and airmass. The same potential issues in input sample representativeness and photometric assumptions discussed previously apply to these measurements, but it is not immediately clear why these issues would manifest in a band-dependent fashion in seeing or why the largest discrepancies occur for an indirect parameter of the images like airmass. These differences may be indicative of features in the transfer function not currently captured by \balrog{} such as PSF modeling errors with unexpected chromatic effects or the unapplied injection zeropoint corrections. Such differences warrant further investigations in preparation for an improved Y6 \balrog{} methodology but do not themselves indicate insufficient consistency for a clustering measurement. We explore this further below.

\subsubsection{Galaxy Clustering Systematics}\label{power-spectra}

Many of the core science cases of interest to cosmology involve measurements of galaxy clustering. To be useful in calibrations for this purpose, it is not enough that the number counts of \balrog{} and Y3 GOLD galaxies follow the same trends with image properties like those shown in Figure \ref{fig:balrog-gold-systematics-compare}. Where the systematic error is independent of the signal (as, for example, variations in the airmass and the true galaxy density on the sky are statistically independent of one another), the resulting variations in survey depth enter, to leading order, as additive systematic errors in the two-point statistics used for cosmology.

Correcting for these observational systematics is critical for unbiased cosmological inference from clustering, and the ability to use \balrog{} as object randoms with realistic measurement biases -- if it sufficiently captures the clustering fluctuations of the data -- offers an ideal calibration method without using the data vector directly which avoids possible overfitting (see \citealt{Suchyta_2016,Choi_2016, GF_SV_magnification}). In addition, direct calibration with \balrog{} would eliminate the need to identify all sufficiently important survey property contributions at a desired precision (and avoid biases from any unidentified systematics) while potentially allowing for measurements on larger scales where the true signal is very small and the corrections have to be \textit{extremely} accurate.

Here we estimate the approximate impact on the clustering signal due to systematic differences between \balrog{} and Y3 GOLD for a sample broadly similar to the {\sc maglim} science sample described in \cite{y3-2x2maglimforecast}, where we cut both the Y3 GOLD and \balrog{} samples to $17.5<i<21.5$ in addition to the previous cuts. We make density maps based on each property map across the full Y3 GOLD footprint by interpolating the trends in \balrog{} and GOLD to fill in cells where we do not have injection samples. These maps are estimates of the {\sc maglim} galaxy number density fluctuations in Y3 if they could be completely described by the survey property in question\footnote{Where only regions with \balrog{} samples are used for the estimate.}.

We then estimate the angular power spectra of both interpolated maps for each survey property using the pseudo-$C_\ell$ estimation code PyMaster \citep{pymaster}. These are then compared to the power spectra of the survey property maps themselves along with a typical nonlinear galaxy power spectrum at $z=0.7$ computed with the \code{CAMB} \citep{camb} implementation of the nonlinear power spectrum described in \cite{halofit_mead}. Finally, we compute the differences in power from the interpolated \balrog{} and Y3 GOLD density maps as a fraction of the galaxy power spectrum at each $\ell$-scale.

\begin{figure*}[htp!]
    \centering
    \includegraphics[width=\textwidth]{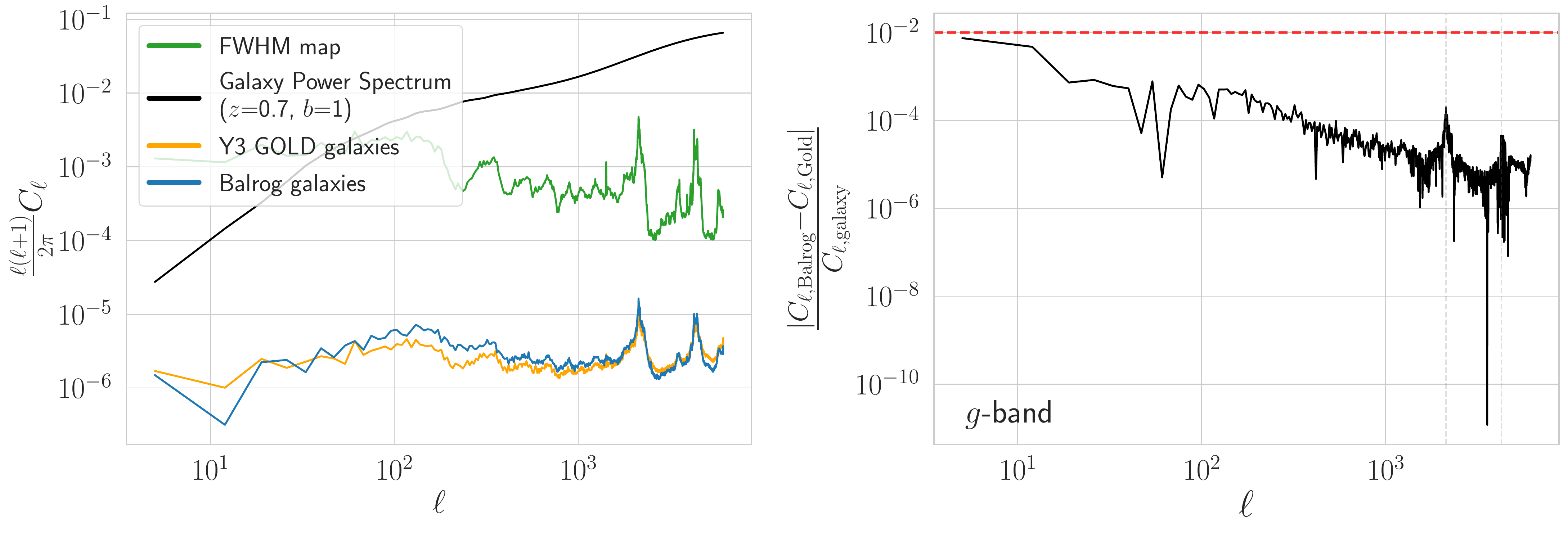}
    \includegraphics[width=\textwidth]{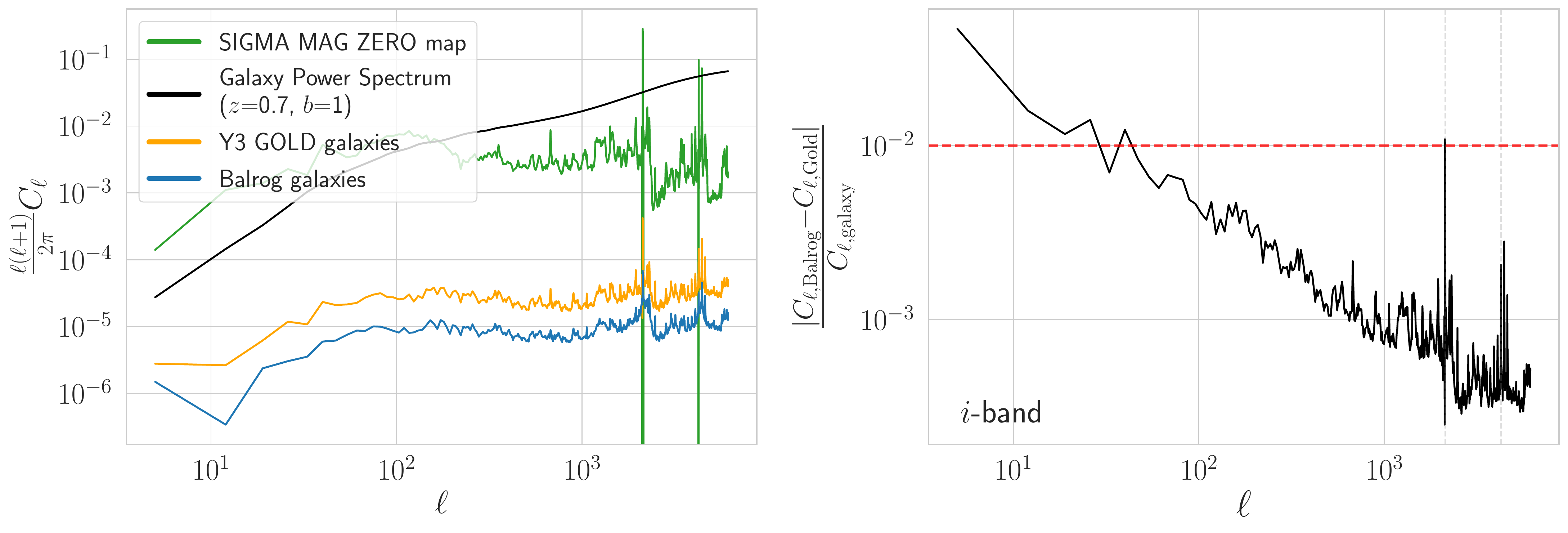}
    \caption{Examples of the survey property maps with the smallest (top row) and largest (bottom row) estimated additive systematic impact on the clustering signal from differences in number density between \protect{\balrog{}} and Y3 GOLD. The left panels show the angular power spectrum of the noted survey property (in green) and the corresponding power spectra of the number densities of the \protect{\balrog{}} (in blue) and Y3 GOLD (in gold) {\sc maglim}-like galaxies across the Y3 footprint using the interpolated trends described in \S \protect{\ref{map-trends}} and \S \protect{\ref{power-spectra}}. The reference galaxy power spectrum in black is \protect{\code{CAMB}}'s implementation of the nonlinear matter power spectrum described in \citet{halofit_mead}, meant to represent a typical cosmological signal at \protect{$z=0.7$} with linear galaxy bias parameter of 1. The right panels show the difference in power between Y3 GOLD and \protect{\balrog{}} as a fraction of the fiducial cosmological power spectrum shown on the left. We draw a red dashed line indicating the 1\% systematic error threshold as reference. Even in the worst case, we find that \protect{\balrog{}} is able to capture the clustering amplitude due to variations in survey properties to better than 1\% for $\ell>50$ (corresponding to \protect{$\theta>{\sim} 3.5$}) deg. Equivalent plots for many other survey property maps in all $griz$-bands are shown in Appendix~\ref{appendix:power_spectra}.}
    \label{fig:power_best_worst}
\end{figure*}

Results for the best ($g$-band PSF FWHM) and worst ($i$-band \code{sig\_zp}) performing map are shown in Figure \ref{fig:power_best_worst}. Angular clustering systematics for the remaining survey properties, generated in the same way, are shown in Appendix~\ref{appendix:power_spectra}. For scales comparable to or smaller than the DECam focal plane (approximately $\ell > 200$), the difference between Y3 GOLD and \balrog{} is in all cases less than $1\%$ of the typical amplitude of the angular clustering of galaxies (plotted in black). For some quantities, such as the $g$-band PSF (shown in the top panel in Figure \ref{fig:power_best_worst}), the differences are several orders of magnitude smaller.

While the differences are small in absolute terms, or as compared to a realistic cosmological signal, the relative deviation between the simulated and real catalogs is in some cases quite large. It is difficult to disentangle the relative contribution to these differences from insufficient sampling across survey property values, issues in the input sample, or missing features in the sampled transfer function (such as the zeropoint corrections discussed in \S \ref{desdm-differences}). We discuss these issues further in Section \ref{discussion}. However, that the absolute additive contributions are well below 1\% at most relevant scales for even a single realization of a 20\% sampling of the footprint gives us confidence that injection simulations like \balrog{} will be crucial for systematics calibration of clustering measurements in Y6 and the next generation of galaxy surveys with even more ambitious precision goals.

Whether \balrog{} is sufficiently similar to Y3 data ultimately depends on the science case and desired measurement precision. In addition, the magnitude of discrepancies can depend strongly on the choice of sample cuts - particularly for those effects related to star-galaxy separation and magnitude limits. However, we find that \balrog{} captures a significant amount of the variation in number density as a function of observing conditions even for \textit{extremely} incomplete samples, and systematics control of well under 1\% for the clustering measurement of a typical cosmology sample. For an additional example of how to estimate the contribution of the intrinsic uncertainty in the \balrog{} methodology to the Y3 photometric redshift calibration error budget, see \cite*{y3-sompz}.

\subsection{Photometric performance of \texorpdfstring{\starsample{}}{delta-Stars}}\label{results-stars}

As discussed in \ref{star-sample}, the injections in \starsample{} consist of pure delta functions convolved with the local PSFEx solution. The extremely high purity of this star sample with realistic transfer properties is unique to injection pipelines such as \balrog{} where we have truth information about the underlying object classification in addition to its photometry - which is not always the case for galaxy samples (discussed further in Section \ref{results-galaxies}). This eliminates the need for a traditional star-galaxy separation metric like \texttt{EXTENDED\_CLASS\_SOF} and (nearly) removes any bias resulting from misclassified objects, though we still cut on $\code{EXTENDED\_CLASS\_SOF}<=1$ to match what is done to create stellar samples in Y3 GOLD. The only contaminants in the main star sample come from ambiguous matches which is why we still cut on $\texttt{match\_flag\_1.5\_asec}<2$. This eliminated 1.9\% of detections for this sample. Here we focus on the photometric performance and leave the discussion on stellar completeness and galaxy contamination in Section \ref{star-gal-sep}. We remind the reader that this sample probes a subtly different measurement likelihood than that of \dfsample{} as we have knowledge of the underlying object classification, as described in \ref{different-likelihoods}.

While the underlying morphology of stellar profiles is not well described by a S\'ersic model, we still use the SOF CModel fits for the stellar sample as there was a systematic calibration offset in the PSF model photometry used in Y3 measurements on the data. This has been corrected for Y6 processing but leaves us without a reliable PSF photometry for our response measurements. However, ultimately this has only a small impact on the recovered photometry for sources smaller than the PSF as these objects are fit with a \code{cm\_T} size near 0 -- effectively eliminating the S\'ersic components.

\subsubsection{SOF CModel Magnitudes}

\begin{figure*}
    \centering
    \includegraphics[width=1.\textwidth]{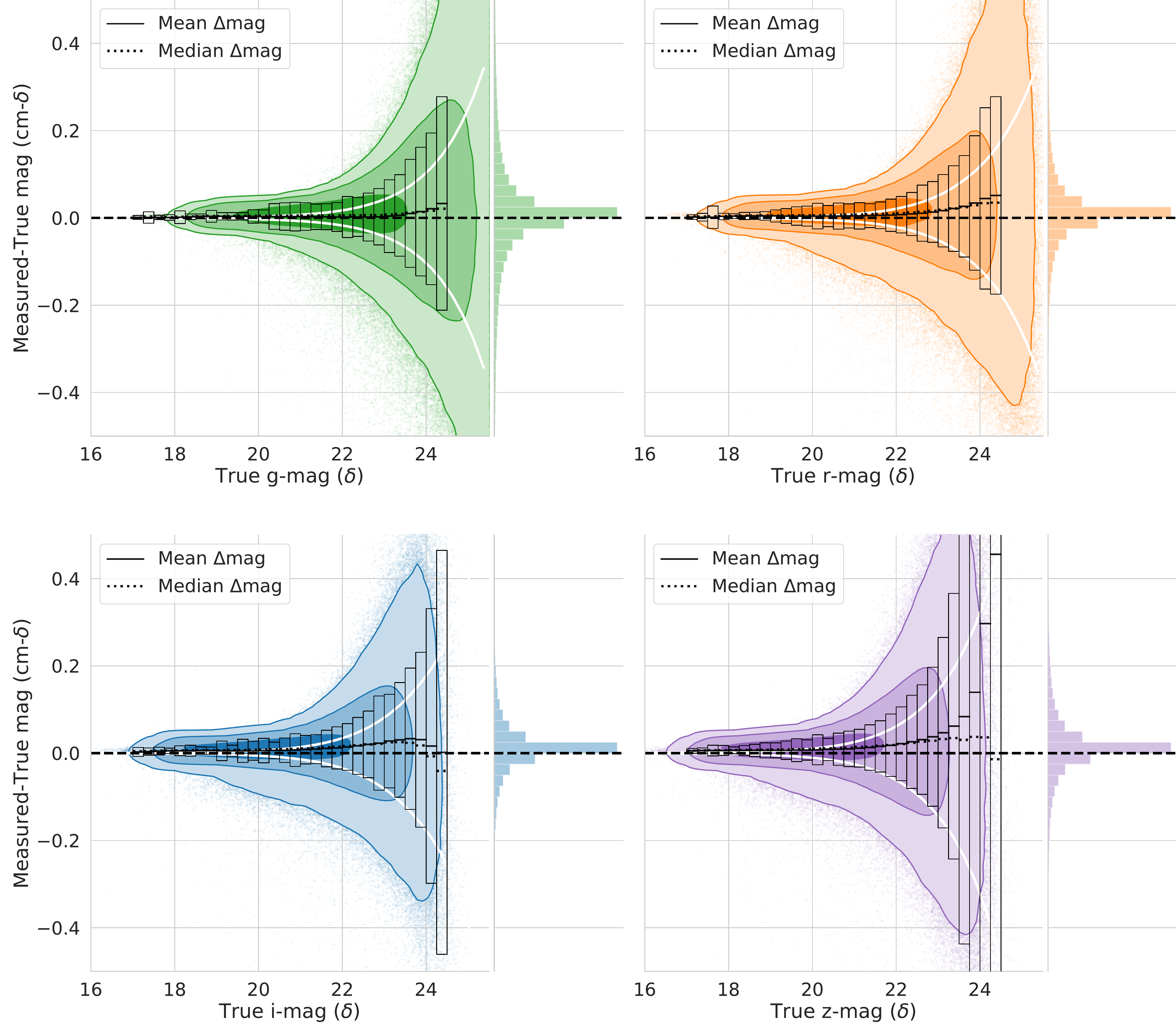}
    \caption{The distribution of differences in recovered \protect{$griz$} SOF CModel magnitude vs the injected \protect{$\delta$}-magnitude (\protect{\dmstar{}}) as a function of input magnitude for the \protect{\starsample{}} sample. The density is overplotted where the contour lines correspond to the percentiles of the first three sigmas of a 2D Gaussian, containing  39.2\%, 86.5\%, and 98.9\% of the data volume respectively. The mean (solid), median (dotted), and standard deviation of the magnitude responses in bins of size 0.25 magnitude are shown in the overlaid black bars. These are compared to the reported SOF CModel errors by the \response{solid} white lines which do not attempt to account for systematic effects. The marginal distributions of \protect{\dmstar{}} are included to highlight the small relative volume of the outlier tails.}
    \label{fig:meas-vs-true-mag-sof-star}
\end{figure*}

\begin{figure*}[ht!]
    \centering
    \includegraphics[width=1.\textwidth]{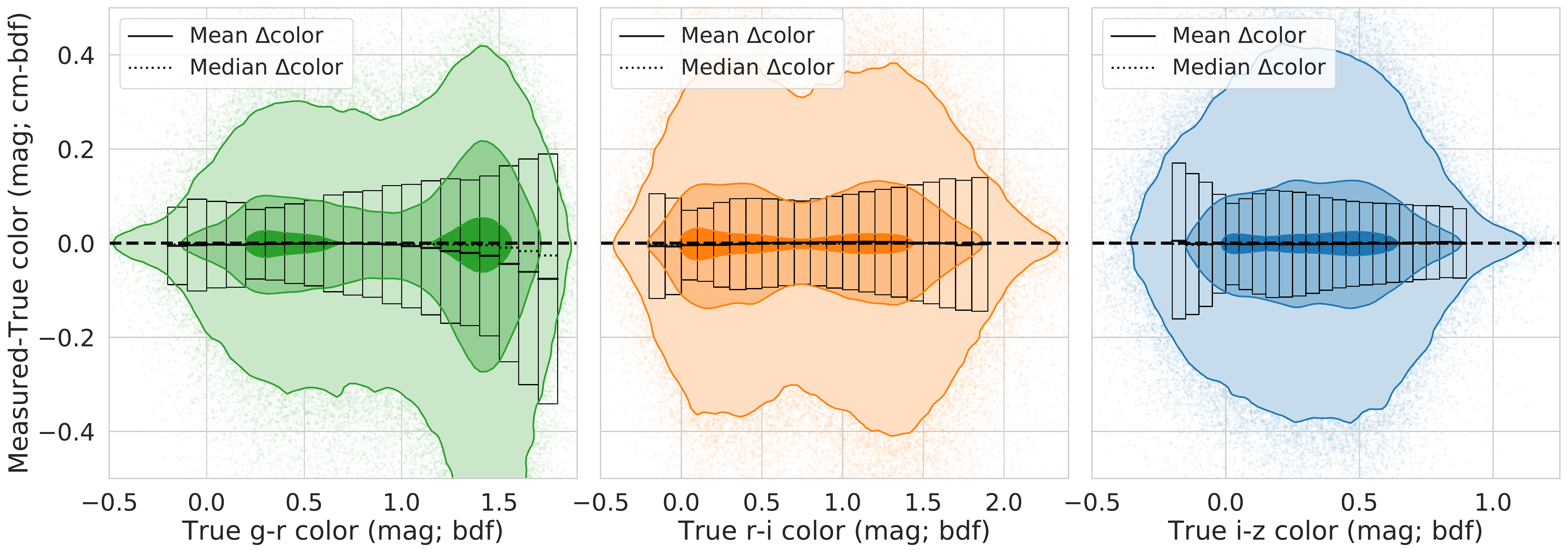}
    \caption{The distribution of differences in measured SOF CModel $g-r$, $r-i$, and $i-z$ color vs. the injected \protect{$\delta$}-color (\protect{\dcstar{}}) as a function of input color for the \protect{\starsample{}} sample. The density is overplotted where the contour lines correspond to the percentiles of the first three sigmas of a 2D Gaussian, containing  39.2\%, 86.5\%, and 98.9\% of the data volume respectively. The mean (solid), median (dotted), and standard deviation of the magnitude responses in bins of size 100 mmag magnitude for $g-r$ and $r-i$ and 50 mmag for $i-z$ are shown in the overlaid black bars.}
    \label{fig:stars-color-response}
\end{figure*}

The difference in recovered CModel magnitude compared to input magnitude \dmstar{} as a function of input magnitude for $griz$ is shown in Figure \ref{fig:meas-vs-true-mag-sof-star}. Density contours are plotted on top of the scatter with percentiles equivalent to the first three sigmas of a 2D Gaussian distribution, corresponding to 39.3\%, 86.5\%, and 98.9\% of the total data volume. The mean response bias \mdmstar{}, median response \medstar{}, and scatter \sigstar{} in truth magnitude bins of size 0.25 magnitudes are over-potted in black bars. These summary statistics provide estimates for the statistical precision and accuracy of the SOF magnitudes, though we stress that the underlying distributions are not Gaussian. These are compared to the mean reported SOF error in the bin indicated by the \response{solid} white curve which does not attempt to account for systematic effects.

The overall calibration of CModel for the stellar sample is quite good, with \mdmstar{} and \medstar{} ranging from 1-10 mmag (or 0.1-0.9\%) across all bands up to an input magnitude of 20 and between 2-15 mmag (0.2-1.4\%) for $20<$\dmstar{}$<22$ except for the final two $z$-band bins. \mdmstar{} stays under 1.5\% for each band in all bins where the number of objects are increasing (input magnitudes of 23.5, 22.5, 22, and 22 respectively) except for the final $z$-band bin which is ${\sim}$1.7\%. The responses are a bit higher than the quoted 3 mmag uniformity of Y3 GOLD stars when compared to the Gaia star catalog (\citealt{y3-gold}, \citealt{gaia}), though the Y3 GOLD uniformity was measured only with respect to Gaia's $G$-band which we find to have the best photometric performance (differences of 0.5-6 mmag) over the quoted magnitude range. The Y3 GOLD measurement used a restricted $0.5<g-i<1.5$ color range as well which eliminates the worst outliers that we still consider here. In addition, the larger discrepancies found here could be the result of the CModel model-misspecification bias discussed previously.

The response bias and scatter increase significantly after these points due to competing systematic effects as the sample becomes progressively more incomplete, with the mean responses rising to ${\sim}1.5-3\%$ as they approach the detection threshold in each band. Small sample sizes and strong selection effects lead to \mdmstar{} and \medstar{} biases of ${\sim}4\%$ for $g$ and $r$ by 24th magnitude, while the biases of the much shallower $i$ and $z$ rise significantly to over 10\%. At the median coadd magnitude limits quoted in Table 2 of \cite{y3-gold} of 24.3, 23.0, 22.6, and 22.2 (corresponding to a S/N of 10), the mean $griz$ biases are measured to be 3.0\%, 4.1\%, 2.5\%, and 2.2\% respectively. The complete set of values for all binned summary statics are included in Table \ref{tab:star-mag-response}. While the underlying measurement likelihood of these objects is non-Gaussian, the morphological simplicity of stars results in these summary statistics qualitatively capturing the response features well when complete. We will return to this point in \ref{results-galaxies} where the situation is significantly more complicated. 

There is evidence of a small band dependence in both the accuracy and precision of the magnitude response. This is most evident when comparing $g$-band, where \dmstar{} is never above 5 mmag (0.5\%) too faint below an input magnitude of 23.25, to the $z$-band \dmstar{} which is exclusively above 5 mmag too faint over the same interval. Unlike the blank image tests in Section \ref{validation}, the \medstar{} values for each band in a bin have a distinct, monotonically increasing shape with the spread between the bands consistently 5-10 mmag brighter than injection magnitudes of 21. However, this effect is much less pronounced when binned by the measured S/N in each band where the detection significance and local sky background is taken into account. Binned in this way, \medstar{} is nearly identical for $i$ and $z$ bands for S/N greater than 20 while $g$ and $r$ are consistently offset by at least 5 and 2 mmag respectively. As this band-dependent response in \medstar{} was not present in the blank image tests, it may suggest issues in the real image calibration such as the estimation of sky background which we discuss more in Sections \ref{results-galaxies} and \ref{undetected-sources}.

\subsubsection{SOF CModel Colors}

\begin{figure*}[ht!]
    \centering
    \includegraphics[width=\textwidth]{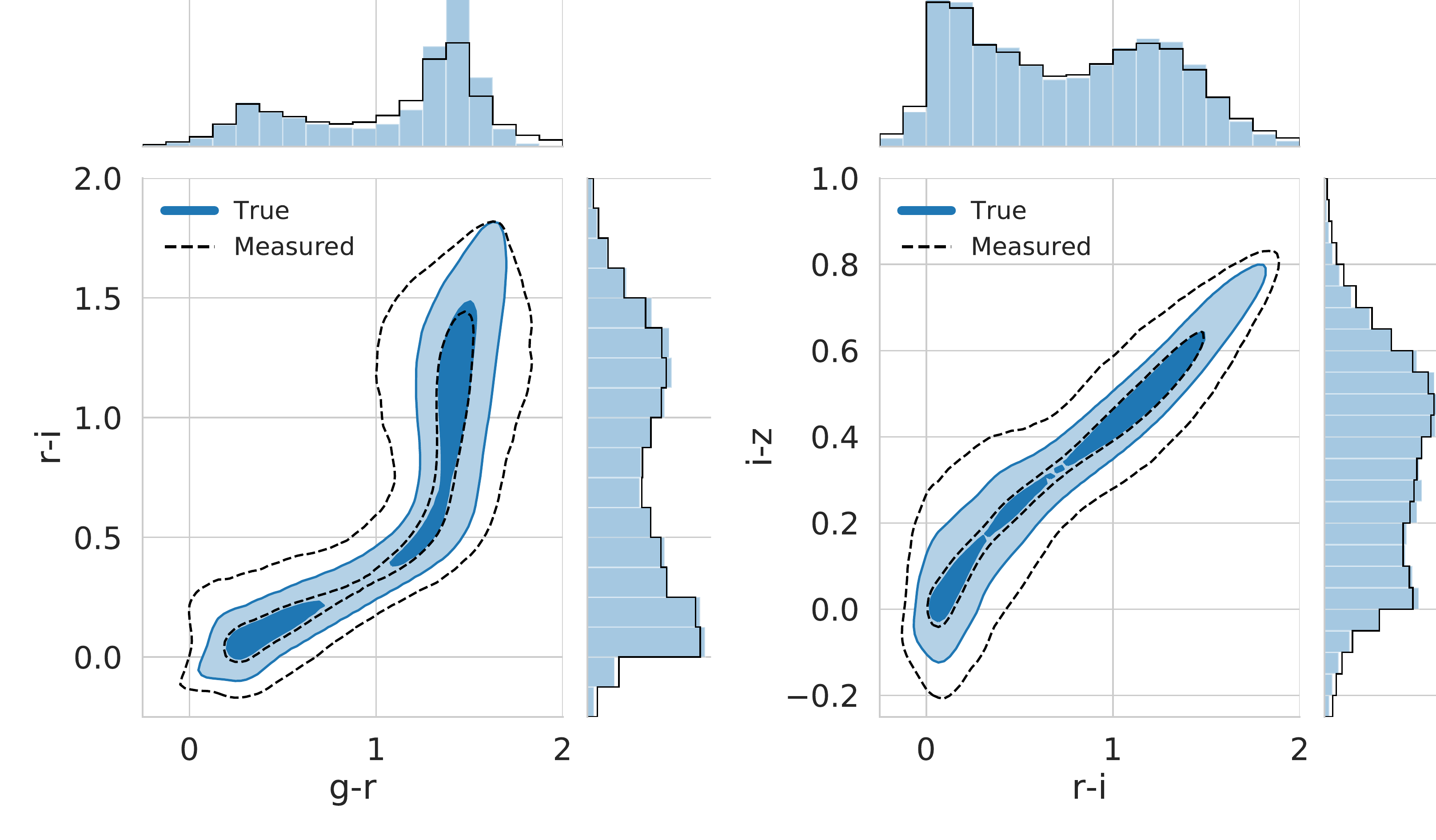}
    \caption{The $g-r$ vs. $r-i$ and $r-i$ vs. $i-z$ color-color distributions for the input colors in blue and measured colors in black. The density contour lines correspond to the percentiles of the first two sigmas of a 2D Gaussian, containing  39.2\% and 86.5\% of the total data volume respectively. The  marginal distributions are included for comparison.}
    \label{fig:stars-color-color-response}
\end{figure*}

Of primary interest is the accuracy of the recovered colors due to their importance for photometric calibration, star-galaxy separation, photometric redshift estimation, and the study of Milky Way structure. We plot the difference in measured SOF CModel $g-r$, $r-i$, and $i-z$ color vs. input $\delta$-color with respect to the input color in Figure \ref{fig:stars-color-response}. The contours and summary statistics are computed in the same way as the magnitudes, though with a bin size of 100 mmag for $g-r$ and $r-i$ and 50 mmag for $i-z$. The color calibration for this sample is excellent. For the three colors examined here, the median color difference \medcstar{} is never greater than 5 mmag (0.5\%) from injected color of -0.25 to 1.25 and is most commonly less than 3 mmag (0.3\%). Beyond 1.25, \medcstar{} grows to a maximum of 25 mmag (2.3\%) too blue for $g-r$ while for $r-i$ it never exceeds an absolute difference of over 3 mmag. The mean responses vary significantly due to extremely long scatter tails in both directions from the magnitude difference and are less reliable estimators of the overall performance in this case. However, they tend to be within a factor of two of the medians except for $g-r$ which increases in absolute size dramatically after 0.75 due to the long tail as can be seen in the figure. The full set of summary statistics are shown in Table \ref{tab:star-mag-response}. Notably we do not find evidence of a systematic chromatic response in CModel color. 

Next we compare the color-color diagrams for $g-r$ vs $r-i$ and $r-i$ vs $i-z$ for the input and recovered samples in Figure \ref{fig:stars-color-color-response}. As expected, the recovered injected colors have broader distributions due to the inherited WF noise as well as moderately large magnitude scatter near the detection threshold. However, the broadening is concentrated outside of the 1-$\sigma$ contours where the agreement is extremely similar.

\subsection{Photometric Performance of \dfsample{}}\label{results-galaxies}

Unlike the synthetic star sample, \dfsample{} objects are sampled from fits to real sources contained in the DES DF. Thus not only are the properties of these injections far more diverse, but we do not have perfect knowledge of their true classification. However, we anticipate that most uses of this \balrog{} sample will be to calibrate galaxy samples used in cosmology analyses. In these cases, we do not care about the true classification as we want to capture the same contamination fraction as the data. For this reason we apply the cut \code{EXTENDED\_CLASS\_SOF > 1} and leave questions of star contamination to Section \ref{star-gal-sep}. Removing ambiguous matches with the cut \code{match\_flag\_1.5\_asec < 2} decreased the sample by just under 1.5\%.

\begin{figure*}[ht!]
    \centering
    \includegraphics[width=1.\textwidth]{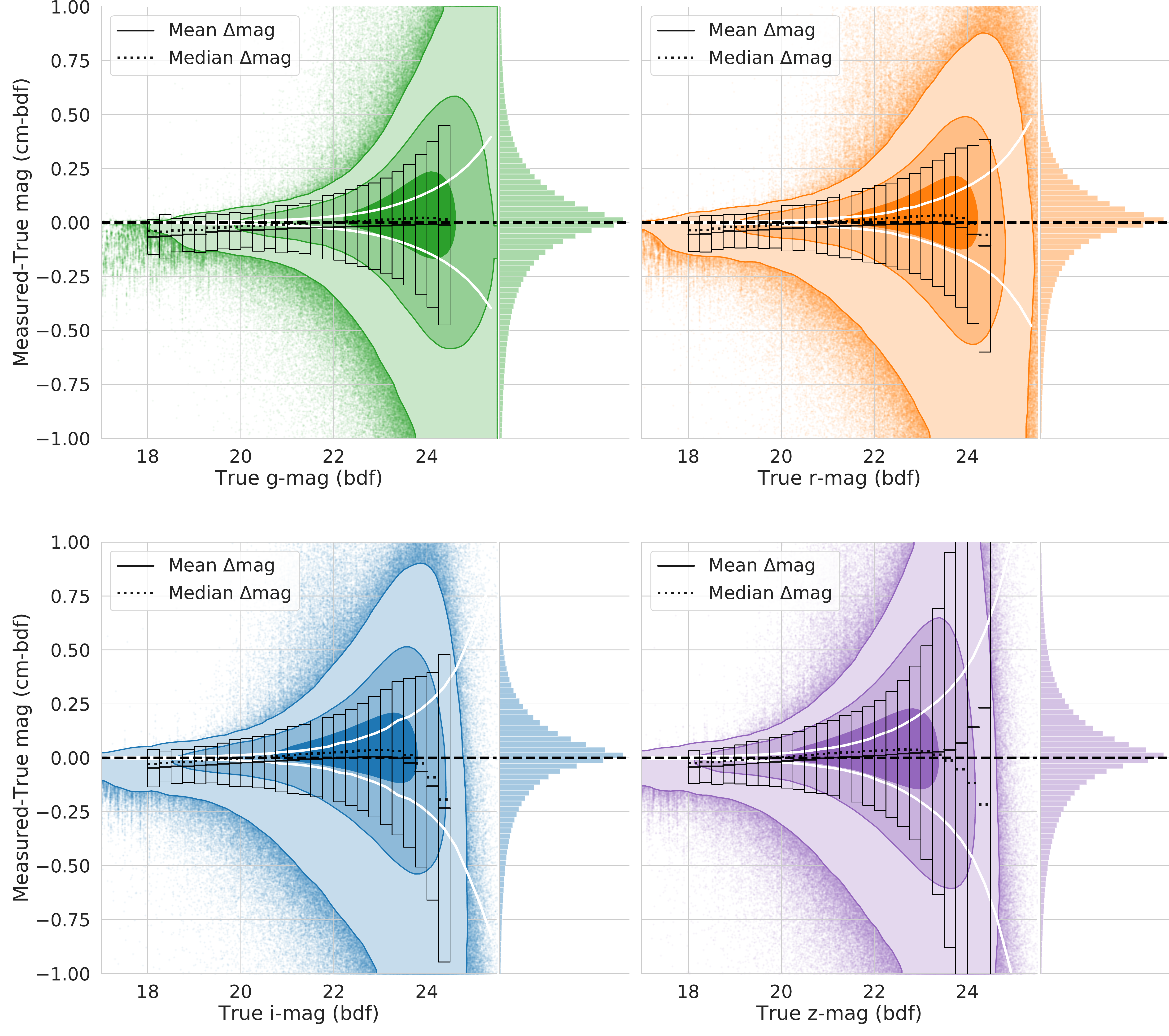}
    \caption{The distribution of differences in recovered \protect{$griz$} SOF CModel magnitude vs the injected DF magnitude (\protect{\dmgal{}}) as a function of input magnitude for the \protect{\dfsample{}} sample. The density is overplotted where the contour lines correspond to the percentiles of the first three sigmas of a 2D Gaussian, containing  39.2\%, 86.5\%, and 98.9\% of the data volume respectively. The mean (solid), median (dotted), and standard deviation of the magnitude responses in bins of size 0.25 magnitude are shown in the overlaid black bars. These are compared to the reported SOF CModel errors by the \response{solid} white lines which do not attempt to account for systematic effects. The marginal distributions of \protect{\dmstar{}} are included to highlight the small relative volume of the outlier tails.}
    \label{fig:meas-vs-true-mag-sof-gal}
\end{figure*}

There are numerous photometries and parameters whose response can be explored with this sample. We restrict ourselves largely to SOF CModel colors, magnitudes, and sizes here for brevity but find similar results for Metacalibration. As with \starsample{}, we include summary statistics of the tabular results in Appendix \ref{appendix:tabular-results}.

\subsubsection{SOF CModel Magnitudes}\label{results-galaxies-mags}

\begin{figure*}
    \centering
    \includegraphics[width=0.95\textwidth]{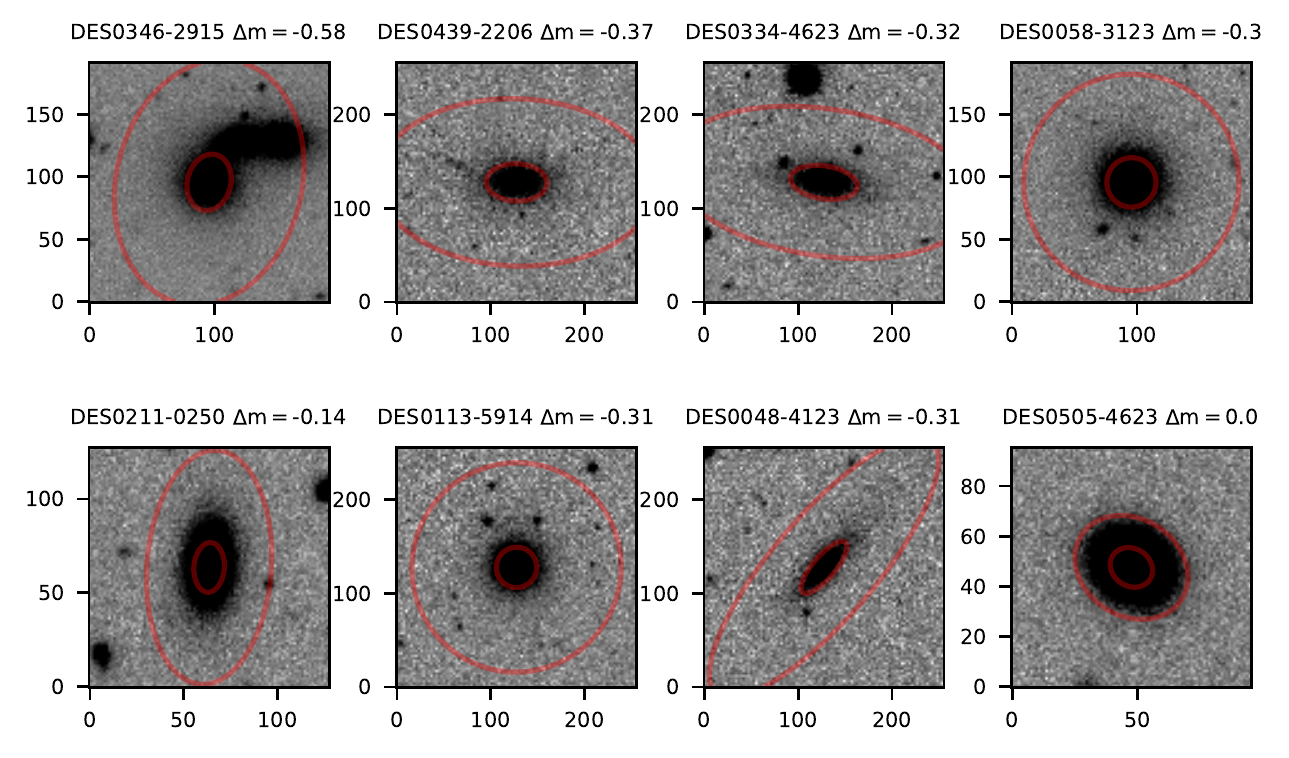}
    \caption{A few examples of injections that contribute to the long scatter tail in magnitude response of bright \protect{\dfsample{}} objects due to blending of extended DF injections discussed in \S \protect{\ref{results-galaxies-mags}}. Each injection had a true \protect{$g$}-band magnitude between 17 and 19, and we include the tilename and magnitude response \protect{$\Delta \text{m}$} at the top of each panel. The red lines correspond to the 50th and 95th percentile flux contours of the \textit{measured} profile. The extended profiles of these injections cause the MEDS image cutout size (based on the fitted \sextractor{} \code{FLUX\_RADIUS} value) to be relatively large which increases the probability of including real neighbors in the MEDS stamp. This in turn can cause SOF to significantly overestimate the \protect{\code{cm\_T}} size which leads to a much larger \protect{$\Delta\text{m}$} than one would naively expect for objects with these bright magnitudes. This is discussed further in \S \ref{catastrophic-fitting}. The final panel shows a typical bright but compact object that is very well calibrated for comparison. Note the presence of a nearby source in the bottom that could have potentially caused the same failure mode if the box size had been slightly larger. The stretch in each panel runs from $-3 \sigma_{\rm sky} \rm to +10 \sigma_{\rm sky}$.}
    \label{fig:bright-blending}
\end{figure*}

We compare the difference in recovered SOF CModel magnitude vs. true DF magnitude \dmgal{} as a function of input magnitude for $griz$ bands in Figure \ref{fig:meas-vs-true-mag-sof-gal}. As with \starsample{}, we characterize the photometric performance of \dfsample{} measured galaxies with the summary statistics \mdmgal{}, \medgal{}, and \siggal{} in bins of truth magnitude overplotted in black bars. Unsurprisingly, the overall scatter in magnitude response for this sample is significantly larger than for the pure stellar injections due to the rich variety of injected morphologies and issues with blending of extended sources. The measured \siggal{}'s reflect this by being an average of over 4 times larger than the corresponding \sigstar{} distribution over the same magnitude range, with the ratio reaching as high as 9 for very bright objects. We then expect the mean response bias \mdmgal{} to be larger as well, but their behaviour is more interesting than the stellar sample. On the bright end below 19th magnitude, the 50th-99th percentile of objects are detected within 30 mmag (or 2.7\%) of truth but there is a clear asymmetric preference for the recovered flux to be too large for the remaining objects. This result is driven by a sizeable fraction of bright, extended injections that are commonly blended with existing Y3 GOLD galaxies and are subsequently measured to have far too large of a size. The measured fluxes of these objects vary significantly depending on local conditions and create visible vertical lines in the response scatter due to their many injection realizations and relatively small population of objects with true magnitude less than 19. Image cutouts for a set of these objects along with the 50th and 95th percentiles of their measured CModel flux profiles are shown in Figure \ref{fig:bright-blending} -- in addition to a more compact, typical injection at the same input magnitude that does not suffer from proximity effects or blending. These examples of large magnitude responses correlated with measured size errors are the first hint of a systematic issue with SOF fits in crowded fields that we investigate in more detail in \S \ref{catastrophic-fitting}.

As in the \starsample{} sample, we detect a relatively small but clear band dependence in the mean and median responses. For all input magnitude bins brighter than 23 where the sample is nearly complete, there is a monotonic increase in the mean and median response in $griz$ with absolute spread of ${\sim}$16 mmag, or about 1.4\% difference between $g$ and $z$. This effect was hinted at in the response of the pure stellar sample but is far more evident here. This chromatic response is diluted but not eliminated when binning in measured S/N rather than input magnitude, with \medgal{} no longer strictly monotonic and with a typical spread of 4-5 mmag for $riz$-bands but 10-20 mmag when including $g$-band for S/N greater than 20.

We believe this chromatic effect is due to a systematic overestimation of the true sky background level in DES (and thus \balrog{}-injected) images. The \sextractor{} sky mode estimator is somewhat susceptible to the presence of neighboring objects in its sky annulus, especially in moderately to highly crowded fields. A mode estimate for the background appropriately allows for the fact that there will be background sources, detections, and undetected sources which is particularly important in the presence of many sources \citep{Stetson87}. As a precise mode estimation was once computationally impractical, traditional codes such as \sextractor{} have in practice used a Pearson-style mode estimator $\rm Mode_{\rm est} = 2.5\cdot Median - 1.5\cdot Mean$ for background estimation. This can result in a slight bias in overestimating the background which becomes larger as the field becomes more crowded and in the neighborhood of bright stars with extended wings (E. Bertin, private communication). This sky overestimation results in too faint a measurement of a galaxy's true magnitude and the effect is stronger when there is more sky noise per object signal. 

\begin{figure*}[ht!]
    \centering
    \includegraphics[width=1\textwidth]{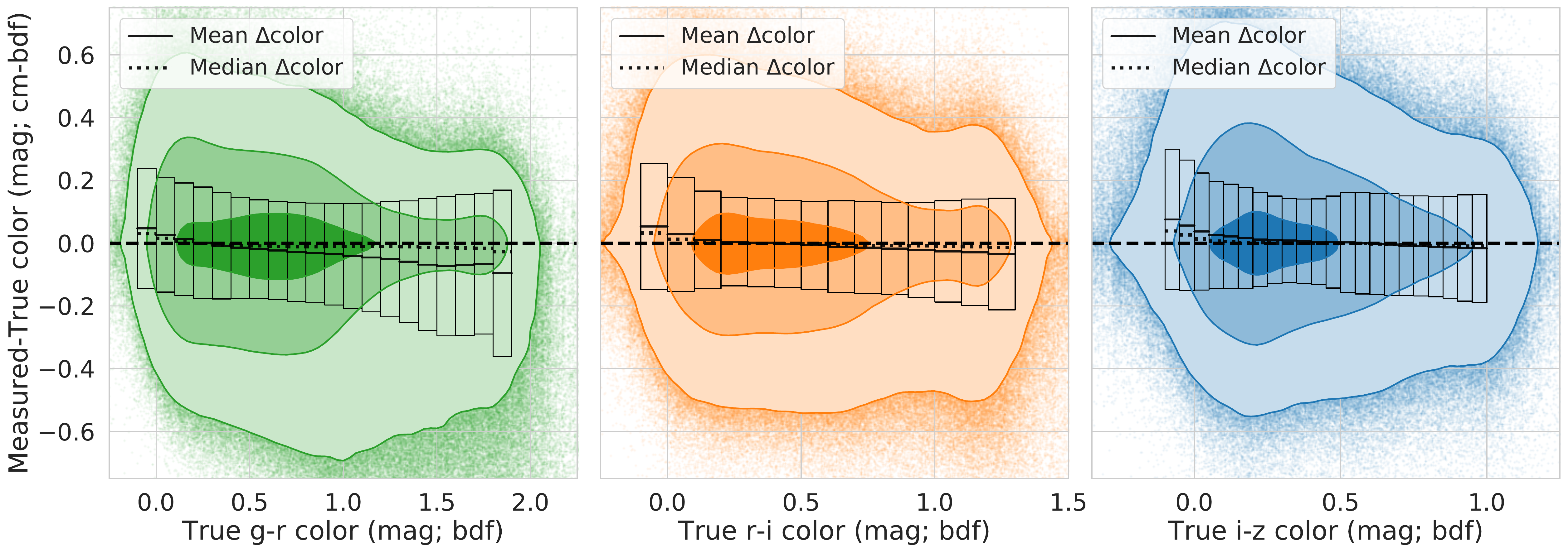}
    \caption{The distribution of differences in measured SOF CModel $g-r$, $r-i$, and $i-z$ color vs. the injected DF color  (\protect{\dcgal{}}) as a function of input color for the \protect{\dfsample{}} sample. The density is overplotted where the contour lines correspond to the percentiles of the first three sigmas of a 2D Gaussian, containing  39.2\%, 86.5\%, and 98.9\% of the data volume respectively. The mean (solid), median (dotted), and standard deviation of the magnitude responses in bins of size 100 mmag magnitude for $g-r$ and $r-i$ and 50 mmag for $i-z$ are shown in the overlaid black bars.}
    \label{fig:gals-color-response}
\end{figure*}

The fact that the sky is more crowded as one moves from bluer ($g, r$) to redder ($i, z$) bands could lead to the chromatic effect described above. That the scale of this effect is lessened by binning objects of similar S/N across bands together supports this conclusion. Note that these offsets are computed with dereddened magnitudes, which has the effect of enhancing the chromatic offset in $g$-band compared to the redder bands. Additionally, \cite{Eckert_2020} analyzed the noise properties of DES images and found that there was a slight positive bias induced in the sky noise level due to faint unresolved sources in the field of essentially all images (see Section \ref{undetected-sources} for more details). The sign of this effect, while smaller, has the same trend and was found to only be significant for $riz$ bands. We plan to investigate this further for the Y6 \balrog{} analysis and potentially propose additional magnitude corrections to account for this effect.

\subsubsection{SOF CModel Colors}\label{results-galaxies-colors}

\begin{figure*}[ht!]
    \centering
    \includegraphics[width=\textwidth]{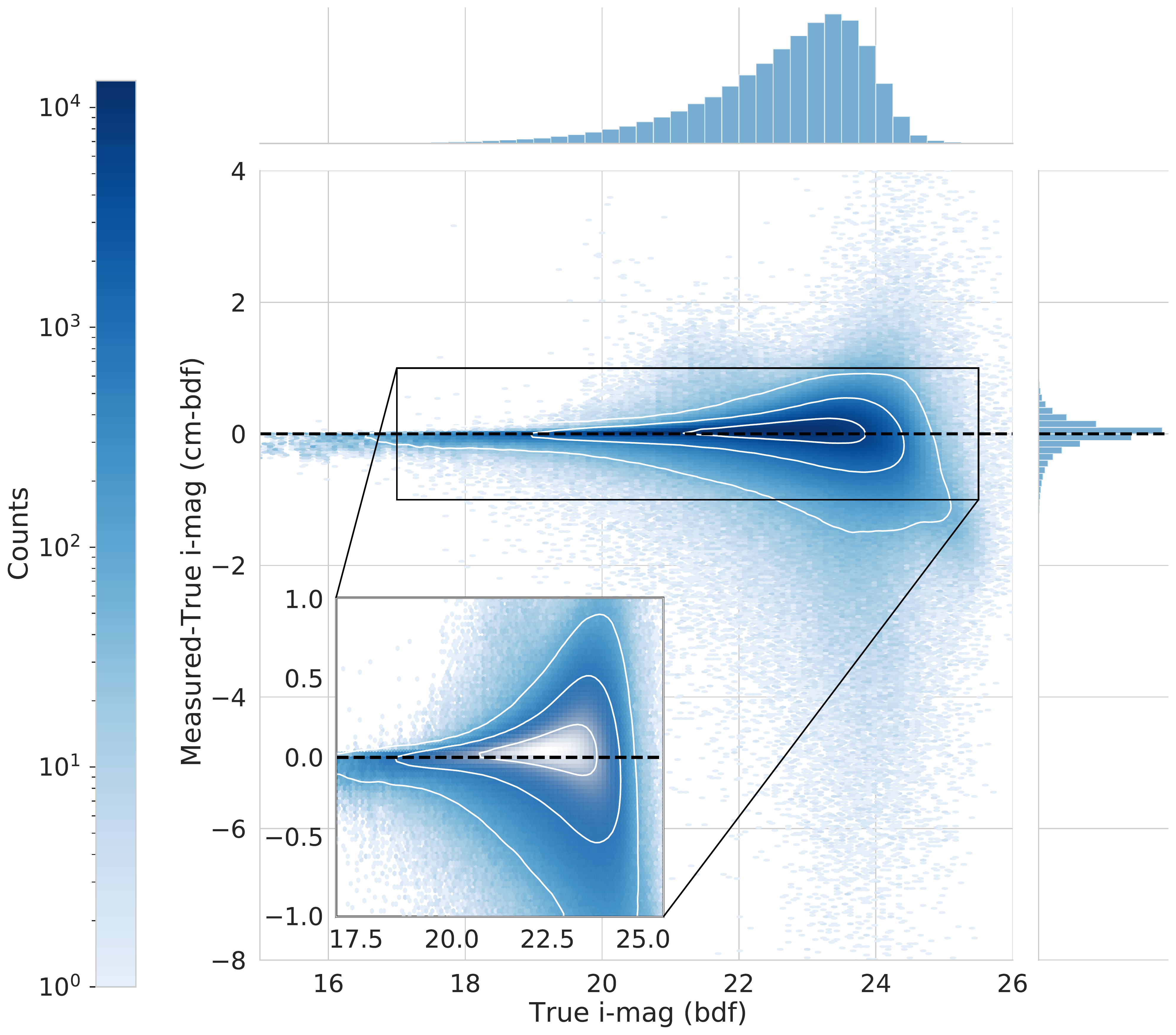}
    \caption{The distribution of differences in recovered \protect{$i$}-band SOF CModel magnitude vs the injected DF magnitude (\protect{\dmgal{}}) as a function of input magnitude. The inset corresponds to the $i$-band panel in Figure \protect{\ref{fig:meas-vs-true-mag-sof-gal}} where the density contours still contain 39.2\%, 86.5\%, and 98.9\% of the data volume respectively. While most of the density is captured in the inset, it misses many of the rich features of the full magnitude response -- particularly the long outlier tail of injections measured to have magnitudes up to 10 greater than truth. We explore some of the causes of this in \S \protect{\ref{catastrophic-fitting}}.}
    \label{fig:meas-vs-true-mag-sof-gal-tails}
\end{figure*}

Next we investigate the color response of \dfsample{} objects in Figure \ref{fig:gals-color-response}, where we plot the difference in measured SOF CModel $g-r$, $r-i$, and $i-z$ colors vs the injected DF colors \dcgal{} against the input colors. The density contours and overplotted summary statistics are defined in the same way as the previous plots. While the color response scatter is significantly larger than in \starsample{}, the overall calibration is still excellent and with less extreme outlier tails than in the individual magnitude responses. The behaviour of the summary statistics is slightly more complex but we find that the median color response \medcgal{} is typically ${\sim}$3 mmag (0.3\%) too faint from -0.25 to 0 and ${\sim}$1-11 mmag too bright between 0 and 1.0 for all three colors. The responses are much noisier outside of these regions due to much smaller sample sizes. \medcgal{} tends to be ${\sim}$15-25 mmag (1.4-2.2\%) too faint below 0.25 and 15-25 mmag too bright beyond 1.0 for all colors (though a bit worse for $r-i$, reaching 12\% too bright near 1.5) while \mdcgal{} differences are about three times as large as \medcgal{} in the same direction depending on the color and bin. As with the stellar injections, individual \mdcgal{} and \medcgal{} bin values can vary significantly due to long scatter tails and we find no evidence of a systematic chromatic response in CModel color. The full color response is summarized in Table \ref{tab:gal-color-response}. 

\subsubsection{Catastrophic Model Fitting}\label{catastrophic-fitting}

\begin{figure*}
    \centering
    \includegraphics[width=\textwidth]{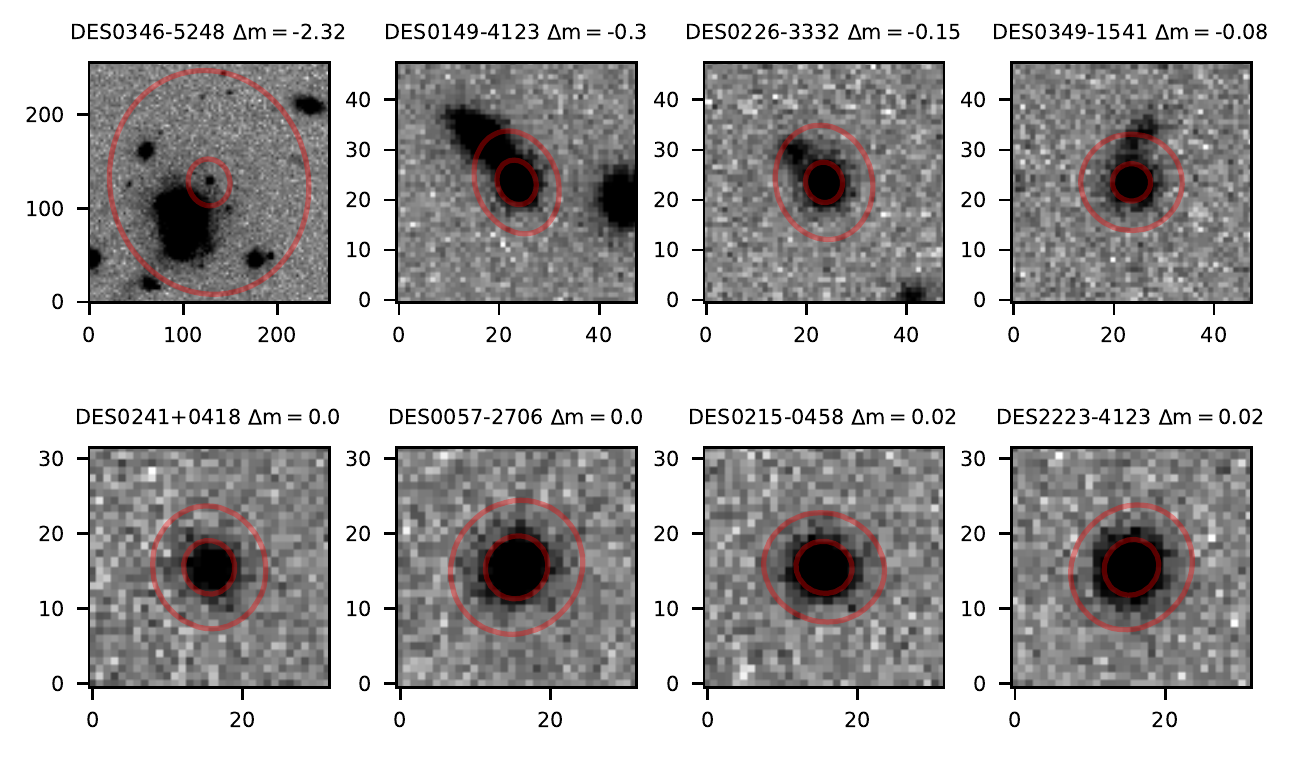}
    \caption{The MEDS image cutouts for a few injection realizations of the same DF object with true \protect{$r$}-magnitude of 21.42 in eight distinct WF tiles (\protect{\code{bal\_id}} of 10034605248852). The red contours give the 50\% and 95\% enclosed light apertures for the injected object as modeled in each tile. The difference between the measured and injected magnitude \protect{$\Delta\text{m}$} is listed next to each tile name, with the cutouts ordered by the magnitude response. The box sizes are in 0.263\protect{\asec{}} pixels. Not all cutouts are the same size, as the box size expands based on the initial \protect{\sextractor{}} \protect{\code{FLUX\_RADIUS}} measurement. The true scale length of the object (after PSF deconvolution) is 0.77\protect{\asec{}}. The fitted profile for the object on tile DES0149-4123 is 1.0\protect{\asec{}} and while that on tile DES0346-5248 is an unrealistic 17\protect{\asec{}}, leading to an overestimate of the object flux corresponding to an error of 2.32 magnitudes. The stretch in each panel runs from \protect{$-3 \sigma_{\rm sky} \rm to +10 \sigma_{\rm sky}$}.}
    \label{fig:eightboxes}
\end{figure*}

\begin{figure*}
  \centering 
  \includegraphics[width=\textwidth]{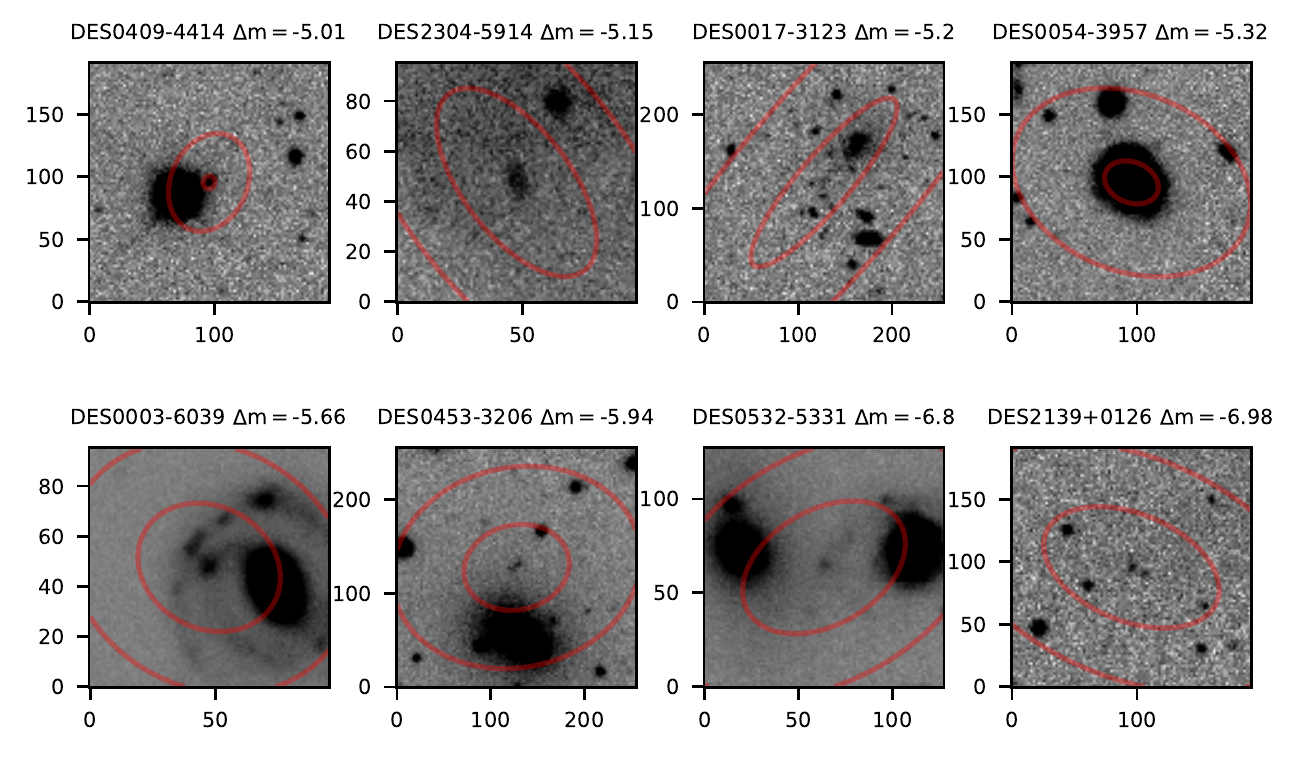}
  \caption{The MEDS image cutouts for eight \balrog{} objects with extremely large differences between the measured and injected magnitude $\Delta\text{m}$. The red lines correspond to the 50th and 95th percentile flux contours of the measured profile. These injections happened to be placed in regions of rapidly varying sky brightness, in the spiral arm of a large spiral galaxy, in a rich cluster, near a stellar diffraction spike, in between two extended galaxies, or simply in crowded fields. In all cases the fitted size is far too large for the source, which in turn leads to an overestimate of the object's flux. This processed is discussed in detail in \S \ref{catastrophic-fitting}. The stretch in each panel runs from 
  \protect{$-3\sigma_{\rm sky} \rm to +10 \sigma_{\rm sky}$}.
  }
  \label{fig:bad-mag-cutouts}
\end{figure*}

While Figure \ref{fig:meas-vs-true-mag-sof-gal} shows that the vast majority of magnitude responses are well calibrated and are typically much less than \dmgal{} of 0.5, it ignores the very long tail of up-scattered outliers that are far larger than the measured photometric errors would predict. The responses of these outliers from blends and catastrophic photometry failures can be over an order of magnitude larger than those previously discussed as shown for $i$-band in Figure \ref{fig:meas-vs-true-mag-sof-gal-tails} where the contours from Figure \ref{fig:meas-vs-true-mag-sof-gal} are overlaid in white.

Here the true complexity of even a small slice of the transfer function is revealed: The many competing effects are often in opposition, with biases in the opposite direction of long, asymmetric tails that vary as a function of truth magnitude in a complex way. Simple Gaussian summary statistics like \mdmgal{} and \siggal{} are not able to appropriately capture the magnitude of these features and we argue that the \balrog{} samples themselves (or at least higher fidelity forms of data compression) should be used for most cosmological analyses that need accurate photometric error modeling. Examples of how the full richness of the transfer function can be used in photometric redshift calibration and the magnification of lens samples are given in Sections \ref{photo-z-calibration} and \ref{magnification} respectively.
  
However, it is reasonable to be skeptical of magnitude responses of \dmgal ${\sim}$2-8 (a factor of 6-1,600 in flux!) by supposedly well-calibrated photometry pipelines. To demonstrate what is causing these extremely large differences in recovered flux, we show in Figure \ref{fig:eightboxes} a set of injections of the same DF object with $r$-band magnitude of 21.42 into eight different WF tiles where the red lines correspond to the 50th and 95th percentile flux contours. In most cases the true magnitude is recovered within the reported errors of a few percent. However, in four instances there is at least one nearby object contained in the MEDS cutout image that interferes with SOF's ability to provide a reliable fit due to either an excess of masked pixels in the cutout or residual light unassociated with the injection. The result is a fitted characteristic size $\code{cm\_T}$ which is much greater than its actual size. For this particular injection, the true size of the object (after deconvolution with the PSF) corresponds to a scale length of 0.77\asec{}. Yet in the four cases with nearby sources the fitted size of the object is at least 1\asec{}, resulting in a flux measurement which is significantly greater than that of the input true flux. In the worst case for tile DES0346-5248, the target object is by chance injected near a very bright pair of merging galaxies and is fitted with a scale length of over 17\asec{} resulting flux 2.32 magnitudes brighter than the input DF value. 

These photometric measurement failures correlated with errors in measured \code{cm\_T} can be even more dramatic. In Figure \ref{fig:bad-mag-cutouts} we show eight examples of catastrophic fitting failures due to crowded fields, nearby bright stars, and unflagged image artifacts. These rare but real environments lead to \balrog{} magnitude responses from 5 to even 7 magnitudes brighter than the injected truth. We emphasize that all of these objects pass the basic Y3 GOLD science catalog quality cuts described in the beginning of Section \ref{photometric-results}.

While the exact causal relationship between complex local environments and extreme magnitude errors requires further analysis, preliminary investigations suggest the following: In crowded fields or areas with unusual image features or artifacts, the \sextractor{} \code{FLUX\_RADIUS} (which defines a circle that contains half of the total corresponding \code{FLUX\_AUTO} value) can get artificially inflated in size as compared to what it would return for an object in an isolated environment. As a source's MEDS cutout image size is rounded up to the next integer multiple of 16, this leads to a MEDS stamp that is significantly larger than what is needed to fit the relevant flux profile in question. This leaves large areas of the stamp with masked pixels when fit with SOF as the algorithm masks rather than models the light of other detected sources within the cutout. The resulting CModel fits then preferentially overestimate \code{cm\_T} for this subpopulation which can greatly increase the inferred flux for a given surface brightness measurement - though we defer investigations into the exact details of the scale and frequency of this effect for a future analysis.

\begin{figure}
    \centering
    \includegraphics[width=0.4725\textwidth]{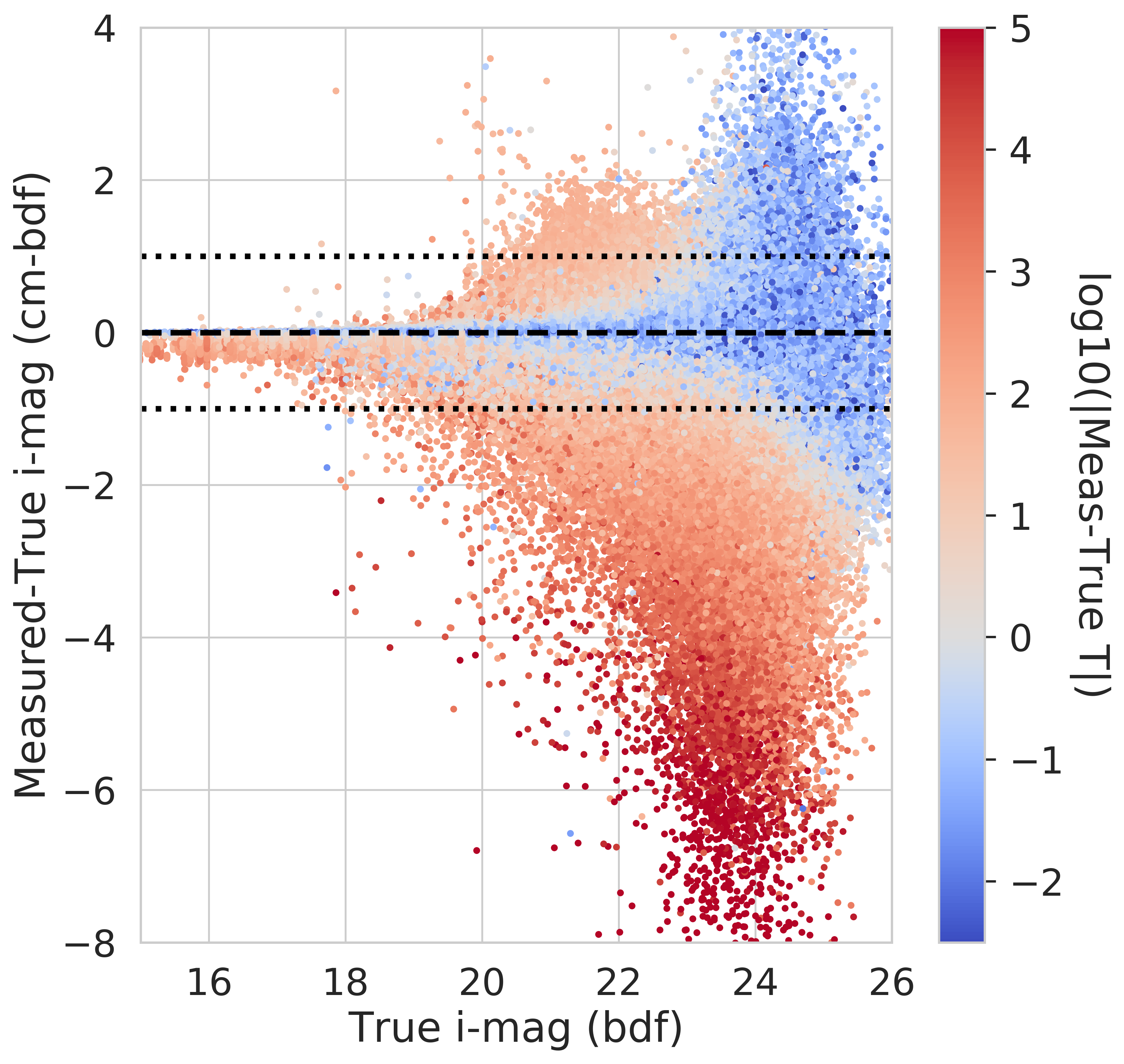}
    \caption{The full $i$-band magnitude response \dmgal{} for \dfsample{} shown in Figure \ref{fig:meas-vs-true-mag-sof-gal-tails} but now colored by the logarithmic absolute error in recovered size parameter \code{cm\_T} vs input size \code{bdf\_T}. The response scatter is largely correlated by error in recovered size; injections with small \dmgal{} values typically have small errors in recovered \code{T} as well (in blue), while nearly all of the extreme magnitude outliers have correspondingly large size errors. The correlation is less strong past the detection threshold at $i{\sim}23$ where other systematic effects increase in importance.}
    \label{fig:mag-scatter-by-T-sof}
\end{figure}

Even without a complete understanding of the underlying cause, the correlation between \dmgal{} and $\Delta\code{T}$ is evident as can be seen in Figure \ref{fig:mag-scatter-by-T-sof}. Here we have plotted the full $i$-band magnitude response of \dfsample{} but colored individual responses by the absolute difference in measured \code{cm\_T} vs. input \code{bdf\_T}. The vast majority of injections with truth $i$-magnitude below 23 with very small \dmgal{} responses have \code{T} differences much less than 1 which are colored blue. Bright objects with responses substantially below the zero line have moderately large errors in recovered \code{T} as we discussed in \S \ref{results-galaxies-mags}, while fainter injections with enormous magnitude errors have correspondingly large errors in \code{T} -- reaching as high as the parameter prior limit of $10^6$ arcsec$^2$ (or scale length of ${\sim}10^3$ arcsec). The situation is more complicated near and past the detection threshold, about 23rd magnitude in $i$-band, where additional systematic effects become important.

Model fitting photometry codes are complex, nonlinear, and sometimes non-local algorithms that can have unexpected consequences -- particularly for low S/N measurements, crowded fields, or when image artifacts are not appropriately weighted or masked. The journey from pixels to catalogs can at times be chaotic, and our modeling of photometric uncertainties should reflect this.

\subsubsection{Scatter from Ambiguous Matches}\label{scatter-ambiguous-matches}

Despite the efforts described in Section \ref{matching} there will always be some ambiguity in the matching to injected sources that can introduce large, non-physical scatter. To check this, we visually inspected hundreds of the MEDS stamps of \balrog{} objects whose absolute magnitude response was greater than 2 -- and in particular the set of objects with large \dmgal{} whose size errors were small. There were a few isolated instances of ambiguous matches where a faint injection landed in the very center of an extremely bright Y3 star whose GAp flux measurement failed. These can easily be accounted for by adapting our ambiguous matching algorithm to reject \balrog{} injections near objects with flagged GAp fluxes but this was not discovered in time to update the catalogs used in downstream measurements. However, this issue has negligible impact as we estimate only a few hundred instances in the total \dfsample{} sample.

\subsection{Star-Galaxy Separation}\label{star-gal-sep}

\begin{figure*}
\centering 
\subfloat[Star Selection]{%
  \includegraphics[width=0.49\textwidth]{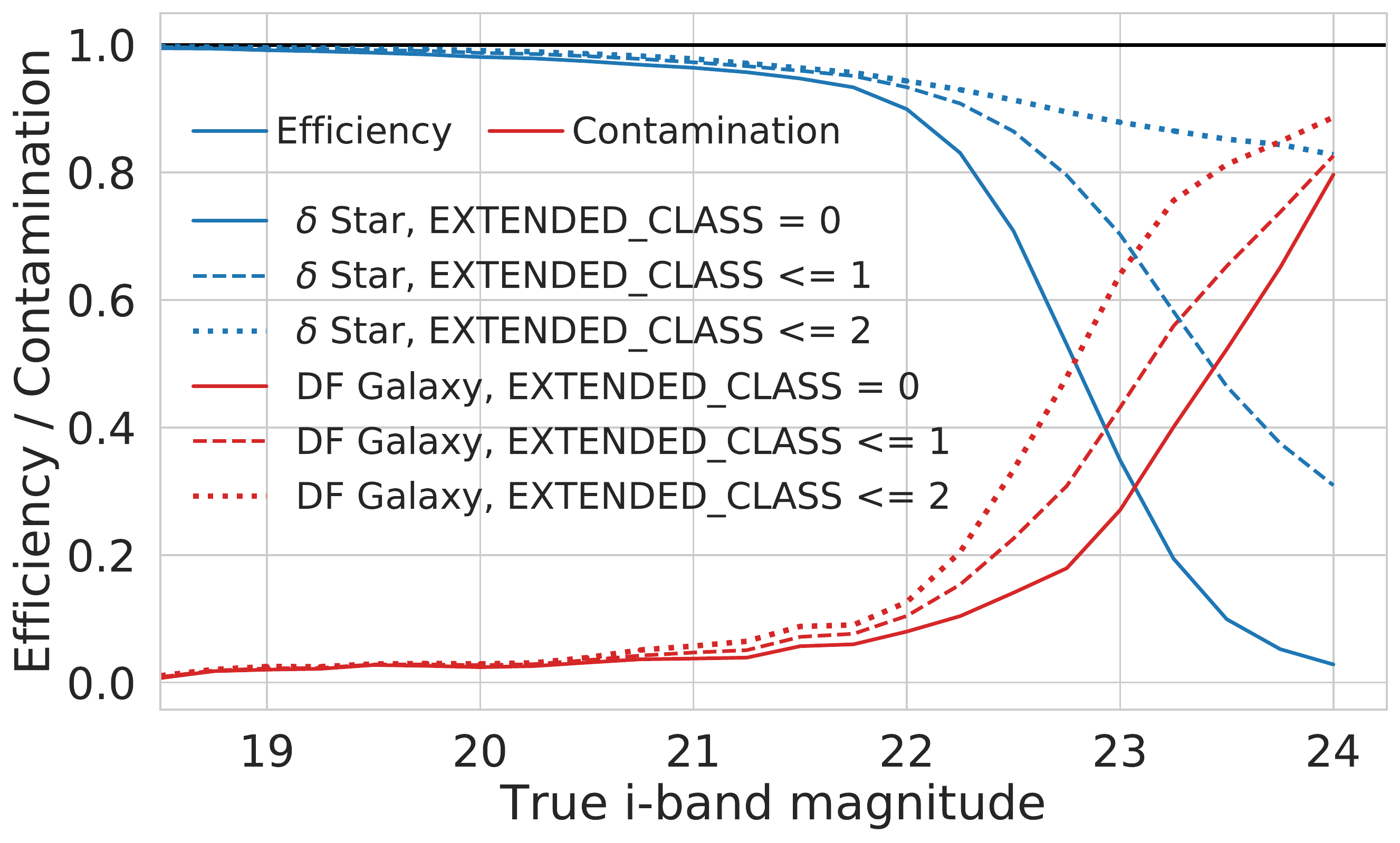}%
  \label{fig:star-eff-cont}%
}
\quad%
\subfloat[Galaxy Selection]{%
  \includegraphics[width=0.49\textwidth]{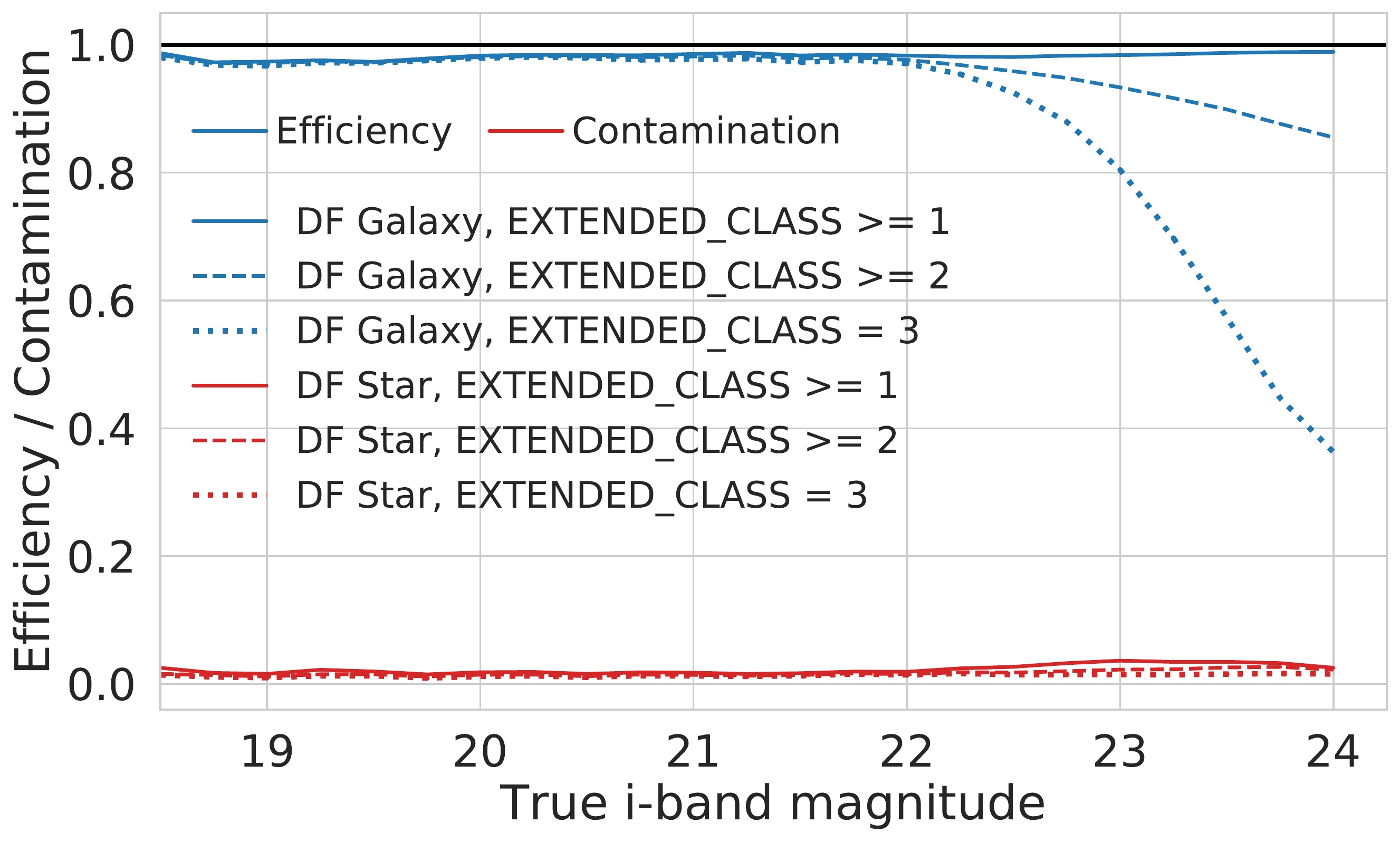}%
  \label{fig:gal-eff-cont}%
}
\caption[Caption with a citation]{The efficiency (in blue) and contamination (in red) of the \balrog{} stellar sample (\protect{\subref*{fig:star-eff-cont}}) and galaxy sample (\protect{\subref*{fig:gal-eff-cont}}). We use the $\delta$ injections of \protect{\starsample{}} as our population of true stars for (\protect{\subref*{fig:star-eff-cont}}) as it is a nearly pure sample, with only ambiguous matches as potential contaminates. We use the DF injections classified as galaxies from the DF \protect{$k$}-nearest neighbor (knn) classifier described in Section \protect{\ref{different-likelihoods}} as our true galaxy sample which has intrinsic uncertainty as detailed in \protect{\cite*{y3-deepfields}}. For (\protect{\subref*{fig:gal-eff-cont}}), we cannot use the $\delta$ injections as the contamination measurement requires a realistic ratio of galaxy and stars sources in the sample so we instead use the classified DF stars. Each line corresponds to the fraction of objects above or below the noted \protect{\code{EXTENDED\_CLASS\_SOF}} threshold value. We do not expect the galaxy efficiency to be 100\% even at magnitudes where complete due to small impurities DF knn classifier.}
\label{fig:efficiency-contamination}
\end{figure*}

We use the $\delta$ injections of \starsample{} to estimate the stellar efficiency (or true positive rate) in blue and the classified DF sources in \dfsample{} for the contamination rate (or false discovery rate) in red for the \balrog{} star sample as a function of injection magnitude in Figure \ref{fig:star-eff-cont}. The solid, dashed, and dotted lines represent the fraction of objects classified as less than or equal to an \texttt{EXTENDED\_CLASS\_SOF} value of 0, 1, or 2 respectively. While \dfsample{} is required to estimate the contamination rate in order to have a realistic relative ratio between star and galaxy counts, we use the \starsample{} sample to compute the efficiency as its truth classifications are nearly noiseless and the measurement does not need any external information about galaxy contaminants. We find that the stars are correctly classified ($\texttt{EXTENDED\_CLASS\_SOF} <= 1$) over 95\% of the time below an $i$-band magnitude of 21.75 and 80\% of the time below magnitude 22.75 before dipping to 70\% efficiency near the detection threshold at $i{\sim}23$. The stellar efficiency quickly drops to below 50\% beyond 23rd magnitude. The efficiency of high confidence stars ($\texttt{EXTENDED\_CLASS\_SOF} == 0$) follows a similar trend but reaches the previously quoted values about 0.5 magnitudes earlier. Alternatively, the rate of DF galaxies misclassified as stars stays below 10\% until 22nd magnitude where there is a sharp increase until the detection limit where at low S/N it is extremely difficult to differentiate between classifications. However, we again note that the stellar efficiency measurement is less noisy due to the higher degree of confidence in accurate classification compared to the DF sample.

We make equivalent measurements for the galaxy efficiency and contamination in Figure \ref{fig:gal-eff-cont} where the solid, dashed, and dotted lines now correspond to the fraction of objects classified as greater than or equal to \code{EXTENDED\_CLASS\_SOF} values of 1, 2, and 3. Here we must use sources in \dfsample{} exclusively as the ratio between stars in the $\delta$-sample and galaxies in the DF sample is not realistic as required by a contamination estimate. The efficiency is slightly lower than the stars on the bright end due to impurities in the DF knn classifier but is quite close to 100\% below 22nd magnitude. The efficiency of high-confidence galaxies ($\texttt{EXTENDED\_CLASS\_SOF} == 3$) decreases sharply near the detection limit, but over 85\% of DF galaxies with assigned classifications are correctly identified ($\texttt{EXTENDED\_CLASS\_SOF} >= 2$) down to 24th magnitude in $i$-band. The contamination rate of stars into the galaxy sample is consistently ${\sim}2$\% until 22nd magnitude where it rises slightly to 4\% at a magnitude of 23. This low level of contamination is largely due to the relatively small number of stars compared to galaxies at these magnitudes and is consistent with the findings quoted in \cite{y3-gold}. A table of the \balrog{} classification (or ``confusion'') matrix as a function of input magnitude is provided in Table \ref{tab:confusion-matrix}.


\section{Applications to DES Y3 Projects}\label{applications}

Below we present some of the most important applications of the Y3 \balrog{} catalogs, particularly those that are relevant for the DES Y3 cosmology analysis. To our knowledge, this is the first time an object injection pipeline has been used for any of the following measurements or played such a critical role in the calibration of a galaxy survey's cosmological constraints.

\subsection{Photometric Redshift Calibration}\label{photo-z-calibration}

Chief among the applications of our results is facilitating a novel inference method for the photometric redshift calibration of weak lensing samples. As shown in \citet*{Buchs_2019}, we can extract information from the DES Y3 DF to break degeneracies in the $riz$\footnote{Only the $riz$ Metacalibration fluxes are used when defining the tomographic bins.} color-redshift relation if we have accurate estimates of the corresponding WF properties of the DF sources. In this inference method, \balrog{} plays the essential role of determining the likelihood of a given deep, many-band color to be observed at a given region of noisier color-magnitude space in DES measurements at Y3 depth. This allows us to rigorously separate the contributions from measurement noise to the true color-redshift relation when estimating the ensemble photometric redshift distribution of the lensing source sample. In practice, this inference method is facilitated by the use of two Self-Organizing Maps (SOM) which classify the galaxies in the deep and wide samples into discrete classes, called \textit{cells}, of color and color-magnitude space. The redshift distribution of a given Y3 source is then given by
\begin{equation}
\label{eqn:sompz_inference}
    p(z|\hat{c},\hat{s},\theta)=\sum_{c} p(z|c)p_{\balrog{}}(c|\hat{c},\hat{s},\theta)p(\hat{c}|\hat{s},\theta),
\end{equation}

\noindent where $z$ is redshift, $c$ is deep SOM cell, $\hat{c}$ is wide SOM cell, $\hat{s}$ is the sample selection function, and $\theta$ is any additional conditions such as position on the sky. The middle factor $p_{\balrog{}}(c|\hat{c},\hat{s},\theta)$, a narrow slice of the full \balrog{} transfer function, expresses the likelihood of a deep color to be observed at a certain region of wide color-magnitude space. This transfer function serves to correctly weight the well-constrained redshift distribution $p(z|c)$ of each deep SOM cell according to the probability of detecting those galaxies. As the SOM cells $\hat{c}$ are determined by Metacalibration magnitude and color, \response{the \balrog{} samples are key to generating} a distribution of observed Metacalibration magnitudes for each injected DF galaxy.

In addition to breaking degeneracies in the color-redshift relation, \balrog{}, by virtue of enabling this scheme, facilitates avoiding otherwise prohibitive selection biases resulting from the use of spectroscopic redshifts for weak lensing redshift calibrations (see, e.g. \citealt{GruenBrimioulle2017}) because it uses spectroscopic redshifts only of galaxies for which 8 bands of DES DF photometry provide relatively well-constrained $p(z)$.

In the first application of this inference scheme to data, \citet*{y3-sompz} found that the intrinsic uncertainty in \balrog{}'s estimation of the transfer function is a negligible contributor to the overall error budget with an uncertainty on the mean redshift in each tomographic bin of $\sigma_{\overline{z}}<10^{-3}$. This is a significant accomplishment as \balrog{} was able to decrease the systematic bias in the photometric redshift estimates without contributing a novel source of intrinsic systematic uncertainty in its sampling of the transfer function, which was not obviously the case a priori. The use of \balrog{} in photometric calibration can be further leveraged in future analyses by incorporating positional-dependent selection effects $\theta$ in the used measurement likelihood $p_{\balrog{}}(c|\hat{c},\hat{s},\theta)$. For further details on this method, we refer the reader to \cite*{y3-sompz}.

\subsection{Magnification Bias on Clustering Samples}\label{magnification}

\response{The magnification of galaxy light profiles induced by gravitational lensing changes both the number of detected sources on the sky as well as their measured properties, such as size. This effect biases measurements of large-scale structure and should be taken into account in the modeling of galaxy clustering and galaxy-galaxy lensing correlation functions \citep{lens-mag-important}.}

\response{We can express the observed galaxy density density fluctuation field $\delta_g^{\text{obs}}$ for a particular redshift bin as}
\begin{equation}\label{delta-mag-def}
    \response{\delta_g^{\text{obs}}=\delta_g^{\text{int}}+\delta_g^{\text{mag}},}
\end{equation}

\noindent \response{where $\delta_g^{\text{int}}$ is the intrinsic number density fluctuation field and $\delta_g^{\text{mag}}$ is the contribution due to magnification. In the weak lensing regime where the magnitude of the convergence $\kappa$ and the shear $\gamma$ are much less than 1, we can approximate the local magnification $\mu$ in terms of the convergence only as}
\begin{equation}\label{mu-def}
    \response{\mu=\frac{1}{(1-\kappa)^2-|\gamma|^2}\approx\frac{1}{1-2\kappa}\approx1+2\kappa.}
\end{equation}

\noindent \response{Under this approximation, the magnification contribution to the number density is proportional to the local convergence}
\begin{equation}\label{delta-mag-prop-to-k}
    \response{\delta_g^{\text{mag}}=C\kappa,}
\end{equation}

\noindent \response{where $C$ encodes the slope of the intrinsic flux distribution of the sample \citep{GF_SV_magnification}, as well as any selection effects.}

\response{Magnification impacts the observed galaxy density field} through two competing effects; a geometric suppression factor $C_{\rm area}$ resulting from the increased sky area for a fixed set of detections, and a boost in the detection efficiency of faint sources which increases the local number density \response{captured by $C_{\rm sample}$:}
\begin{equation}
    \delta_g^{\text{obs}}\approx\delta_g^{\text{int}}+[C_{\rm sample}+C_{\rm area}]\cdot\delta\kappa.
\end{equation}

\noindent \response{In addition, magnification may change the measured properties of sources that would be detected without this effect -- such as their size or even color as the blending rate is increased  -- which may alter their selection into clustering samples or their assigned tomographic redshift bin.}
    
While a simple argument in \cite{y3-2x2ptmagnification} shows $C_{\rm area}$ to be -2, the contribution by $C_{\rm sample}$ for even a simple linear response to $\delta\kappa$ depends on the \response{the change in observed number density} $\response{n^\text{obs}}$ \response{for a given $\delta\kappa$ compared to the intrinsic number density $n^\text{int}$ as a function of measured object fluxes $F$:}
\begin{equation}\label{eq:csample}
    \response{C_{\rm sample}\cdot\delta\kappa = \frac{n^{\text{obs}}(F;\kappa+\delta\kappa)-n^{\text{obs}}(F;\kappa)}{n^{\text{int}}(F)},}
\end{equation}

\noindent \response{which is extremely difficult to model as} $\response{n^\text{obs}}$ \response{implicitly depend on complex detection and measurement systematics.}

To aid in this effort, supplemental runs to \runtwo{} and \runtwoa{} (designated as \runtwomag{} and \runtwoamag{} respectively) were created where the same input objects were injected with identical simulation configuration except for an additional \galsim{} \code{magnify} call that was applied to all objects uniformly. Each object was given a lensing magnification \response{corresponding to} $\response{\delta\kappa=0.01}$, effectively increasing the flux and area of objects by \response{about 2\%}. A given galaxy sample selection can be applied to both the magnified \response{$(\kappa=\delta_\kappa)$} and unmagnified \response{($\kappa=0$)} runs and Equation \ref{eq:csample} can be used to estimate the magnification bias $C_{\rm sample}$. This estimate will include not only the impact of magnification on galaxy fluxes but any selection bias (e.g. on size) introduced by \response{the altered images on the downstream fitted photometry.}

\begin{figure}[t!]
    \centering
     \includegraphics[width=0.4725\textwidth]{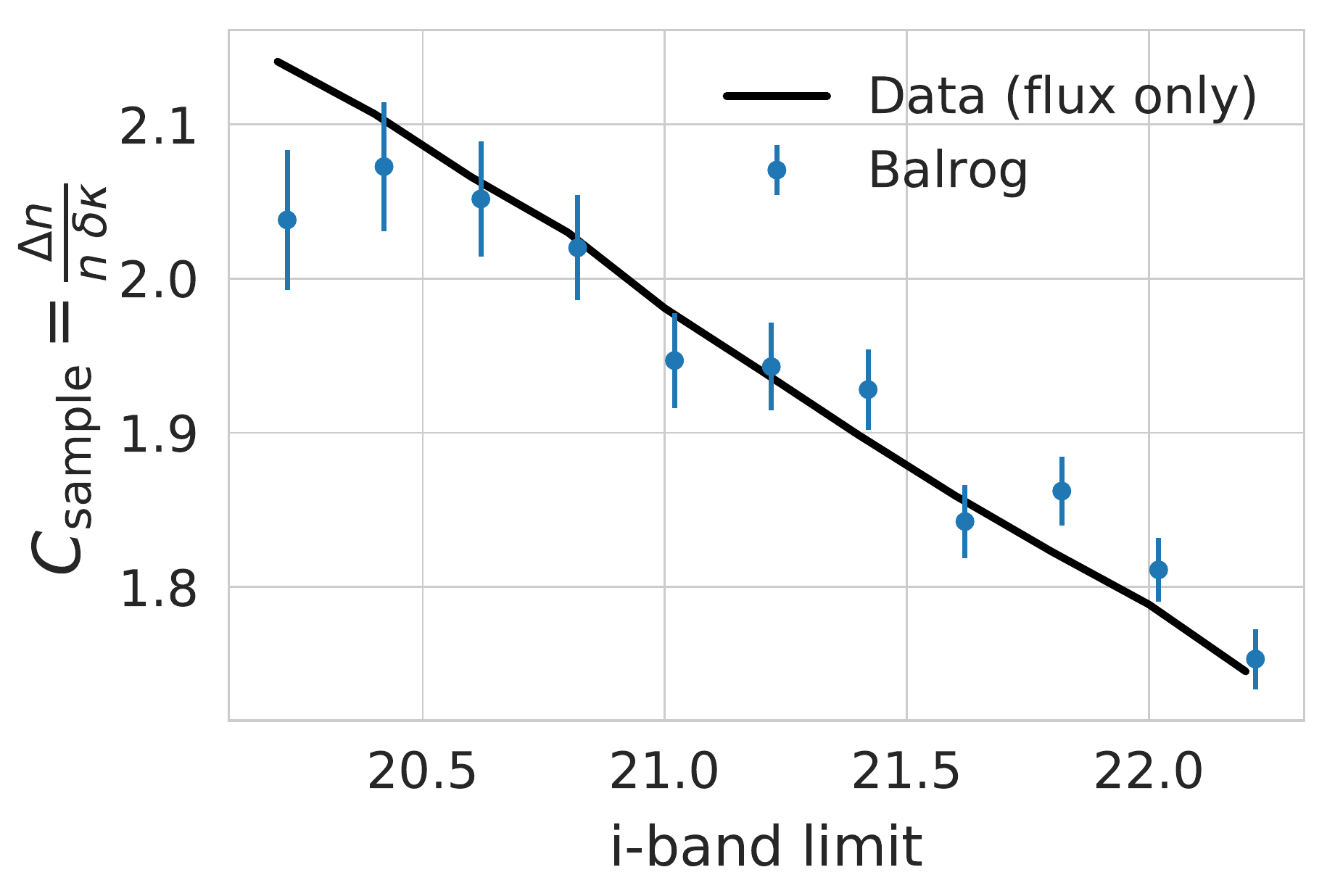}
    \caption[Caption with a citation]{The magnification bias from boosted detection efficiency \protect{$C_{\text{sample}}$} estimated on samples with a uniform $i$-band flux magnitude limit. The \protect{\balrog{}} estimates in blue use the magnified \protect{\balrog{}} runs with a 2\% magnification applied to every input object. The data estimate applies the same artificial magnification to the galaxy magnitudes in the data and reapplies the selection. The \protect{\balrog{}} estimates of \protect{$C_{\text{sample}}$} are consistent with the flux-only estimates for this very simple selection, indicating that the contributions from additional selection effects are small. However, the \balrog{} \protect{$C_{\text{sample}}$} estimates are systematically lower than the simple data estimate for the real Y3 samples used in \cite{y3-2x2ptmagnification} where the selections are significantly more complicated.} \label{fig:magnification_flat_flux}
\end{figure}

Figure \ref{fig:magnification_flat_flux} shows the $C_{\rm sample}$ estimates from \balrog{} for samples with a constant $i$-band flux limit and a simple galaxy section criteria of 

\begin{center}
\begin{tabular}{ll}
\small
     & \texttt{EXTENDED\_CLASS\_SOF = 3} \\
    \texttt{AND} & \texttt{FLAGS\_GOLD\_SOF\_ONLY \& 126.}
\end{tabular}
\end{center}

\noindent The same process is also applied to the real data where magnification is applied only to the galaxy fluxes. For this very simple selection the \balrog{} estimates are consistent with the data flux-only estimates, indicating any contribution from size selection or other systematics is small. 

In \cite{y3-2x2ptmagnification} this \balrog{} methodology is applied to the lens samples used in the DES Y3 analysis including more complex color cuts and tomographic redshift binning. In this analysis, the {\sc maglim} lens sample \citep{y3-2x2maglimforecast}, which has a redshift dependent magnitude limit and tomographic binning, is found to have a $C_{\rm sample}$ from approximately 2 to 5 from low to high redshift. The redMaGiC lens sample \citep{redmagic}, which is a Luminous Red Galaxy (LRG) selection, has $C_{\rm sample}$ from values consistent with 0 to approximately 4 at high redshift. The \balrog{} estimates of $C_{\rm sample}$ are systematically lower than the flux-only estimates due to the additional selection effects captured by the full \balrog{} transfer function. See \cite{y3-2x2ptmagnification} for additional details.

\subsection{Noise from Undetected Sources} \label{undetected-sources}

\begin{figure}
    \centering
     \includegraphics[width=0.48\textwidth]{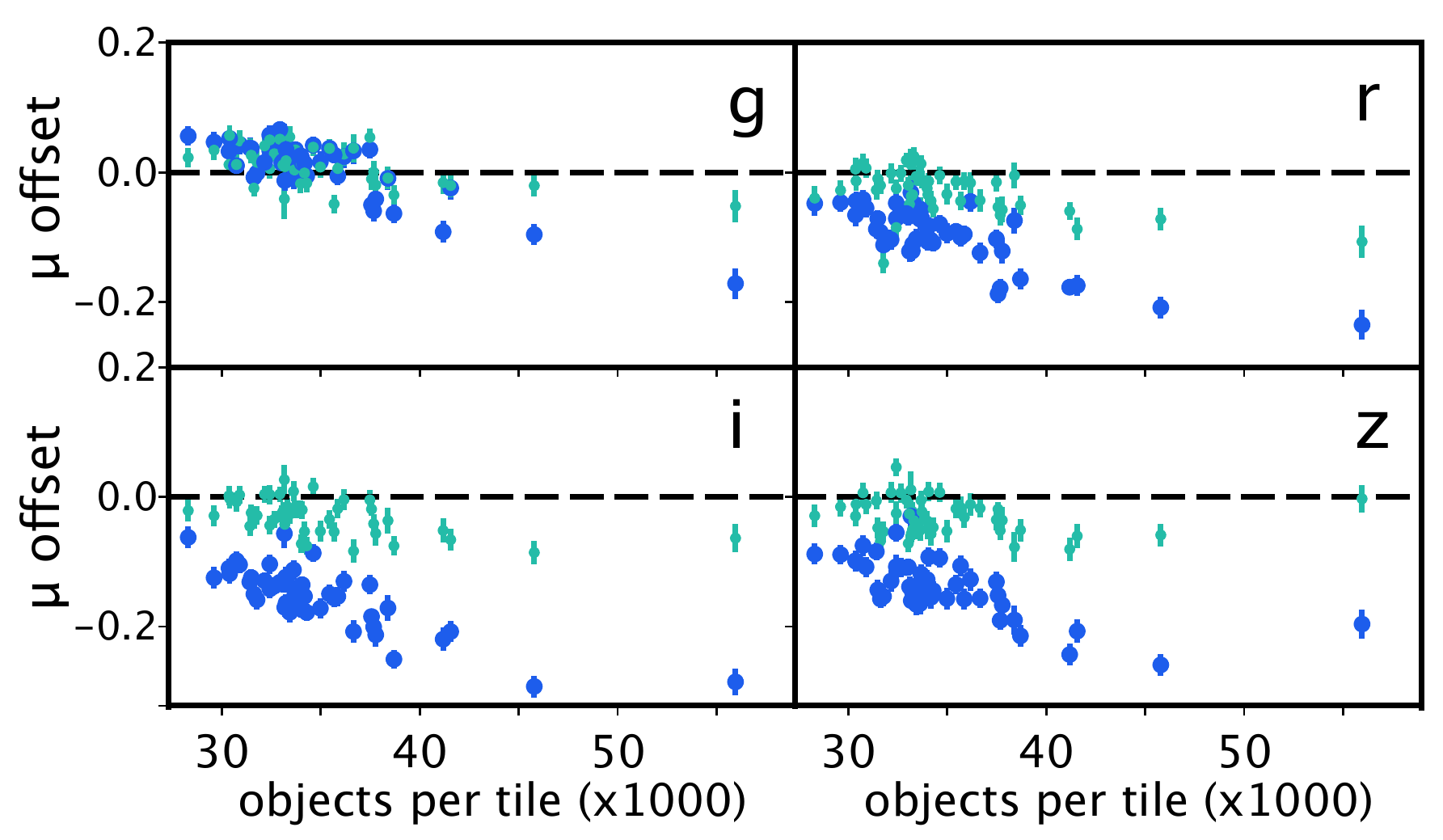}
    \caption[Caption with a citation]{Reproduced from \cite{Eckert_2020}, Figure 3. The Gaussian mean offset $\mu$ in the \response{BFD} flux moment pull distribution as a function of object density for the 48 used \balrog{} tiles in blue. The green points show the mean offset for the tiles after a local sky subtraction which mitigates the flux bias. While $g$-band is relatively unaffected, the redder $riz$-bands show statistically significant sky oversubtraction that is correlated with object density.} \label{fig:bkg-oversubtract}
\end{figure}

It is important to accurately characterize image noise to get unbiased estimates of an object's photometric properties and image moments. While Poisson noise is dominant for calibrated images, there are other less-studied contributions to the image noise including undetected sources (US). Using the Bayesian Fourier Domain (BFD) method described in \cite{BA2014} on \balrog{} detections across 48 tiles, the variance of measured galaxy moments was found to be up to 30\% in excess of Poisson predictions in \cite{Eckert_2020}. Furthermore, an over-subtraction of the background was detected in the $riz$-bands leading to a bias in the zeroth moment flux estimator as shown in Figure \ref{fig:bkg-oversubtract}. The blue points show the mean $\mu$ of the Gaussian fit to the pull distribution of BDF flux moments for each tile as a function of object density where a clear correlation can be seen, particularly for the redder bands. The green points are the same measurements after making a local estimate of the background in each postage stamp.

In order to determine if the excess noise was due to US, a slight variant on the \balrog{} injection procedure was followed in which we injected zero-flux objects into 39 tiles at random positions and then made cutout postage stamps of these random patches of sky. The cross-power spectra of distinct exposures of the ``dark" injections in $griz$ was then computed, which would yield zero signal if the noise is Poisson or read noise. A clear detection of US noise is made in each band. This empirical approach allows computed BFD moments to calibrate the moment covariance matrix on the survey images rather than relying on simulations of unknown fidelity, and naturally includes the contribution by US as a source of noise within the Bayesian calculation. See \cite{Eckert_2020} for further details.

\subsection{Accurate Joint Redshift - Stellar Mass Probability Distributions with Random Forests}\label{joint-redshift-mass-pdf}

In \cite{Mucesh_2020}, \balrog{} is used together with the Random Forest machine learning algorithm to produce well-calibrated joint redshift-stellar mass probability distributions at a fraction of the speed of traditional template-fitting methods. This was made possible because \balrog{} produces an ideal training sample: it captures both the realistic noise properties of DES WF measurements as well as the redshift and mass information of the DF injections from the COSMOS2015 catalog \citep{Laigle_2016}.

\subsection{Photometric Response Near Galaxy Clusters}\label{clusters}

\begin{figure*}
    \centering
    \includegraphics[width=\textwidth]{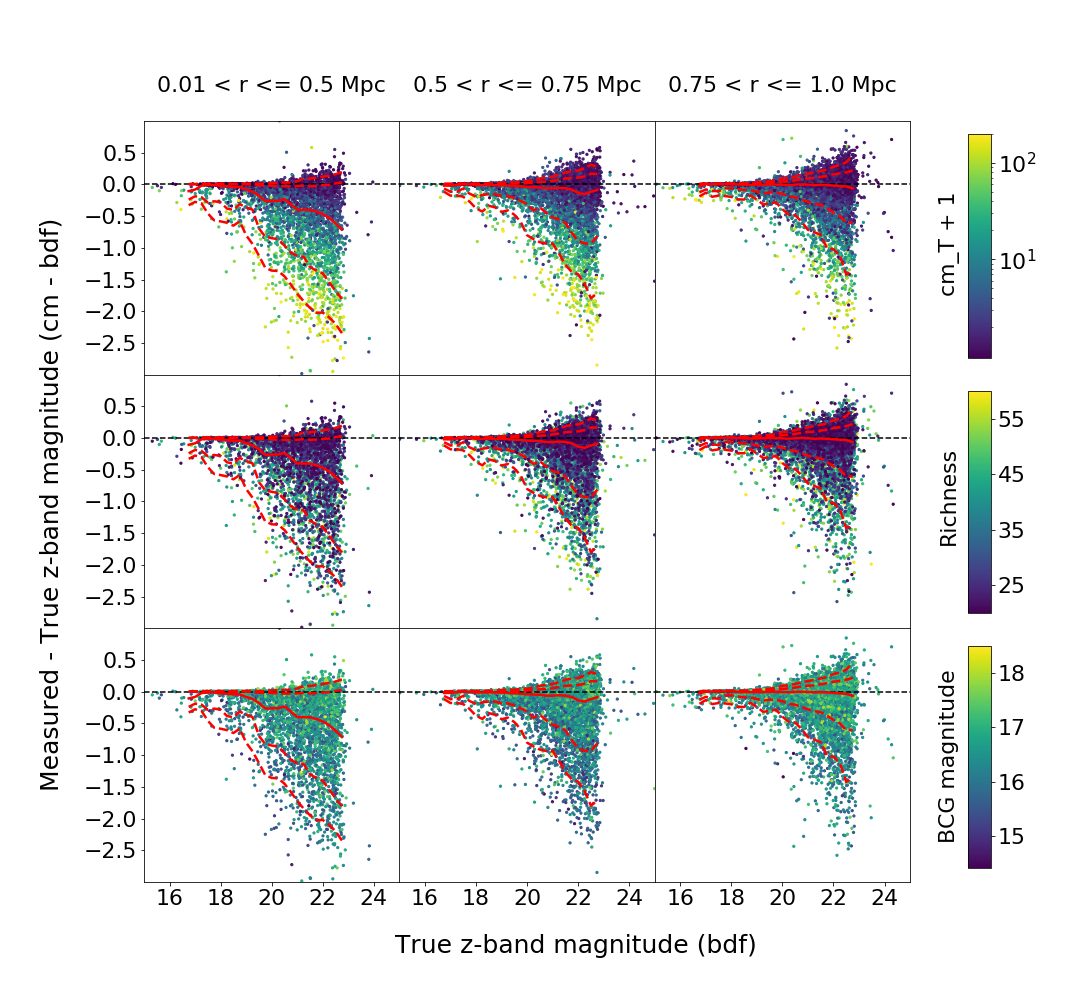}
    \caption{The difference in measured CModel \protect{$z$}-band magnitude vs. the injected DF magnitude \protect{\dmgal{}} as a function of input magnitude for the high-density \code{clusters} run \response{in a redshift range of 0.2 to 0.3}. The three columns present the \protect{\dmgal{}} responses binned by their radial distances to nearby cluster centers as specified at the top of the columns. The median response biases across the range of the injected magnitude are displayed as solid red lines, with the first and second \protect{\siggal{}} contours indicated by the dashed lines above and below. As the injections approach the center of a cluster, the median bias becomes increasingly negative indicating that the objects are measured to be progressively brighter than injected truth the closer they are to the bright central galaxy (BCG). In the three rows we color the magnitude responses as a function of (a) measured object size \protect{\code{cm\_T}}, (b) cluster richness, $\lambda$, and (c) the magnitude of the cluster's BCG. The measured object size appears to have the strongest influence over magnitude bias among the three quantities, though richer clusters also show larger \protect{\dmgal{}} responses. We use \code{cm\_T}+1 for the color scale as the \protect{\code{ngmix}} \protect{\code{T}} sizes are allowed to be slightly negative. This preliminary analysis will be followed up in more detail in Masegian \& Zhang et al. (in preparation).}
    \label{fig:cluster_radius_offset}
\end{figure*}

Clusters of galaxies -- especially rich, crowded clusters -- are known to present \response{additional} obstacles in the accurate detection and characterization of cluster members. These member galaxies often have higher detection incompleteness and significant photometric biases because of the increased rate of proximity effects. Detected sources in or near galaxy clusters in the sky can be further biased because of blending with member galaxies or contamination from intra-cluster light \citep{intra-cluster-light}. To aid in studies of these difficult measurement biases and selection effects, a high-density \balrog{} run was performed targeting areas near rich galaxy clusters.

A sample of 900 tiles, each containing a galaxy cluster with optical richness $\lambda > 35$ (see \citealt{sv-redmapper} for a description of richness and the DES cluster catalog), were injected with a similar DF galaxy sample as used in \dfsample{} at a lattice separation of 10\asec{} resulting in four times the injection density of the main cosmology runs. This higher injection density was needed to properly sample the effects of clusters on the transfer function as a function of radius from a cluster center given the number of tiles used\footnote{While this increases the probability of unwanted proximity effects from other \balrog{} injections, we estimate that the chances of two neighboring injections with $\code{bdf\_T} > 10$ (or ${\sim}$ 3.3\asec{}) in this run to be less than 0.25\%.}. \response{Additionally, we used a more restrictive $riz$ detection magnitude of 23 to increase the fraction of detected objects for this analysis.}

 The magnitude responses of the injected galaxies were measured as well as their distances to the center of the nearby clusters. The sample was further subdivided by the host cluster's richness, the measured object size \code{cm\_T}, and the magnitude of the cluster's brightest cluster galaxy (BCG) to see how these parameters affect the magnitude bias of the inserted objects. Preliminary results of this analysis \response{for clusters in a redshift range from 0.2 to 0.3} are shown in Figure \ref{fig:cluster_radius_offset} and will be followed up in Masegian \& Zhang et al. (in preparation) including a careful study of the detection efficiency as a function of various parameters including radial distance. Unsurprisingly, the magnitude responses become more negatively biased closer to cluster centers where the complex environments make accurate photometric measurements difficult and faint sources are up-scattered by the abundant residual light.

We find a similar correlation between an object's measured size and magnitude response as seen in \S \ref{catastrophic-fitting}. The proximity effects that cause asymmetric overestimates of \code{cm\_T} are amplified in the very crowded cluster environments, a trend that grows even stronger closer to the cluster centers. Correlation between magnitude bias and the other examined parameters, cluster richness and the BCG magnitude, is weaker but still present -- particularly for richness. All correlations appear to bias the recovered magnitudes in the same direction. The scale of these effects increases as the injections approach cluster centers. Taken together, the proximity to cluster centers, cluster richness, and BCG brightness artificially increases the number of observed objects near clusters above a fixed brightness threshold which, in turn, can collectively bias cluster measurements from a corresponding increase in cluster member galaxies. We plan on accounting for these correlations in future DES cluster analyses.

\section{Current Methodological Limitations and Future Directions}\label{discussion}

While this latest iteration of \balrog{} has made great advances in its ability to precisely quantify difficult measurement systematics, there remain many challenges to overcome if we are to reach the level of precision required by upcoming Stage IV surveys like LSST where the increased depth, pipeline complexity, and blending rate will otherwise limit the constraining power on cosmological parameters. Some of these challenges, such as properly accounting for per-object chromatic corrections at injection time or pushing the injection step further upstream in the measurement pipeline to account for more systematic effects in the image calibration, are largely technical barriers that can be addressed with more development time. Our ambiguous matching scheme can be improved by incorporating pixel-level information on the overlap between injected and real sources similar to the blending parameter introduced in \cite{Huang_2018}. In addition, many of the complexities and additional development time needed for careful emulation of a survey's measurement pipeline can be nearly eliminated by having injection pipelines placed directly in the software stack of the fiducial data processing runs. While this was not possible in DES, this approach is now taken in HSC with \code{SynPipe} and planned for LSST. However, there are more fundamental barriers in leveraging injection pipelines to their full potential.

A primary challenge is increasing the representativeness of the input catalog. Using the DECam observations of sources in the DES DF as the basis for the input object photometry rectified many of the input sample issues described in \cite{Suchyta_2016} -- particularly the discrepancy in recovered \balrog{} colors as compared to Y1 GOLD that arose from interpolating the spectral energy distribution (SED) of COSMOS galaxies to match DECam filters. However, Figure \ref{fig:balrog-gold-1d-compare} shows that we have further work to do. While it is difficult to disentangle intrinsic errors in the emulation of the DESDM pipeline from the input sample representativeness, there are some clear avenues for improvement. The conceptually simplest is to sample a wider population of deep objects across more deep patches of sky in order to incorporate greater cosmic variance in the injection sample. However, these deep observations are very expensive which limits the practicality of this approach. It may be possible to combine with external deep datasets, though this comes at the expense of a return to SED interpolations to match DECam filters. In addition, more detailed studies of the difficult PSF modeling in the DF may yield a stellar population more similar to the WF measurements and resolve some of the largest discrepancies between \balrog{} and Y3 GOLD for bright, PSF-like objects.

Another possibility is that the discrepancies between the recovered WF sample and Y3 GOLD are driven, at least in part, by the inability of CModel profiles to accurately capture the full diversity of galaxy morphologies. True galaxy profiles have many complex features such as spiral arms, star knots, and long asymmetric disruptions from mergers that we are not currently capturing with our DF injections. The most direct solution to this problem is to inject the MEDS image cutouts of the DF sources. We have already built the basic infrastructure to do so with \balrog{}, as described in \S \ref{input-sample}, but there are new issues to consider. The image cutouts can include artifacts, excessive masking, truncated profiles of nearby objects, or even be blended with other sources. This may be rectified in the future by using machine learning methods such as non-negative matrix factorization or generative adversarial networks to handle the required pixel-level deblending of sources in the stamps (see \cite{scarlet} and \cite{deblend-gan} for examples respectively).

However, using the image cutouts directly would introduce undesired noise when injecting into single-epoch exposures that had better seeing conditions than the composite PSF of the single-chip DF coadd and remains an unresolved issue. In addition, precisely defining the ``truth'' properties of the stamps is less straightforward than for model fit injections. This will likely be handled by making accurate measurements of each relevant WF photometry type on the stamps which would eliminate inheriting non-physical parameter biases from small profile definition differences such as the resulting magnitude bias from differences in \code{fracdev} prior shown in Figure \ref{fig:grid-test-plots}.

\response{The} most difficult challenge to overcome is the high computational cost of injection pipelines. The new single-epoch processing and additional photometric measurements in Y3 \balrog{} has increased the total mean CPU time per recovered injection to ${\sim}$80 seconds; about 12 times greater than in \cite{Suchyta_2016}. This large increase in runtime is only at Y3 depth corresponding to ${\sim}$4-6 epochs per injection \response{and made it unfeasible to directly calibrate many of the other key aspects of the Y3 cosmological analysis for which \balrog{} would otherwise be the ideal measurement tool. For example, to achieve the equivalent statistical precision on how the blending of galaxies at different redshifts affects the multiplicative shear bias as measured in \cite{des-imsims-y3}, we estimate that we would have to run the equivalent of over a dozen \dfsample{} samples to sufficiently capture how an identical injection population responds to an input shear signal that varies with redshift. In addition, the original goal of using \balrog{} to directly calibrate the spatially-dependent measurement biases and completeness inhomogeneities in the galaxy clustering measurement as outlined in \cite{Suchyta_2016} would require many more injections than sources for the estimation of the angular correlation function; a daunting prospect in light of the ${\sim}$40\% Y3 density achieved in this analysis. The situation will become significantly worse for much deeper surveys like LSST where we can expect hundreds of exposures for each object.}

\response{Perhaps even more consequential than the low number density realistically achievable, the high cost of running \balrog{} led to only a single injection realization across just 20\% of the total footprint area. This limited the calibrations from \balrog{} in Y3 to either be based on mean measurements, such as those described in Section \ref{applications}, or required a reduction in the considered footprint such as the clustering measurement presented in Section \ref{power-spectra}. While even the relatively low sampling of \dfsample{} was sufficient to capture systematics variations in the clustering amplitude to better than 1\% for a {\sc maglim}-like sample in the overlap area, reaching this threshold (or beyond) for highly incomplete samples or for accurate calibrations of large-scale fluctuations may require orders of magnitude more injections. Despite an expected significant increase in the total tiles sampled for Y6, achieving the many realizations of full footprint coverage required for the most ambitious \balrog{} measurements, such as providing realistic random properties for clustering and shear two-point measurements, will likely remain impractical without a dramatic increase in computational investment.}

One promising solution that we plan to explore is the use of the \balrog{} samples as a training set for an emulator that predicts additional realizations conditional on the survey property maps. A somewhat similar approach is taken in \cite{organized_randoms} where they mitigate galaxy clustering systematics by producing ``organized'' random catalogs with fluctuations in number density imprinted from a SOM approach that trained on maps of the variations in KiDS observing properties. Using an injection catalog like \balrog{} directly as a training sample for this approach would leverage our very high fidelity measurements of the survey transfer function to include unknown systematics not fully captured by the identified survey properties. While still more computationally expensive than a machine learning-only approach, this will allow us to build an efficient way of creating accurate random samples tuned for the desired measurement without increasing the total survey pipeline computational cost by more than a factor of two. We plan to use the presented \balrog{} catalogs to gauge the accuracy and feasibility of this approach in an upcoming analysis.

\section{Summary and Conclusion}\label{summary}

We have presented here the suite of DES Y3 \balrog{} simulations and resulting object catalogs used in downstream Y3 analyses. Like its Y1 predecessor, this current iteration of \balrog{} directly samples the DES transfer function by injecting an ensemble of realistic sources into real survey images to make precise measurements of the inherited systematic biases in the photometric response. However, the updated methodology (and entirely new coding framework) for Y3 \balrog{} makes significant strides beyond \cite{Suchyta_2016} in replicating many of the more complex features of the DESDM pipeline including the coaddition of single-epoch images and multi-epoch photometric measurements from SOF and Metacalibration in order to probe more aspects of the true measurement likelihood. In addition, we used a more realistic input sample based on the DES DF source catalog with observations using DECam filters which eliminated the need for template fitting to COSMOS galaxies and incorporated more cosmic variance in object properties. We also implemented a novel ambiguous matching scheme to capture many of the impacts of source blending while largely eliminating the contributions from undesired dropouts that happened to land on top of existing bright sources.

This effort culminated in tens of millions of Monte Carlo samples of the DES transfer function at high fidelity across 20\% of the full DES footprint to Y3 depth, capturing systematic biases from more variations in observing conditions than any previous \balrog{} analysis. The improved methodology resulted in the injected objects matching Y3 GOLD photometric properties and capturing clustering systematics correlated with survey property maps to better than 1\% accuracy for a typical cosmology sample on relevant scales. Additionally, we find that \balrog{} captures the clustering amplitudes of these systematics within a few percent for even \textit{highly} incomplete samples -- an encouraging first step for future analyses that wish to leverage more of our hard-earned (and often expensive) photons.

We quantified the photometric responses of \balrog{} injections through the Y3 DESDM measurement pipeline, particularly for magnitudes, colors, and morphology. We find that the magnitudes of most injections are well calibrated until selection effects near the detection threshold become significant, although we have found a clear asymmetric bias for objects in crowded fields or near image artifacts to have moderately to severely overestimated sizes which correlate with large negative magnitude biases. These biases are fairly common for bright, extended objects and can become extremely large (up to \dmgal{}${\sim}$8) at fainter magnitudes; though they constitute a much larger relative fraction of objects on the bright end. We demonstrated that these catastrophic photometry failures are real effects and often pass science cuts. We plan on exploring the causal relationship of this photometric failure mode further in a future analysis. While these magnitude response biases can cause significant discrepancies from more naive error estimates, fortunately their effect appears to have little impact on the recovered colors where we find typical median response biases of ${\sim}$1-3 mmag for stars and ${\sim}$5-10 mmag for galaxies in the densest regions of parameter space -- an effective median color calibration offset of less than 1\%.

Finally, we discussed a few of the most important applications of the presented \balrog{} catalogs to the Y3 cosmology analysis and other DES science measurements. In particular, we provided a realistic measurement likelihood in the calibration of photometric redshifts to reduce systematic biases in one of the highest sources of uncertainty in the cosmological measurement without contributing any additional uncertainty to the overall error budget. Unexpected findings such as the noise contributions from undetected sources in DES images and sky over-subtraction in the $riz$-bands described in \cite{Eckert_2020}, in addition to the moderate band-dependence in magnitude response and discovery of a new class of catastrophic photometry failures correlated with measured size, are indicative of the diagnostic power of object injection pipelines like \balrog{} in modern galaxy surveys.

We believe that this paper only scratches the surface in cosmological calibration potential and the identification of new systematics with injection pipelines such as \balrog{}. In particular, the combination of direct Monte Carlo sampling of the transfer function with an emulator to boost the total statistical power has the potential to facilitate many of the most difficult measurements in modern galaxy surveys. It is clear that we have yet to dig too deep.

\section*{Acknowledgements}

SE and TJ acknowledge support from the U.S. Department of Energy, Office of Science, Office of High Energy Physics, under Award Numbers DESC0010107 and A00-1465-001.

Funding for the DES Projects has been provided by the U.S. Department of Energy, the U.S. National Science Foundation, the Ministry of Science and Education of Spain, 
the Science and Technology Facilities Council of the United Kingdom, the Higher Education Funding Council for England, the National Center for Supercomputing 
Applications at the University of Illinois at Urbana-Champaign, the Kavli Institute of Cosmological Physics at the University of Chicago, 
the Center for Cosmology and Astro-Particle Physics at the Ohio State University,
the Mitchell Institute for Fundamental Physics and Astronomy at Texas A\&M University, Financiadora de Estudos e Projetos, 
Funda{\c c}{\~a}o Carlos Chagas Filho de Amparo {\`a} Pesquisa do Estado do Rio de Janeiro, Conselho Nacional de Desenvolvimento Cient{\'i}fico e Tecnol{\'o}gico and 
the Minist{\'e}rio da Ci{\^e}ncia, Tecnologia e Inova{\c c}{\~a}o, the Deutsche Forschungsgemeinschaft and the Collaborating Institutions in the Dark Energy Survey. 

The Collaborating Institutions are Argonne National Laboratory, the University of California at Santa Cruz, the University of Cambridge, Centro de Investigaciones Energ{\'e}ticas, 
Medioambientales y Tecnol{\'o}gicas-Madrid, the University of Chicago, University College London, the DES-Brazil Consortium, the University of Edinburgh, 
the Eidgen{\"o}ssische Technische Hochschule (ETH) Z{\"u}rich, 
Fermi National Accelerator Laboratory, the University of Illinois at Urbana-Champaign, the Institut de Ci{\`e}ncies de l'Espai (IEEC/CSIC), 
the Institut de F{\'i}sica d'Altes Energies, Lawrence Berkeley National Laboratory, the Ludwig-Maximilians Universit{\"a}t M{\"u}nchen and the associated Excellence Cluster Universe, 
the University of Michigan, NFS's NOIRLab, the University of Nottingham, The Ohio State University, the University of Pennsylvania, the University of Portsmouth, 
SLAC National Accelerator Laboratory, Stanford University, the University of Sussex, Texas A\&M University, and the OzDES Membership Consortium.

Based in part on observations at Cerro Tololo Inter-American Observatory at NSF's NOIRLab (NOIRLab Prop. ID 2012B-0001; PI: J. Frieman), which is managed by the Association of Universities for Research in Astronomy (AURA) under a cooperative agreement with the National Science Foundation.

The DES data management system is supported by the National Science Foundation under Grant Numbers AST-1138766 and AST-1536171.
The DES participants from Spanish institutions are partially supported by MICINN under grants ESP2017-89838, PGC2018-094773, PGC2018-102021, SEV-2016-0588, SEV-2016-0597, and MDM-2015-0509, some of which include ERDF funds from the European Union. IFAE is partially funded by the CERCA program of the Generalitat de Catalunya.
Research leading to these results has received funding from the European Research
Council under the European Union's Seventh Framework Program (FP7/2007-2013) including ERC grant agreements 240672, 291329, and 306478.
We  acknowledge support from the Brazilian Instituto Nacional de Ci\^encia
e Tecnologia (INCT) do e-Universo (CNPq grant 465376/2014-2).

This manuscript has been authored by Fermi Research Alliance, LLC under Contract No. DE-AC02-07CH11359 with the U.S. Department of Energy, Office of Science, Office of High Energy Physics.



\section*{Data Availability}

The DES Y3 data products used in this work will be made publicly available following publication at the URL: \url{https://des.ncsa.illinois.edu/releases}.



\response{
\software{
Astropy \citep{astropy:2013, astropy:2018},
CAMB \citep{camb},
corner \citep{corner},
fitsio (\url{https://github.com/esheldon/fitsio}),
GalSim \citep{galsim},
Matplotlib \citep{Hunter:2007},
meds \citep{meds},
ngmix \citep{Sheldon_2014},
ngmixer (\url{https://github.com/esheldon/ngmixer}),
PSFEx \citep{psfex},
PyMaster \citep{pymaster},
SExtractor \citep{sextractor},
SWarp \citep{swarp},
seaborn \citep{seaborn}
}
}




\bibliographystyle{yahapj_twoauthor}
\bibliography{bibliography}

\begin{thebibliography}{}
\providecommand\natexlab[1]{#1}
\providecommand\JournalTitle[1]{#1}

\bibitem[{{Abbott} {et~al.}(2009){Abbott}, {Abbott} {et~al.}}]{ligo-overview}
{Abbott}, B.~P., {Abbott}, R., {Adhikari}, R., {et~al.} 2009,
  \href{http://dx.doi.org/10.1088/0034-4885/72/7/076901}{\JournalTitle{Reports
  on Progress in Physics}, 72, 076901}

\bibitem[{{Aihara} {et~al.}(2011){Aihara}, {Allende Prieto} {et~al.}}]{sdssDR8}
{Aihara}, H., {Allende Prieto}, C., {An}, D., {et~al.} 2011,
  \href{http://dx.doi.org/10.1088/0067-0049/193/2/29}{\JournalTitle{\apjs},
  193, 29}

\bibitem[{{Aihara} {et~al.}(2018){Aihara}, {Arimoto} {et~al.}}]{hsc_overview}
{Aihara}, H., {Arimoto}, N., {Armstrong}, R., {et~al.} 2018,
  \href{http://dx.doi.org/10.1093/pasj/psx066}{\JournalTitle{\pasj}, 70, S4}

\bibitem[{{Akerib} {et~al.}(2013){Akerib}, {Bai} {et~al.}}]{lux-overview}
{Akerib}, D.~S., {Bai}, X., {Bedikian}, S., {et~al.} 2013,
  \href{http://dx.doi.org/10.1016/j.nima.2012.11.135}{\JournalTitle{Nuclear
  Instruments and Methods in Physics Research A}, 704, 111}

\bibitem[{{Akerib} {et~al.}(2017){Akerib}, {Alsum} {et~al.}}]{lux-salt}
{Akerib}, D.~S., {Alsum}, S., {Ara{\'u}jo}, H.~M., {et~al.} 2017,
  \href{http://dx.doi.org/10.1103/PhysRevLett.118.021303}{\JournalTitle{\prl},
  118, 021303}

\bibitem[{{Alonso} {et~al.}(2019){Alonso}, {Sanchez}, {Slosar} \& {LSST Dark
  Energy Science Collaboration}}]{pymaster}
{Alonso}, D., {Sanchez}, J., {Slosar}, A., \& {LSST Dark Energy Science
  Collaboration}. 2019,
  \href{http://dx.doi.org/10.1093/mnras/stz093}{\JournalTitle{\mnras}, 484,
  4127}

\bibitem[{{Amiaux} {et~al.}(2012){Amiaux}, {Scaramella}
  {et~al.}}]{euclid_overview}
{Amiaux}, J., {Scaramella}, R., {Mellier}, Y., {et~al.} 2012,
  \href{http://dx.doi.org/10.1117/12.926513}{in Society of Photo-Optical
  Instrumentation Engineers (SPIE) Conference Series, Vol. 8442, Space
  Telescopes and Instrumentation 2012: Optical, Infrared, and Millimeter Wave,
  ed. M.~C. {Clampin}, G.~G. {Fazio}, H.~A. {MacEwen}, \& J.~{Oschmann},
  Jacobus~M.}, 84420Z

\bibitem[{{Antilogus} {et~al.}(2014){Antilogus}, {Astier}, {Doherty},
  {Guyonnet} \& {Regnault}}]{brighter-fatter}
{Antilogus}, P., {Astier}, P., {Doherty}, P., {Guyonnet}, A., \& {Regnault}, N.
  2014,
  \href{http://dx.doi.org/10.1088/1748-0221/9/03/C03048}{\JournalTitle{Journal
  of Instrumentation}, 9, C03048}

\bibitem[{{Astropy Collaboration}(2013){Astropy Collaboration}, {Robitaille}
  {et~al.}}]{astropy:2013}
{Astropy Collaboration}. 2013,
  \href{http://dx.doi.org/10.1051/0004-6361/201322068}{\JournalTitle{\aap},
  558, A33}

\bibitem[{{Astropy Collaboration}(2018){Astropy Collaboration}, {Price-Whelan}
  {et~al.}}]{astropy:2018}
{Astropy Collaboration}. 2018,
  \href{http://dx.doi.org/10.3847/1538-3881/aabc4f}{\JournalTitle{\aj}, 156,
  123}

\bibitem[{{Bernstein} \& {Armstrong}(2014)}]{BA2014}
{Bernstein}, G.~M. \& {Armstrong}, R. 2014,
  \href{http://dx.doi.org/10.1093/mnras/stt2326}{\JournalTitle{\mnras}, 438,
  1880}

\bibitem[{{Bertin}(2011)}]{psfex}
{Bertin}, E. 2011, in Astronomical Society of the Pacific Conference Series,
  Vol. 442, Astronomical Data Analysis Software and Systems XX, ed. I.~N.
  {Evans}, A.~{Accomazzi}, D.~J. {Mink}, \& A.~H. {Rots}, 435

\bibitem[{{Bertin} \& {Arnouts}(1996)}]{sextractor}
{Bertin}, E. \& {Arnouts}, S. 1996,
  \href{http://dx.doi.org/10.1051/aas:1996164}{\JournalTitle{\aaps}, 117, 393}

\bibitem[{{Bertin} {et~al.}(2002){Bertin}, {Mellier}, {Radovich}, {Missonnier},
  {Didelon} \& {Morin}}]{swarp}
{Bertin}, E., {Mellier}, Y., {Radovich}, M., {et~al.} 2002, in Astronomical
  Society of the Pacific Conference Series, Vol. 281, Astronomical Data
  Analysis Software and Systems XI, ed. D.~A. {Bohlender}, D.~{Durand}, \&
  T.~H. {Handley}, 228

\bibitem[{{Bienaym{\'e}} {et~al.}(2018){Bienaym{\'e}}, {Leca} \&
  {Robin}}]{BLR18}
{Bienaym{\'e}}, O., {Leca}, J., \& {Robin}, A.~C. 2018,
  \href{http://dx.doi.org/10.1051/0004-6361/201833395}{\JournalTitle{\aap},
  620, A103}

\bibitem[{{Biwer} {et~al.}(2017){Biwer}, {Barker} {et~al.}}]{ligo-fakes}
{Biwer}, C., {Barker}, D., {Batch}, J.~C., {et~al.} 2017,
  \href{http://dx.doi.org/10.1103/PhysRevD.95.062002}{\JournalTitle{\prd}, 95,
  062002}

\bibitem[{{Blake} {et~al.}(2010){Blake}, {Brough} {et~al.}}]{wiggleZselection}
{Blake}, C., {Brough}, S., {Colless}, M., {et~al.} 2010,
  \href{http://dx.doi.org/10.1111/j.1365-2966.2010.16747.x}{\JournalTitle{\mnras},
  406, 803}

\bibitem[{{Blumenthal} {et~al.}(1984){Blumenthal}, {Faber}, {Primack} \&
  {Rees}}]{clustering_1984}
{Blumenthal}, G.~R., {Faber}, S.~M., {Primack}, J.~R., \& {Rees}, M.~J. 1984,
  \href{http://dx.doi.org/10.1038/311517a0}{\JournalTitle{\nat}, 311, 517}

\bibitem[{{Brainerd} {et~al.}(1996){Brainerd}, {Blandford} \&
  {Smail}}]{wl_1996}
{Brainerd}, T.~G., {Blandford}, R.~D., \& {Smail}, I. 1996,
  \href{http://dx.doi.org/10.1086/177537}{\JournalTitle{\apj}, 466, 623}

\bibitem[{{Buchs} {et~al.}(2019){Buchs}, {Davis} {et~al.}}]{Buchs_2019}
{Buchs}, R., {Davis}, C., {Gruen}, D., {et~al.} 2019,
  \href{http://dx.doi.org/10.1093/mnras/stz2162}{\JournalTitle{\mnras}, 489,
  820}

\bibitem[{Bunce(1980)}]{hybrid-monte-carlo}
Bunce, G. 1980,
  \href{http://dx.doi.org/https://doi.org/10.1016/0029-554X(80)90348-1}{\JournalTitle{Nuclear
  Instruments and Methods}, 172, 553}

\bibitem[{{Chang} {et~al.}(2015){Chang}, {Busha} {et~al.}}]{chang_2015}
{Chang}, C., {Busha}, M.~T., {Wechsler}, R.~H., {et~al.} 2015,
  \href{http://dx.doi.org/10.1088/0004-637X/801/2/73}{\JournalTitle{\apj}, 801,
  73}

\bibitem[{{Choi} {et~al.}(2016){Choi}, {Heymans} {et~al.}}]{Choi_2016}
{Choi}, A., {Heymans}, C., {Blake}, C., {et~al.} 2016,
  \href{http://dx.doi.org/10.1093/mnras/stw2241}{\JournalTitle{\mnras}, 463,
  3737}

\bibitem[{Connolly {et~al.}(2010)Connolly, Peterson {et~al.}}]{connolly_2010}
Connolly, A.~J., Peterson, J., Jernigan, J.~G., {et~al.} 2010,
  \href{http://dx.doi.org/10.1117/12.857819}{in Modeling, Systems Engineering,
  and Project Management for Astronomy IV, ed. G.~Z. Angeli \& P.~Dierickx,
  Vol. 7738}, International Society for Optics and Photonics (SPIE), 612

\bibitem[{Dawson {et~al.}(2015)Dawson, Schneider, Tyson \& Jee}]{Dawson_2015}
Dawson, W.~A., Schneider, M.~D., Tyson, J.~A., \& Jee, M.~J. 2015,
  \href{http://dx.doi.org/10.3847/0004-637x/816/1/11}{\JournalTitle{The
  Astrophysical Journal}, 816, 11}

\bibitem[{{de Jong} {et~al.}(2013){de Jong}, {Verdoes Kleijn}, {Kuijken} \&
  {Valentijn}}]{kids_overview}
{de Jong}, J. T.~A., {Verdoes Kleijn}, G.~A., {Kuijken}, K.~H., \& {Valentijn},
  E.~A. 2013,
  \href{http://dx.doi.org/10.1007/s10686-012-9306-1}{\JournalTitle{Experimental
  Astronomy}, 35, 25}

\bibitem[{{de Vaucouleurs}(1948)}]{devacouleurs}
{de Vaucouleurs}, G. 1948, \JournalTitle{Annales d'Astrophysique}, 11, 247

\bibitem[{{Drlica-Wagner} {et~al.}(2018){Drlica-Wagner}, {Sevilla-Noarbe}
  {et~al.}}]{y1-gold}
{Drlica-Wagner}, A., {Sevilla-Noarbe}, I., {Rykoff}, E.~S., {et~al.} 2018,
  \href{http://dx.doi.org/10.3847/1538-4365/aab4f5}{\JournalTitle{\apjs}, 235,
  33}

\bibitem[{{Eckert} {et~al.}(2020){Eckert}, {Bernstein} {et~al.}}]{Eckert_2020}
{Eckert}, K., {Bernstein}, G.~M., {Amara}, A., {et~al.} 2020,
  \href{http://dx.doi.org/10.1093/mnras/staa2133}{\JournalTitle{\mnras}, 497,
  2529}

\bibitem[{{Elvin-Poole} {et~al.}(2018){Elvin-Poole}, {Crocce}
  {et~al.}}]{clustering_y1}
{Elvin-Poole}, J., {Crocce}, M., {Ross}, A.~J., {et~al.} 2018,
  \href{http://dx.doi.org/10.1103/PhysRevD.98.042006}{\JournalTitle{\prd}, 98,
  042006}

\bibitem[{{Elvin-Poole} {et~al.}(2021)}]{y3-2x2ptmagnification}
{Elvin-Poole}, J. {et~al.} 2021, \JournalTitle{To be submitted to MNRAS}

\bibitem[{{Fenech Conti} {et~al.}(2017){Fenech Conti}, {Herbonnet}, {Hoekstra},
  {Merten}, {Miller} \& {Viola}}]{kidsSimulation}
{Fenech Conti}, I., {Herbonnet}, R., {Hoekstra}, H., {et~al.} 2017,
  \href{http://dx.doi.org/10.1093/mnras/stx200}{\JournalTitle{\mnras}, 467,
  1627}

\bibitem[{{Flaugher} {et~al.}(2015){Flaugher}, {Diehl} {et~al.}}]{decam}
{Flaugher}, B., {Diehl}, H.~T., {Honscheid}, K., {et~al.} 2015,
  \href{http://dx.doi.org/10.1088/0004-6256/150/5/150}{\JournalTitle{\aj}, 150,
  150}

\bibitem[{Foreman-Mackey(2016)}]{corner}
Foreman-Mackey, D. 2016,
  \href{http://dx.doi.org/10.21105/joss.00024}{\JournalTitle{The Journal of
  Open Source Software}, 1, 24}

\bibitem[{{Gaia Collaboration}(2018){Gaia Collaboration}, {Brown}
  {et~al.}}]{gaia}
{Gaia Collaboration}. 2018,
  \href{http://dx.doi.org/10.1051/0004-6361/201833051}{\JournalTitle{\aap},
  616, A1}

\bibitem[{{Garcia-Fernandez} {et~al.}(2018){Garcia-Fernandez}, {Sanchez}
  {et~al.}}]{GF_SV_magnification}
{Garcia-Fernandez}, M., {Sanchez}, E., {Sevilla-Noarbe}, I., {et~al.} 2018,
  \href{http://dx.doi.org/10.1093/mnras/sty282}{\JournalTitle{\mnras}, 476,
  1071}

\bibitem[{{G{\'o}rski} {et~al.}(2005){G{\'o}rski}, {Hivon} {et~al.}}]{healpix}
{G{\'o}rski}, K.~M., {Hivon}, E., {Banday}, A.~J., {et~al.} 2005,
  \href{http://dx.doi.org/10.1086/427976}{\JournalTitle{\apj}, 622, 759}

\bibitem[{{Gruen} \& {Brimioulle}(2017)}]{GruenBrimioulle2017}
{Gruen}, D. \& {Brimioulle}, F. 2017,
  \href{http://dx.doi.org/10.1093/mnras/stx471}{\JournalTitle{\mnras}, 468,
  769}

\bibitem[{{Hartley} {et~al.}(2020){Hartley}, {Choi} {et~al.}}]{y3-deepfields}
{Hartley}, W.~G., {Choi}, A., {Amon}, A., {et~al.} 2020, \JournalTitle{arXiv
  e-prints}, arXiv:2012.12824

\bibitem[{{Hildebrandt}(2016)}]{2016MNRAS.455.3943H}
{Hildebrandt}, H. 2016,
  \href{http://dx.doi.org/10.1093/mnras/stv2575}{\JournalTitle{\mnras}, 455,
  3943}

\bibitem[{{Huang} {et~al.}(2018){Huang}, {Leauthaud} {et~al.}}]{Huang_2018}
{Huang}, S., {Leauthaud}, A., {Murata}, R., {et~al.} 2018,
  \href{http://dx.doi.org/10.1093/pasj/psx126}{\JournalTitle{\pasj}, 70, S6}

\bibitem[{{Huff} \& {Mandelbaum}(2017)}]{metacal-theory}
{Huff}, E. \& {Mandelbaum}, R. 2017, \JournalTitle{arXiv e-prints},
  arXiv:1702.02600

\bibitem[{Hunter(2007)}]{Hunter:2007}
Hunter, J.~D. 2007,
  \href{http://dx.doi.org/10.1109/MCSE.2007.55}{\JournalTitle{Computing in
  Science \& Engineering}, 9, 90}

\bibitem[{{Huterer} {et~al.}(2006){Huterer}, {Takada}, {Bernstein} \&
  {Jain}}]{huterer_2006}
{Huterer}, D., {Takada}, M., {Bernstein}, G., \& {Jain}, B. 2006,
  \href{http://dx.doi.org/10.1111/j.1365-2966.2005.09782.x}{\JournalTitle{\mnras},
  366, 101}

\bibitem[{{Ivezi{\'c}} {et~al.}(2019){Ivezi{\'c}}, {Kahn}
  {et~al.}}]{lsst_overview}
{Ivezi{\'c}}, {\v Z}., {Kahn}, S.~M., {Tyson}, J.~A., {et~al.} 2019,
  \href{http://dx.doi.org/10.3847/1538-4357/ab042c}{\JournalTitle{\apj}, 873,
  111}

\bibitem[{{Jarvis} {et~al.}(2016){Jarvis}, {Sheldon} {et~al.}}]{meds}
{Jarvis}, M., {Sheldon}, E., {Zuntz}, J., {et~al.} 2016,
  \href{http://dx.doi.org/10.1093/mnras/stw990}{\JournalTitle{\mnras}, 460,
  2245}

\bibitem[{{Jarvis} {et~al.}(2021){Jarvis}, {Bernstein} {et~al.}}]{y3-piff}
{Jarvis}, M., {Bernstein}, G.~M., {Amon}, A., {et~al.} 2021,
  \href{http://dx.doi.org/10.1093/mnras/staa3679}{\JournalTitle{\mnras}, 501,
  1282}

\bibitem[{{Johnston} {et~al.}(2020){Johnston}, {Wright}
  {et~al.}}]{organized_randoms}
{Johnston}, H., {Wright}, A.~H., {Joachimi}, B., {et~al.} 2020,
  \JournalTitle{arXiv e-prints}, arXiv:2012.08467

\bibitem[{{Kong} {et~al.}(2020){Kong}, {Burleigh} {et~al.}}]{obiwan}
{Kong}, H., {Burleigh}, K.~J., {Ross}, A., {et~al.} 2020,
  \href{http://dx.doi.org/10.1093/mnras/staa2742}{\JournalTitle{\mnras}, 499,
  3943}

\bibitem[{{Kuijken, K.}(2008)}]{gaap}
{Kuijken, K.} 2008,
  \href{http://dx.doi.org/10.1051/0004-6361:20066601}{\JournalTitle{A\&A}, 482,
  1053}

\bibitem[{{Laigle} {et~al.}(2016){Laigle}, {McCracken} {et~al.}}]{Laigle_2016}
{Laigle}, C., {McCracken}, H.~J., {Ilbert}, O., {et~al.} 2016,
  \href{http://dx.doi.org/10.3847/0067-0049/224/2/24}{\JournalTitle{\apjs},
  224, 24}

\bibitem[{{Leistedt} {et~al.}(2016){Leistedt}, {Peiris}
  {et~al.}}]{2016ApJS..226...24L}
{Leistedt}, B., {Peiris}, H.~V., {Elsner}, F., {et~al.} 2016,
  \href{http://dx.doi.org/10.3847/0067-0049/226/2/24}{\JournalTitle{\apjs},
  226, 24}

\bibitem[{{Lewis} {et~al.}(2000){Lewis}, {Challinor} \& {Lasenby}}]{camb}
{Lewis}, A., {Challinor}, A., \& {Lasenby}, A. 2000,
  \href{http://dx.doi.org/10.1086/309179}{\JournalTitle{\apj}, 538, 473}

\bibitem[{{Lv} \& {Liu}(2010)}]{misspecification}
{Lv}, J. \& {Liu}, J.~S. 2010, \JournalTitle{arXiv e-prints}, arXiv:1005.5483

\bibitem[{{MacCrann} {et~al.}(2020){MacCrann}, {Becker}
  {et~al.}}]{des-imsims-y3}
{MacCrann}, N., {Becker}, M.~R., {McCullough}, J., {et~al.} 2020,
  \JournalTitle{arXiv e-prints}, arXiv:2012.08567

\bibitem[{{Mandelbaum}(2018)}]{wl_overview}
{Mandelbaum}, R. 2018,
  \href{http://dx.doi.org/10.1146/annurev-astro-081817-051928}{\JournalTitle{\araa},
  56, 393}

\bibitem[{{Martini} {et~al.}(2018){Martini}, {Bailey} {et~al.}}]{desi}
{Martini}, P., {Bailey}, S., {Besuner}, R.~W., {et~al.} 2018,
  \href{http://dx.doi.org/10.1117/12.2313063}{in Society of Photo-Optical
  Instrumentation Engineers (SPIE) Conference Series, Vol. 10702, Ground-based
  and Airborne Instrumentation for Astronomy VII, ed. C.~J. {Evans},
  L.~{Simard}, \& H.~{Takami}}, 107021F

\bibitem[{{Massey} {et~al.}(2013){Massey}, {Hoekstra}
  {et~al.}}]{2013MNRAS.429..661M}
{Massey}, R., {Hoekstra}, H., {Kitching}, T., {et~al.} 2013,
  \href{http://dx.doi.org/10.1093/mnras/sts371}{\JournalTitle{\mnras}, 429,
  661}

\bibitem[{{McClure} {et~al.}(1985){McClure}, {Hesser}, {Stetson} \&
  {Stryker}}]{cluster-e3}
{McClure}, R.~D., {Hesser}, J.~E., {Stetson}, P.~B., \& {Stryker}, L.~L. 1985,
  \href{http://dx.doi.org/10.1086/131586}{\JournalTitle{\pasp}, 97, 665}

\bibitem[{{McCracken} {et~al.}(2012){McCracken}, {Milvang-Jensen}
  {et~al.}}]{ultravista}
{McCracken}, H.~J., {Milvang-Jensen}, B., {Dunlop}, J., {et~al.} 2012,
  \href{http://dx.doi.org/10.1051/0004-6361/201219507}{\JournalTitle{\aap},
  544, A156}

\bibitem[{{Mead} {et~al.}(2015){Mead}, {Peacock}, {Heymans}, {Joudaki} \&
  {Heavens}}]{halofit_mead}
{Mead}, A.~J., {Peacock}, J.~A., {Heymans}, C., {Joudaki}, S., \& {Heavens},
  A.~F. 2015,
  \href{http://dx.doi.org/10.1093/mnras/stv2036}{\JournalTitle{\mnras}, 454,
  1958}

\bibitem[{{Melchior} {et~al.}(2018){Melchior}, {Moolekamp} {et~al.}}]{scarlet}
{Melchior}, P., {Moolekamp}, F., {Jerdee}, M., {et~al.} 2018,
  \href{http://dx.doi.org/10.1016/j.ascom.2018.07.001}{\JournalTitle{Astronomy
  and Computing}, 24, 129}

\bibitem[{{Morganson} {et~al.}(2018){Morganson}, {Gruendl}
  {et~al.}}]{des-image-pipeline}
{Morganson}, E., {Gruendl}, R.~A., {Menanteau}, F., {et~al.} 2018,
  \href{http://dx.doi.org/10.1088/1538-3873/aab4ef}{\JournalTitle{\pasp}, 130,
  074501}

\bibitem[{{Mucesh} {et~al.}(2020){Mucesh}, {Hartley} {et~al.}}]{Mucesh_2020}
{Mucesh}, S., {Hartley}, W.~G., {Palmese}, A., {et~al.} 2020,
  \JournalTitle{arXiv e-prints}, arXiv:2012.05928

\bibitem[{{Myles} {et~al.}(2021){Myles}, {Alarcon} {et~al.}}]{y3-sompz}
{Myles}, J., {Alarcon}, A., {Amon}, A., {et~al.} 2021,
  \href{http://dx.doi.org/10.1093/mnras/stab1515}{\JournalTitle{\mnras}, 505,
  4249}

\bibitem[{{Porredon} {et~al.}(2021){Porredon}, {Crocce}
  {et~al.}}]{y3-2x2maglimforecast}
{Porredon}, A., {Crocce}, M., {Fosalba}, P., {et~al.} 2021,
  \href{http://dx.doi.org/10.1103/PhysRevD.103.043503}{\JournalTitle{\prd},
  103, 043503}

\bibitem[{{Pujol} {et~al.}(2020){Pujol}, {Sureau}, {Bobin}, {Courbin},
  {Gentile} \& {Kilbinger}}]{misspecification-shear}
{Pujol}, A., {Sureau}, F., {Bobin}, J., {et~al.} 2020,
  \href{http://dx.doi.org/10.1051/0004-6361/202038657}{\JournalTitle{\aap},
  641, A164}

\bibitem[{{Reiman} \& {G{\"o}hre}(2019)}]{deblend-gan}
{Reiman}, D.~M. \& {G{\"o}hre}, B.~E. 2019,
  \href{http://dx.doi.org/10.1093/mnras/stz575}{\JournalTitle{\mnras}, 485,
  2617}

\bibitem[{{Rodr{\'\i}guez-Monroy} {et~al.}(2021){Rodr{\'\i}guez-Monroy},
  {Weaverdyck} {et~al.}}]{y3-galaxyclustering}
{Rodr{\'\i}guez-Monroy}, M., {Weaverdyck}, N., {Elvin-Poole}, J., {et~al.}
  2021, \JournalTitle{arXiv e-prints}, arXiv:2105.13540

\bibitem[{{Ross} {et~al.}(2012){Ross}, {Percival}
  {et~al.}}]{2012MNRAS.424..564R}
{Ross}, A.~J., {Percival}, W.~J., {S{\'a}nchez}, A.~G., {et~al.} 2012,
  \href{http://dx.doi.org/10.1111/j.1365-2966.2012.21235.x}{\JournalTitle{\mnras},
  424, 564}

\bibitem[{{Rowe} {et~al.}(2015){Rowe}, {Jarvis} {et~al.}}]{galsim}
{Rowe}, B.~T.~P., {Jarvis}, M., {Mandelbaum}, R., {et~al.} 2015,
  \href{http://dx.doi.org/10.1016/j.ascom.2015.02.002}{\JournalTitle{Astronomy
  and Computing}, 10, 121}

\bibitem[{{Rozo} {et~al.}(2016){Rozo}, {Rykoff} {et~al.}}]{redmagic}
{Rozo}, E., {Rykoff}, E.~S., {Abate}, A., {et~al.} 2016,
  \href{http://dx.doi.org/10.1093/mnras/stw1281}{\JournalTitle{\mnras}, 461,
  1431}

\bibitem[{{Rykoff} {et~al.}(2016){Rykoff}, {Rozo} {et~al.}}]{sv-redmapper}
{Rykoff}, E.~S., {Rozo}, E., {Hollowood}, D., {et~al.} 2016,
  \href{http://dx.doi.org/10.3847/0067-0049/224/1/1}{\JournalTitle{\apjs}, 224,
  1}

\bibitem[{Schlegel {et~al.}(1998)Schlegel, Finkbeiner \& Davis}]{Schlegel_1998}
Schlegel, D.~J., Finkbeiner, D.~P., \& Davis, M. 1998,
  \href{http://dx.doi.org/10.1086/305772}{\JournalTitle{The Astrophysical
  Journal}, 500, 525}

\bibitem[{{Scoville} {et~al.}(2007){Scoville}, {Aussel} {et~al.}}]{cosmos}
{Scoville}, N., {Aussel}, H., {Brusa}, M., {et~al.} 2007,
  \href{http://dx.doi.org/10.1086/516585}{\JournalTitle{\apjs}, 172, 1}

\bibitem[{{Sevilla-Noarbe} {et~al.}(2021){Sevilla-Noarbe}, {Bechtol}
  {et~al.}}]{y3-gold}
{Sevilla-Noarbe}, I., {Bechtol}, K., {Carrasco Kind}, M., {et~al.} 2021,
  \href{http://dx.doi.org/10.3847/1538-4365/abeb66}{\JournalTitle{\apjs}, 254,
  24}

\bibitem[{{Sheldon}(2014)}]{Sheldon_2014}
{Sheldon}, E.~S. 2014,
  \href{http://dx.doi.org/10.1093/mnrasl/slu104}{\JournalTitle{\mnras}, 444,
  L25}

\bibitem[{{Sheldon} \& {Huff}(2017)}]{metacal-practice}
{Sheldon}, E.~S. \& {Huff}, E.~M. 2017,
  \href{http://dx.doi.org/10.3847/1538-4357/aa704b}{\JournalTitle{\apj}, 841,
  24}

\bibitem[{{Smith} {et~al.}(1986){Smith}, {McClure}, {Stetson}, {Hesser} \&
  {Bell}}]{palomar-5}
{Smith}, G.~H., {McClure}, R.~D., {Stetson}, P.~B., {Hesser}, J.~E., \& {Bell},
  R.~A. 1986, \href{http://dx.doi.org/10.1086/114063}{\JournalTitle{\aj}, 91,
  842}

\bibitem[{{Smol{\v{c}}i{\'c}} {et~al.}(2004){Smol{\v{c}}i{\'c}}, {Ivezi{\'c}}
  {et~al.}}]{m_dwarf}
{Smol{\v{c}}i{\'c}}, V., {Ivezi{\'c}}, {\v{Z}}., {Knapp}, G.~R., {et~al.} 2004,
  \href{http://dx.doi.org/10.1086/426475}{\JournalTitle{\apjl}, 615, L141}

\bibitem[{{Stetson}(1987)}]{Stetson87}
{Stetson}, P.~B. 1987,
  \href{http://dx.doi.org/10.1086/131977}{\JournalTitle{\pasp}, 99, 191}

\bibitem[{{Suchyta} {et~al.}(2016){Suchyta}, {Huff} {et~al.}}]{Suchyta_2016}
{Suchyta}, E., {Huff}, E.~M., {Aleksi{\'c}}, J., {et~al.} 2016,
  \href{http://dx.doi.org/10.1093/mnras/stv2953}{\JournalTitle{\mnras}, 457,
  786}

\bibitem[{{Tanoglidis} {et~al.}(2020){Tanoglidis}, {Drlica-Wagner}
  {et~al.}}]{low-sb}
{Tanoglidis}, D., {Drlica-Wagner}, A., {Wei}, K., {et~al.} 2020,
  \JournalTitle{arXiv e-prints}, arXiv:2006.04294

\bibitem[{{Tegmark} {et~al.}(2006){Tegmark}, {Eisenstein}
  {et~al.}}]{clustering_sdss}
{Tegmark}, M., {Eisenstein}, D.~J., {Strauss}, M.~A., {et~al.} 2006,
  \href{http://dx.doi.org/10.1103/PhysRevD.74.123507}{\JournalTitle{\prd}, 74,
  123507}

\bibitem[{{The Dark Energy Survey Collaboration}(2005)}]{des-2005}
{The Dark Energy Survey Collaboration}. 2005, \JournalTitle{arXiv e-prints},
  astro

\bibitem[{Tolkien(1954)}]{fellowship}
Tolkien, J. 1954, The Fellowship of the Ring (Allen \& Unwin)

\bibitem[{{Troxel} {et~al.}(2018){Troxel}, {MacCrann} {et~al.}}]{wl_y1}
{Troxel}, M.~A., {MacCrann}, N., {Zuntz}, J., {et~al.} 2018,
  \href{http://dx.doi.org/10.1103/PhysRevD.98.043528}{\JournalTitle{\prd}, 98,
  043528}

\bibitem[{{Unruh} {et~al.}(2020){Unruh}, {Schneider}, {Hilbert}, {Simon},
  {Martin} \& {Puertas}}]{lens-mag-important}
{Unruh}, S., {Schneider}, P., {Hilbert}, S., {et~al.} 2020,
  \href{http://dx.doi.org/10.1051/0004-6361/201936915}{\JournalTitle{\aap},
  638, A96}

\bibitem[{Waskom(2021)}]{seaborn}
Waskom, M.~L. 2021,
  \href{http://dx.doi.org/10.21105/joss.03021}{\JournalTitle{Journal of Open
  Source Software}, 6, 3021}

\bibitem[{{Weaverdyck} \& {Huterer}(2020)}]{mitigating-lss-contam}
{Weaverdyck}, N. \& {Huterer}, D. 2020, \JournalTitle{arXiv e-prints},
  arXiv:2007.14499

\bibitem[{{Zhang} {et~al.}(2019){Zhang}, {Yanny}
  {et~al.}}]{intra-cluster-light}
{Zhang}, Y., {Yanny}, B., {Palmese}, A., {et~al.} 2019,
  \href{http://dx.doi.org/10.3847/1538-4357/ab0dfd}{\JournalTitle{\apj}, 874,
  165}

\bibitem[{{Zuntz} {et~al.}(2018){Zuntz}, {Sheldon} {et~al.}}]{y1-wl}
{Zuntz}, J., {Sheldon}, E., {Samuroff}, S., {et~al.} 2018,
  \href{http://dx.doi.org/10.1093/mnras/sty2219}{\JournalTitle{\mnras}, 481,
  1149}

\end{thebibliography}


\appendix

\numberwithin{figure}{section}
\numberwithin{table}{section}

\section{Injection Software}\label{appendix:injection}

\response{Here we describe a few of the most relevant configuration options when running the new injection framework, as well as templates for custom injection classes defined by the user for more advanced interfacing; see the code repository\footnote{\url{https://github.com/sweverett/Balrog-GalSim}} for more details on running the simulations.}

\subsection{Injection Configuration}\label{appendix:configuration}

\response{Configuration settings specific to a typical \balrog{} run have been wrapped into custom \galsim{} \code{image} and \code{stamp} types, both called \code{Balrog}:}

\begin{itemize}
    \item \response{\textbf{\code{image: Balrog}} - This image type is required for a full \balrog{} run. It parses all novel configuration entries and defines how to add \galsim{} objects to an existing image with consistent noise properties. It also allows the \balrog{} framework to be run on blank images for testing.}
    \item \response{\textbf{\code{stamp: Balrog}} - An optional stamp type that allows \galsim{} to skip objects whose fast Fourier transform (FFT) grid sizes are extremely large and can occasionally cause memory errors when using photometric model fits to DES DF objects.}
\end{itemize}

\response{We also provide a much simpler \code{image} class called \code{AddOn} which adds any simulated images onto an initial image without the full \balrog{} machinery.  Some configuration details can also be set on the command-line call to \code{balrog\_injection.py} for ease of use as long as they do not conflict with any settings in the configuration file.}

\subsection{Input Sample and Object Profiles}\label{appendix:input-sample}

\response{In principle any native \galsim{} input and object type can be used for injection. However, the object sampling, truth property updating, and truth catalog generation steps require knowledge about underlying structure of the input data (e.g. parametric models vs. image cutouts). We handle this ambiguity through the use of \code{BalInput} and \code{BalObject} parent classes that define the necessary implementation details to connect \galsim{} to \balrog{}. These classes can be used to register any needed injection types to \balrog{} including custom \galsim{} classes. Subclasses provided for injection types used in DES Y3 runs are described below:}

\begin{itemize}
    \item \response{\textbf{\code{ngmixGalaxy}}: 
    Described in \S \ref{input-sample}. A sum of \galsim{} \code{Gaussian} objects that represent a Gaussian mixture model fit to a source by the measurement software \ngmix{}. \balrog{} can currently inject the following \ngmix{} profile model types: a single Gaussian (\code{gauss}), a composite model (\code{cm}) that combines an exponential disk with a de Vaucouleurs’ profile, and a modified CModel with fixed size ratio between the two components (\code{bdf})). As \ngmix{} allows for objects with negative size before convolution with a PSF, these negative values are clipped to a small non-zero value (\code{T}=$10^{-6}$, corresponding to a size scale of ${\sim} 10^{-3}$ arcsec) to avoid rendering failures.}
    \item \response{\textbf{\code{DESStar}}: A synthetic star sample with realistic density and property distributions across the DES footprint was created to a depth of $27$ magnitude in $g$. These objects are treated as delta functions convolved with the local PSF. These magnitudes are referenced as $\delta$-mag in later figures. Further details about this star catalog are described in Section \ref{star-sample}.}
    \item \response{\textbf{\code{MEDSGalaxy}}: Single-epoch image cutouts of detected DES objects are stored in MEDS files for each band. These image cutouts can be used directly for injection after deconvolving with the original PSF solution and re-convolving with the local injection PSF.}
\end{itemize}

\response{\balrog{} can inject multiple object types in the same run by setting the \code{gal} field in the configuration as a \code{List} type; this is identical to \galsim{} configuration behaviour. The relative fraction of each injection type is then set in the \code{pos\_sampling} field described below.}

\subsection{Updating Truth Properties and Optional Transformations}\label{appendix:updating-truth}

\response{Most \balrog{} runs sample objects from an existing catalog. Some of the object properties are modified to fit the needs of the simulation such as the positions, orientations, and fluxes. Updates to positions and orientations are automatically applied to the output truth catalogs while flux corrections due to local extinction and zeropoint offsets are not, though we save the applied extinction factor. Different behaviour for these quantities as well as any additional changes can be defined when creating the relevant \code{BalObject} subclass.}

\response{Position sampling is determined by the configuration parameter \code{pos\_sampling} and can be set to \code{Uniform} for spherical random sampling or one of the following grid choices that are regularly-spaced in image space: \code{RectGrid} for a rectangular lattice, \code{HexGrid} for a hexagonal lattice, and \code{MixedGrid} for one of the previous grid choices that mixes multiple injection object types on the same grid with a set relative abundance \code{inj\_frac}. The user has control over the grid spacing as well as whether to apply random translations and/or rotations of the grid for each tile in addition to random rotations of the object profiles themselves with \code{rotate\_objs}.}

\response{In addition to flux scaling to match the zeropoint of each image, an additional extinction factor can be applied with the configuration option \code{extinct\_objs}. If set, extinction factors in $griz$ for each tile are loaded and applied to object fluxes. Incorporating more sophisticated per-object, SED-dependent extinction implementations based on the maps provided in \cite{Schlegel_1998} is planned for a future code release but are currently applied at the tile level. Any of the native \galsim{} noise models can be added to the injection stamps with the Poisson component ignoring the existing image pixel values as long as the \balrog{} (or \code{AddOn}) image type is used. Optional transformations such as a constant shear or magnification factor that are uniform across a tile can be added in the injection configuration with the same syntax as a typical \galsim{} configuration, while per-object effects need to be implemented into the relevant \code{BalObject} subclass.}

\subsection{Configuration Example}\label{appendix:config}

Here we show the high-level configuration settings used for \runtwo{} and \runtwoa{}, where capitalized quantities in \{\} refer to local file paths:

\begin{lstlisting}[basicstyle=\small]
modules:
  - galsim.des,
  - injector,
  - ngmix_catalog,
  - des_star_catalog
  
input:
  des_star_catalog:
    base_dir: {INPUT_DIR}
    data_version: y3v02
    model_type: Model_16.5-26.5

  ngmix_catalog:
    catalog_type: bdf
    de_redden: True
    dir: {INPUT_DIR}
    file_name: {INPUT_FILENAME}
    t_max: 100

gal:
  type: List
  items:
  - # Inject CModel fit to DF sources
    type: ngmixGalaxy
  - # Inject synthetic stars
    type: desStar
    
psf:
  type: DES_PSFEx
  
stamp:
  draw_method: no_pixel
  gsparams:
    maximum_fft_size: 16384
  type: Balrog

image:
  bands: griz
  extinct_objs: True
  rotate_objs: True
  n_realizations: 1
  noise: {} # Turn on Poisson noise
  nproc: 16
  pos_sampling:
    des_star_catalog:
        type: MixedGrid
        inj_frac: 0.1
    ngmix_catalog
        type: MixedGrid
        grid_spacing: 20
        grid_type: HexGrid
        inj_frac: 0.9
        offset: Random
        rotate: Random
  random_seed: {SEED}
  run_name: {main/aux}
  type: Balrog
  version: y3v02
  wcs:
    type: Fits
  xsize: 2048
  ysize: 4096
\end{lstlisting}

\section{Angular Clustering Systematics}\label{appendix:power_spectra}

Section \ref{power-spectra} introduced a method for translating the differences between the \balrog{} and Y3 GOLD catalogs into a predicted systematic error in the angular clustering of galaxies. We first choose a sample selection which is applied to both catalogs. We then measure the dependence of galaxy counts fluctuations in both selected \balrog{} and Y3 GOLD samples on several measured image quality indicators, as in Figure \ref{fig:balrog-gold-systematics-compare}. Finally, for each data quality indicator, we interpolate the density fluctuation trends to the full survey area and estimate the angular clustering that these trends imply. As small systematic variations in the survey depth enter, to leading order, as additive power in the measured clustering signal, a comparison of the power we measure in these interpolated maps offers a direct estimate of the importance of any deviation between our injection catalogs and the real data.

Here we show the same maps as Figure \ref{fig:power_best_worst} for six measured survey properties in all bands, for the $17.5<i<21.5$ sample selection meant to emulate the Y3 {\sc maglim} sample. With the exception of a negligible spike in power in a few of the \code{SIGMA\_MAG\_ZERO} maps, the measured systematic errors are less than $1\%$ of the fiducial galaxy clustering signal (calculated as described in Figure \ref{fig:balrog-gold-systematics-compare}) on scales below approximately $1^{\circ}$ ($\ell > 180$).


\begin{figure*}
    \centering
    \includegraphics[width=.9865\textwidth]{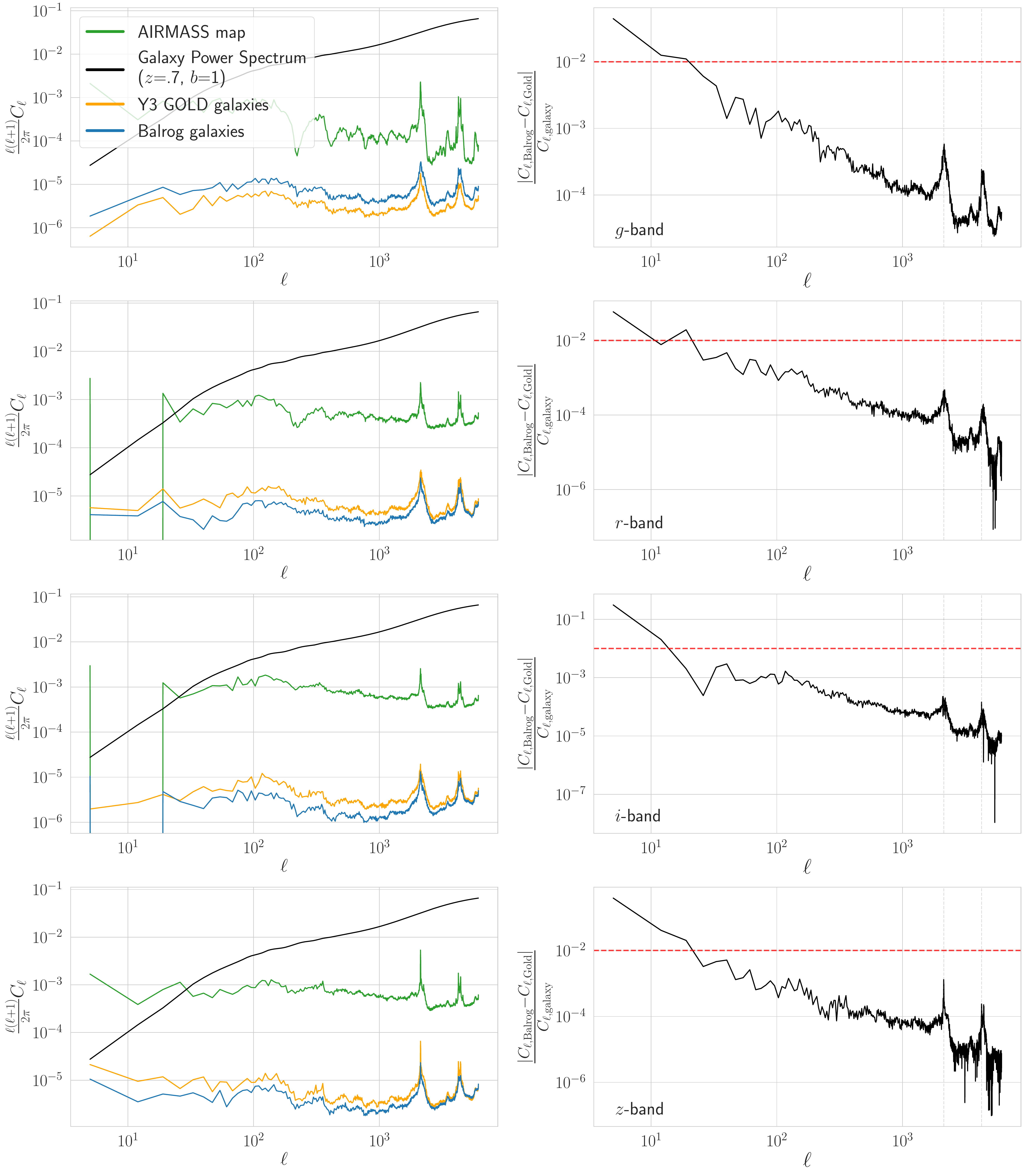}
    \caption{Power spectra of the mean airmass, and associated interpolated \protect{\balrog{}} and Y3 GOLD galaxy count variations, as in Figure~\ref{fig:power_best_worst}. The left panels show the angular power spectrum of the noted survey property (in green) and the corresponding power spectra of the number densities of the \protect{\balrog{}} (in blue) and Y3 GOLD (in gold) {\sc maglim}-like galaxies across the Y3 footprint using the interpolated trends described in \S \protect{\ref{map-trends}} and \S \protect{\ref{power-spectra}}. \response{The reference \code{CAMB} nonlinear matter power spectrum in black is at \protect{$z=0.7$} with a linear galaxy bias parameter of 1.} The right panels show the difference in power between Y3 GOLD and \protect{\balrog{}} as a fraction of the fiducial cosmological power spectrum shown on the left.}
    \label{fig:power_airmass}
\end{figure*}

\begin{figure*}
    \centering
    \includegraphics[width=.98605\textwidth]{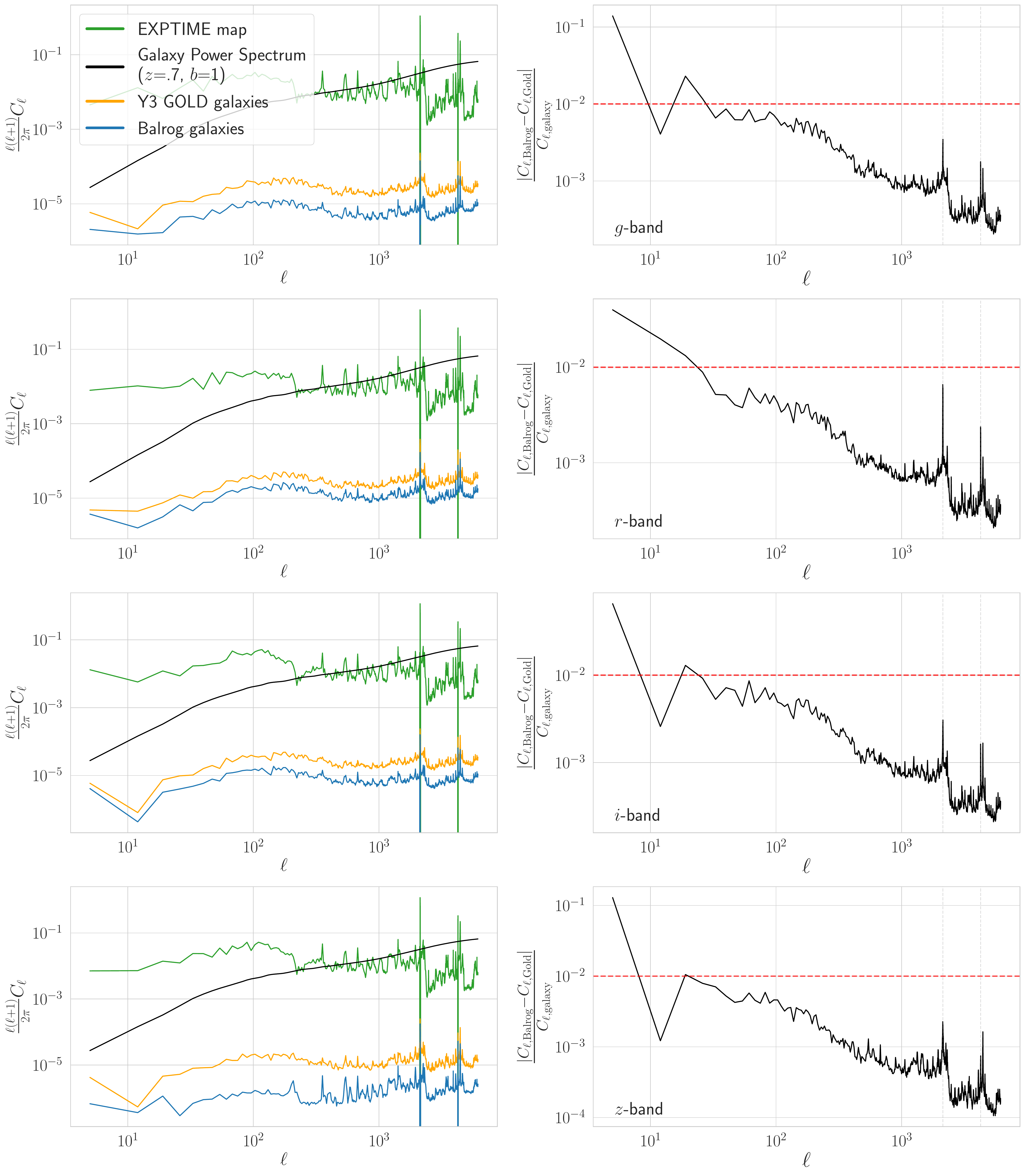}
    \caption{Power spectra of the mean exposure time, and associated interpolated \protect{\balrog{}} and Y3 GOLD galaxy count variations, as in Figure~\ref{fig:power_best_worst}. The left panels show the angular power spectrum of the noted survey property (in green) and the corresponding power spectra of the number densities of the \protect{\balrog{}} (in blue) and Y3 GOLD (in gold) {\sc maglim}-like galaxies across the Y3 footprint using the interpolated trends described in \S \protect{\ref{map-trends}} and \S \protect{\ref{power-spectra}}. \response{The reference \code{CAMB} nonlinear matter power spectrum in black is at \protect{$z=0.7$} with a linear galaxy bias parameter of 1.} The right panels show the difference in power between Y3 GOLD and \protect{\balrog{}} as a fraction of the fiducial cosmological power spectrum shown on the left.}
    \label{fig:power_exptime}
\end{figure*}

\begin{figure*}
    \centering
    \includegraphics[width=.9865\textwidth]{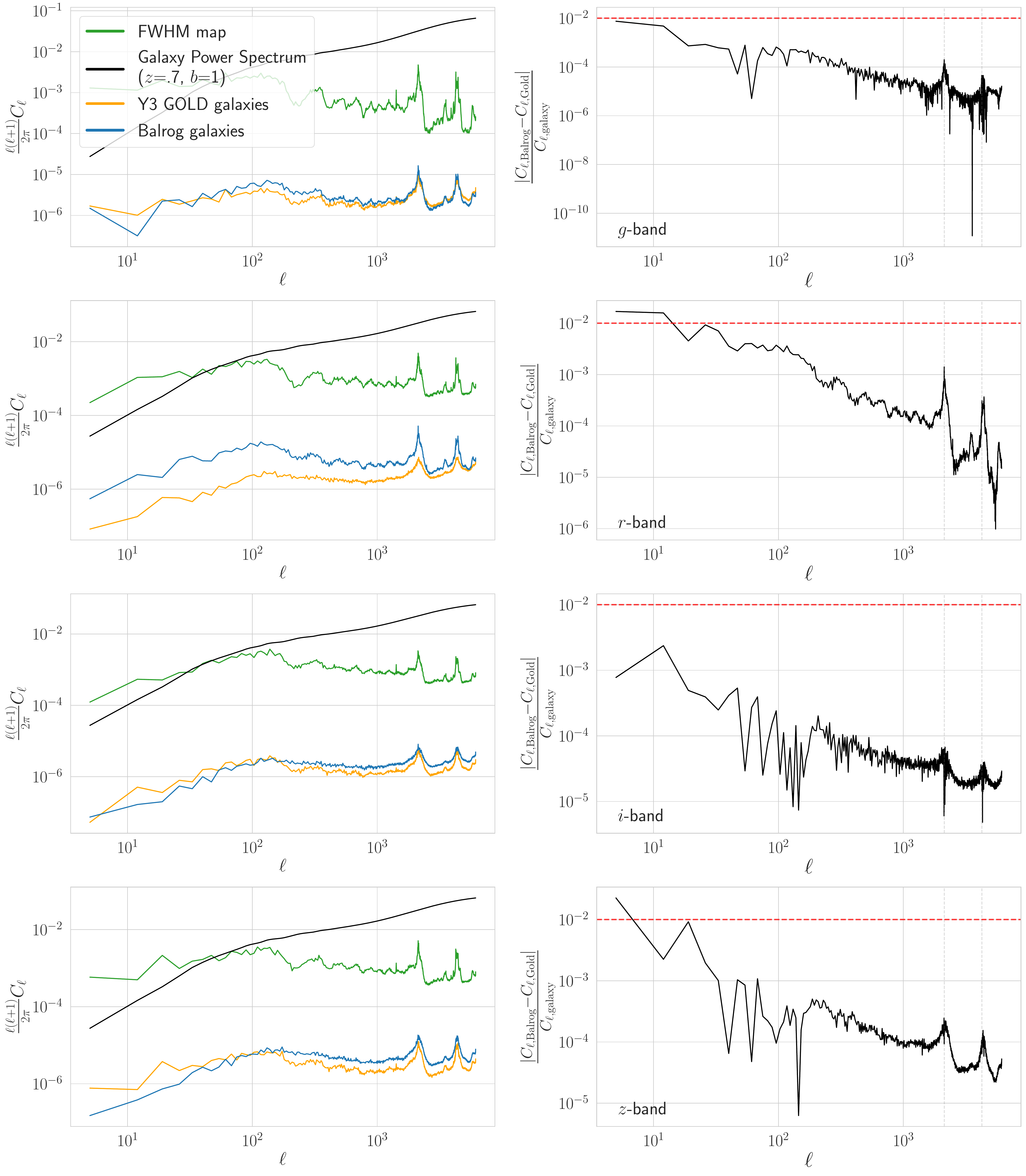}
    \caption{Power spectra of the mean PSF FWHM, and associated interpolated \protect{\balrog{}} and Y3 GOLD galaxy count variations, as in Figure~\ref{fig:power_best_worst}. The left panels show the angular power spectrum of the noted survey property (in green) and the corresponding power spectra of the number densities of the \protect{\balrog{}} (in blue) and Y3 GOLD (in gold) {\sc maglim}-like galaxies across the Y3 footprint using the interpolated trends described in \S \protect{\ref{map-trends}} and \S \protect{\ref{power-spectra}}. \response{The reference \code{CAMB} nonlinear matter power spectrum in black is at \protect{$z=0.7$} with a linear galaxy bias parameter of 1.} The right panels show the difference in power between Y3 GOLD and \protect{\balrog{}} as a fraction of the fiducial cosmological power spectrum shown on the left.}
    \label{fig:power_fwhm}
\end{figure*}

\begin{figure*}
    \centering
    \includegraphics[width=.98655\textwidth]{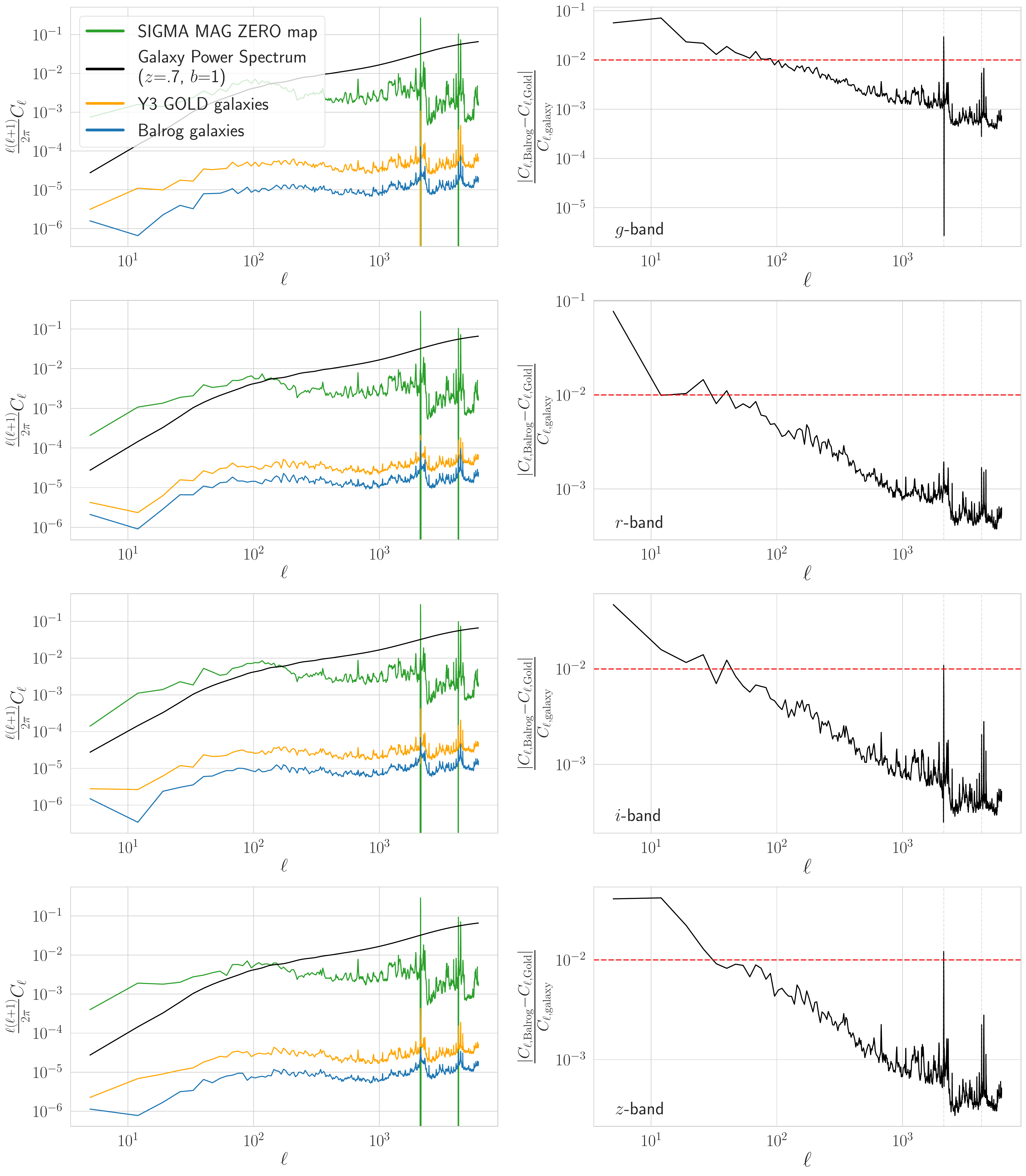}
    \caption{Power spectra of the mean error on the grey zeropoint correction, and associated interpolated \protect{\balrog{}} and Y3 GOLD galaxy count variations, as in Figure~\ref{fig:power_best_worst}. The left panels show the angular power spectrum of the noted survey property (in green) and the corresponding power spectra of the number densities of the \protect{\balrog{}} (in blue) and Y3 GOLD (in gold) {\sc maglim}-like galaxies across the Y3 footprint using the interpolated trends described in \S \protect{\ref{map-trends}} and \S \protect{\ref{power-spectra}}. \response{The reference \code{CAMB} nonlinear matter power spectrum in black is at \protect{$z=0.7$} with a linear galaxy bias parameter of 1.} The right panels show the difference in power between Y3 GOLD and \protect{\balrog{}} as a fraction of the fiducial cosmological power spectrum shown on the left.}
    \label{fig:power_sigma_mag_zero}
\end{figure*}

\begin{figure*}
    \centering
    \includegraphics[width=.98635\textwidth]{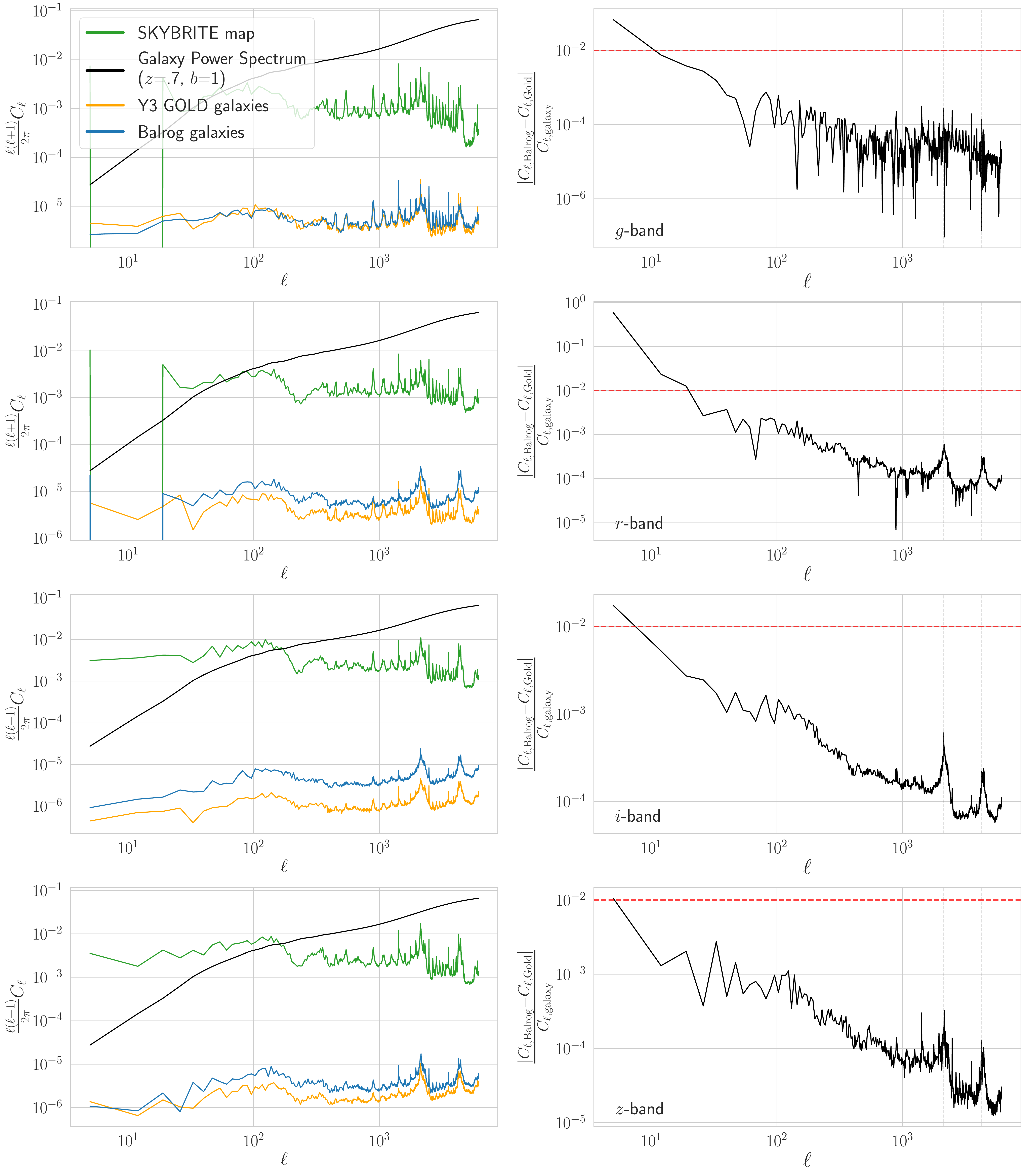}
    \caption{Power spectra of the mean sky brightness, and associated interpolated \protect{\balrog{}} and Y3 GOLD galaxy count variations, as in Figure~\ref{fig:power_best_worst}. The left panels show the angular power spectrum of the noted survey property (in green) and the corresponding power spectra of the number densities of the \protect{\balrog{}} (in blue) and Y3 GOLD (in gold) {\sc maglim}-like galaxies across the Y3 footprint using the interpolated trends described in \S \protect{\ref{map-trends}} and \S \protect{\ref{power-spectra}}. \response{The reference \code{CAMB} nonlinear matter power spectrum in black is at \protect{$z=0.7$} with a linear galaxy bias parameter of 1.} The right panels show the difference in power between Y3 GOLD and \protect{\balrog{}} as a fraction of the fiducial cosmological power spectrum shown on the left.}
    \label{fig:power_skybrite}
\end{figure*}

\begin{figure*}
    \centering
    \includegraphics[width=.98605\textwidth]{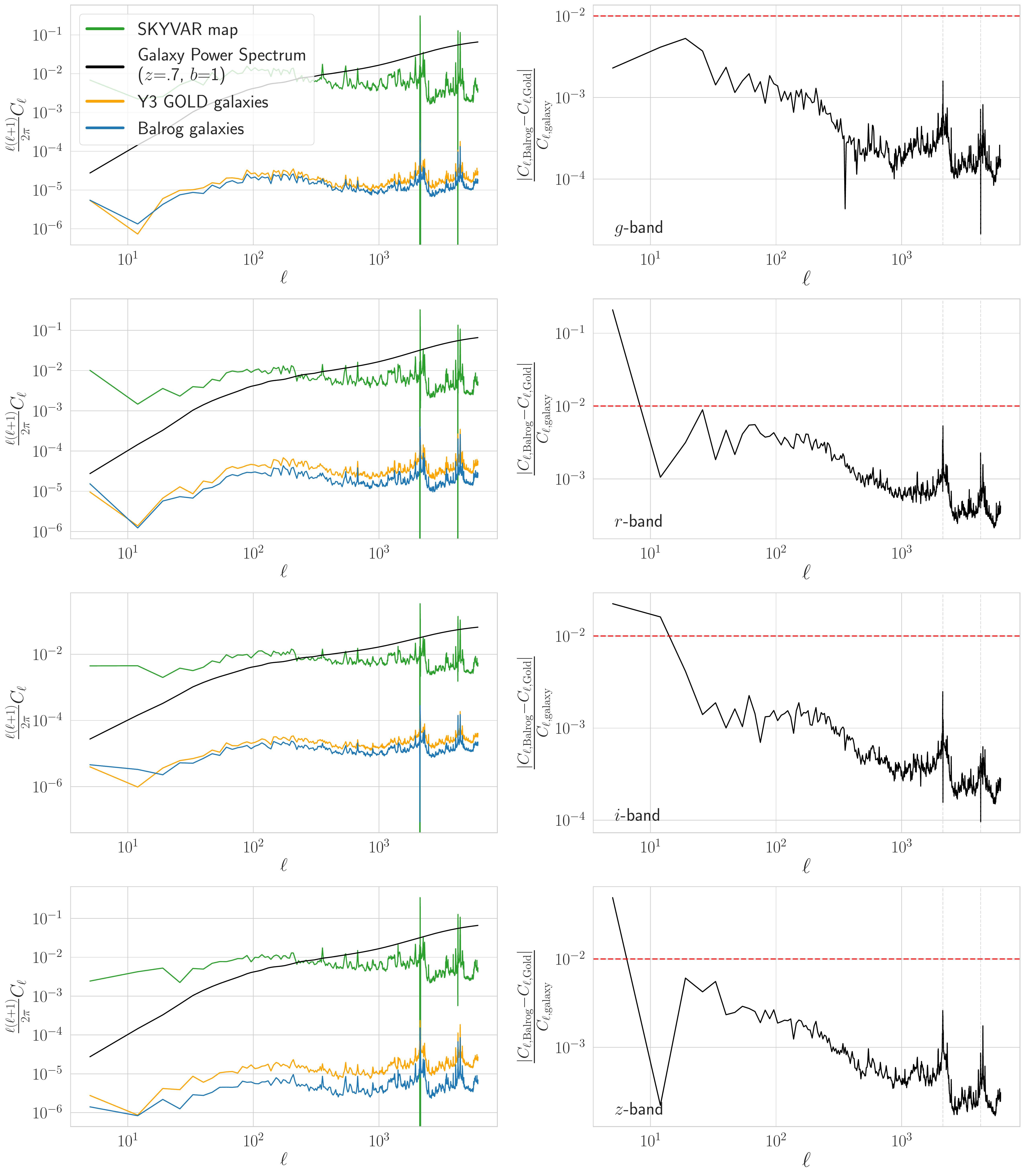}
    \caption{Power spectra of the variance from sky background, and associated interpolated \protect{balrog{}} and Y3 GOLD galaxy count variations, as in Figure~\ref{fig:power_best_worst}. The left panels show the angular power spectrum of the noted survey property (in green) and the corresponding power spectra of the number densities of the \protect{\balrog{}} (in blue) and Y3 GOLD (in gold) {\sc maglim}-like galaxies across the Y3 footprint using the interpolated trends described in \S \protect{\ref{map-trends}} and \S \protect{\ref{power-spectra}}. \response{The reference \code{CAMB} nonlinear matter power spectrum in black is at \protect{$z=0.7$} with a linear galaxy bias parameter of 1.} The right panels show the difference in power between Y3 GOLD and \protect{\balrog{}} as a fraction of the fiducial cosmological power spectrum shown on the left.}
    \label{fig:power_skyvar}
\end{figure*}

\section{Tabular Results}\label{appendix:tabular-results}

Here we present the tabular results of many of the plots shown in Section \ref{photometric-results}. The mean (${<}\Delta{>}$), median ($\widetilde{\Delta}$), and standard deviation ($\sigma$) of the \balrog{} $griz$ magnitude responses binned in injection magnitude for the \starsample{} and \dfsample{} samples are shown in Tables \ref{tab:star-mag-response} and \ref{tab:gal-mag-response} respectively. The equivalent quantities for the color responses are shown in Tables \ref{tab:star-color-response} and \ref{tab:gal-color-response}. Measurements of the \balrog{} classification, or ``confusion'', matrix described in \S \ref{star-gal-sep} are shown in Table \ref{tab:confusion-matrix}.

\begin{table*}[tb!]
\vspace{1.55in}
\centering
\begin{tabular}{c|ccc|ccc|ccc|ccc}
\toprule
True Mag &
  $\langle \Delta g\rangle $ &
  $\widetilde{\Delta g}$ &
  $\sigma_{g}$ &
  $\langle \Delta r\rangle $ &
  $\widetilde{\Delta r}$ &
  $\sigma_{r}$ &
  $\langle \Delta i\rangle $ &
  $\widetilde{\Delta i}$ &
  $\sigma_{i}$ &
  $\langle \Delta z\rangle $ &
  $\widetilde{\Delta z}$ &
  $\sigma_{z}$ \\ 
  & (mag) & (mag) & (mag) & (mag) & (mag) & (mag) & (mag) & (mag)  & (mag) & (mag) & (mag) & (mag) \\
  \midrule
17.00  & 0.001 & 0.000 & 0.004 & 0.002 & 0.003 & 0.005 & 0.003 & 0.004  & 0.010 & 0.005 & 0.006  & 0.006 \\
17.25 & 0.001 & 0.003 & 0.013 & 0.002 & 0.003 & 0.009 & 0.004 & 0.005  & 0.008 & 0.006 & 0.006  & 0.006 \\
17.50  & 0.001 & 0.002 & 0.005 & 0.002 & 0.004 & 0.026 & 0.005 & 0.006  & 0.009 & 0.006 & 0.007  & 0.012 \\
17.75 & 0.001 & 0.002 & 0.006 & 0.003 & 0.004 & 0.007 & 0.005 & 0.006  & 0.006 & 0.006 & 0.007  & 0.011 \\
18.00  & 0.002 & 0.003 & 0.015 & 0.004 & 0.005 & 0.006 & 0.006 & 0.006  & 0.011 & 0.007 & 0.007  & 0.008 \\
18.25 & 0.003 & 0.003 & 0.006 & 0.005 & 0.005 & 0.008 & 0.006 & 0.007  & 0.013 & 0.006 & 0.007  & 0.011 \\
18.50  & 0.004 & 0.004 & 0.006 & 0.005 & 0.006 & 0.008 & 0.006 & 0.007  & 0.009 & 0.007 & 0.008  & 0.014 \\
18.75 & 0.004 & 0.004 & 0.014 & 0.005 & 0.006 & 0.010 & 0.006 & 0.007  & 0.008 & 0.007 & 0.008  & 0.017 \\
19.00  & 0.004 & 0.004 & 0.008 & 0.005 & 0.006 & 0.014 & 0.005 & 0.007  & 0.022 & 0.007 & 0.008  & 0.017 \\
19.25 & 0.004 & 0.005 & 0.008 & 0.004 & 0.006 & 0.017 & 0.006 & 0.007  & 0.013 & 0.007 & 0.009  & 0.021 \\
19.50  & 0.004 & 0.005 & 0.008 & 0.004 & 0.006 & 0.021 & 0.005 & 0.007  & 0.026 & 0.007 & 0.009  & 0.026 \\
19.75 & 0.004 & 0.005 & 0.015 & 0.004 & 0.006 & 0.023 & 0.006 & 0.008  & 0.022 & 0.007 & 0.010  & 0.023 \\
20.00  & 0.003 & 0.005 & 0.015 & 0.004 & 0.006 & 0.029 & 0.006 & 0.008  & 0.028 & 0.008 & 0.010  & 0.034 \\
20.25 & 0.003 & 0.005 & 0.029 & 0.004 & 0.007 & 0.024 & 0.006 & 0.009  & 0.019 & 0.009 & 0.011  & 0.025 \\
20.50  & 0.003 & 0.005 & 0.031 & 0.004 & 0.007 & 0.030 & 0.007 & 0.010  & 0.028 & 0.009 & 0.012  & 0.037 \\
20.75 & 0.003 & 0.005 & 0.033 & 0.005 & 0.008 & 0.028 & 0.007 & 0.010  & 0.038 & 0.010 & 0.013  & 0.041 \\
21.00  & 0.003 & 0.005 & 0.032 & 0.005 & 0.008 & 0.030 & 0.008 & 0.011  & 0.033 & 0.012 & 0.014  & 0.043 \\
21.25 & 0.003 & 0.006 & 0.029 & 0.005 & 0.009 & 0.027 & 0.009 & 0.012  & 0.033 & 0.013 & 0.015  & 0.052 \\
21.50  & 0.003 & 0.006 & 0.030 & 0.006 & 0.010 & 0.031 & 0.011 & 0.014  & 0.042 & 0.016 & 0.017  & 0.059 \\
21.75 & 0.002 & 0.006 & 0.033 & 0.006 & 0.010 & 0.037 & 0.012 & 0.015  & 0.048 & 0.018 & 0.019  & 0.072 \\
22.00  & 0.002 & 0.006 & 0.047 & 0.007 & 0.011 & 0.042 & 0.014 & 0.016  & 0.053 & 0.022 & 0.021  & 0.085 \\
22.25 & 0.002 & 0.006 & 0.044 & 0.009 & 0.013 & 0.049 & 0.017 & 0.018  & 0.065 & 0.026 & 0.024  & 0.107 \\
22.50  & 0.002 & 0.006 & 0.055 & 0.011 & 0.014 & 0.064 & 0.020 & 0.020  & 0.076 & 0.031 & 0.026  & 0.126 \\
22.75 & 0.003 & 0.006 & 0.065 & 0.014 & 0.016 & 0.070 & 0.023 & 0.022  & 0.108 & 0.038 & 0.028  & 0.159 \\
23.00  & 0.004 & 0.008 & 0.083 & 0.017 & 0.019 & 0.083 & 0.028 & 0.025  & 0.107 & 0.047 & 0.033  & 0.218 \\
23.25 & 0.006 & 0.009 & 0.093 & 0.022 & 0.022 & 0.098 & 0.031 & 0.025  & 0.131 & 0.062 & 0.035  & 0.304 \\
23.50  & 0.010 & 0.012 & 0.124 & 0.027 & 0.024 & 0.116 & 0.033 & 0.024  & 0.162 & 0.084 & 0.031  & 0.521 \\
23.75 & 0.015 & 0.014 & 0.147 & 0.034 & 0.029 & 0.154 & 0.031 & 0.018  & 0.200 & 0.140 & 0.037  & 0.876 \\
24.00  & 0.021 & 0.016 & 0.174 & 0.045 & 0.034 & 0.208 & 0.017 & -0.007 & 0.315 & 0.297 & 0.036  & 1.571 \\
24.25 & 0.033 & 0.021 & 0.245 & 0.052 & 0.035 & 0.226 & 0.002 & -0.041 & 0.463 & 0.456 & -0.014 & 2.170 \\ \bottomrule
\end{tabular}
\caption{The mean ($\langle \Delta\rangle $), median ($\widetilde{\Delta}$), and standard deviation ($\sigma$) of the \balrog{} $griz$ magnitude responses binned in injection magnitude for the \protect{\starsample} sample. The quoted magnitudes correspond to the left bin edge. Simple Gaussian statistics do not fully capture the complexity of the responses -- see Figure \ref{fig:meas-vs-true-mag-sof-star}.}
\label{tab:star-mag-response}
\end{table*}

\begin{table*}[tb!]
\centering
\begin{tabular}{c|ccc|ccc|ccc|ccc}
\toprule
True Mag &
  $\langle \Delta g\rangle $ &
  $\widetilde{\Delta g}$ &
  $\sigma_{g}$ &
  $\langle \Delta r\rangle $ &
  $\widetilde{\Delta r}$ &
  $\sigma_{r}$ &
  $\langle \Delta i\rangle $ &
  $\widetilde{\Delta i}$ &
  $\sigma_{i}$ &
  $\langle \Delta z\rangle $ &
  $\widetilde{\Delta z}$ &
  $\sigma_{z}$ \\ 
  & (mag) & (mag) & (mag) & (mag) & (mag) & (mag) & (mag) & (mag)  & (mag) & (mag) & (mag) & (mag) \\
  \midrule
18.00  & -0.066 & -0.039 & 0.081 & -0.055 & -0.035 & 0.081 & -0.048 & -0.029 & 0.087 & -0.043 & -0.024 & 0.076 \\
18.25 & -0.063 & -0.042 & 0.101 & -0.052 & -0.033 & 0.084 & -0.042 & -0.024 & 0.069 & -0.039 & -0.020 & 0.076 \\
18.50  & -0.059 & -0.036 & 0.077 & -0.046 & -0.028 & 0.079 & -0.039 & -0.019 & 0.079 & -0.040 & -0.020 & 0.083 \\
18.75 & -0.055 & -0.036 & 0.078 & -0.039 & -0.020 & 0.076 & -0.039 & -0.020 & 0.077 & -0.034 & -0.014 & 0.083 \\
19.00  & -0.055 & -0.033 & 0.083 & -0.041 & -0.021 & 0.077 & -0.035 & -0.015 & 0.086 & -0.031 & -0.010 & 0.090  \\
19.25 & -0.044 & -0.023 & 0.084 & -0.036 & -0.018 & 0.079 & -0.031 & -0.011 & 0.085 & -0.026 & -0.006 & 0.101 \\
19.50  & -0.040 & -0.022 & 0.078 & -0.033 & -0.013 & 0.087 & -0.027 & -0.006 & 0.096 & -0.022 & -0.002 & 0.105 \\
19.75 & -0.040 & -0.020 & 0.085 & -0.030 & -0.009 & 0.088 & -0.025 & -0.003 & 0.109 & -0.019 & 0.002  & 0.115 \\
20.00  & -0.035 & -0.015 & 0.078 & -0.026 & -0.006 & 0.105 & -0.022 & 0.000  & 0.110 & -0.016 & 0.005  & 0.125 \\
20.25 & -0.035 & -0.015 & 0.098 & -0.024 & -0.003 & 0.105 & -0.020 & 0.003  & 0.119 & -0.012 & 0.009  & 0.134 \\
20.50  & -0.032 & -0.012 & 0.090 & -0.023 & 0.000  & 0.109 & -0.016 & 0.006  & 0.126 & -0.008 & 0.013  & 0.153 \\
20.75 & -0.030 & -0.009 & 0.110 & -0.020 & 0.002  & 0.122 & -0.013 & 0.009  & 0.145 & -0.003 & 0.017  & 0.161 \\
21.00  & -0.027 & -0.006 & 0.107 & -0.018 & 0.005  & 0.133 & -0.010 & 0.013  & 0.155 & 0.001  & 0.021  & 0.174 \\
21.25 & -0.026 & -0.005 & 0.116 & -0.016 & 0.008  & 0.148 & -0.007 & 0.017  & 0.163 & 0.003  & 0.025  & 0.194 \\
21.50  & -0.023 & -0.002 & 0.127 & -0.014 & 0.010  & 0.157 & -0.005 & 0.020  & 0.176 & 0.006  & 0.028  & 0.211 \\
21.75 & -0.022 & 0.000  & 0.147 & -0.012 & 0.014  & 0.171 & -0.002 & 0.023  & 0.189 & 0.008  & 0.031  & 0.228 \\
22.00  & -0.020 & 0.002  & 0.154 & -0.010 & 0.017  & 0.181 & -0.001 & 0.026  & 0.203 & 0.011  & 0.034  & 0.254 \\
22.25 & -0.019 & 0.005  & 0.171 & -0.009 & 0.020  & 0.192 & 0.001  & 0.030  & 0.222 & 0.015  & 0.036  & 0.291 \\
22.50  & -0.017 & 0.007  & 0.187 & -0.007 & 0.024  & 0.212 & 0.003  & 0.033  & 0.248 & 0.020  & 0.039  & 0.339 \\
22.75 & -0.017 & 0.010  & 0.200 & -0.005 & 0.028  & 0.231 & 0.005  & 0.036  & 0.279 & 0.022  & 0.037  & 0.403 \\
23.00  & -0.014 & 0.013  & 0.220 & -0.004 & 0.031  & 0.259 & 0.004  & 0.036  & 0.314 & 0.024  & 0.030  & 0.496 \\
23.25 & -0.012 & 0.017  & 0.247 & -0.004 & 0.034  & 0.293 & -0.002 & 0.031  & 0.355 & 0.028  & 0.014  & 0.663 \\
23.50  & -0.011 & 0.020  & 0.279 & -0.008 & 0.033  & 0.329 & -0.023 & 0.013  & 0.391 & 0.037  & -0.013 & 0.916 \\
23.75 & -0.009 & 0.022  & 0.323 & -0.023 & 0.021  & 0.369 & -0.064 & -0.026 & 0.442 & 0.069  & -0.053 & 1.312 \\
24.00  & -0.009 & 0.020  & 0.383 & -0.055 & -0.007 & 0.413 & -0.132 & -0.091 & 0.528 & 0.142  & -0.115 & 1.874 \\
24.25 & -0.012 & 0.014  & 0.463 & -0.108 & -0.057 & 0.492 & -0.233 & -0.194 & 0.713 & 0.232  & -0.217 & 2.463 \\ \bottomrule
\end{tabular}
\caption{The mean ($\langle \Delta\rangle $), median ($\widetilde{\Delta}$), and standard deviation ($\sigma$) of the \balrog{} $griz$ magnitude responses binned in injection magnitude for the \protect{\dfsample} sample. The quoted magnitudes correspond to the left bin edge. Simple Gaussian statistics do not fully capture the complexity of the responses -- see Figure \ref{fig:meas-vs-true-mag-sof-gal}.}
\label{tab:gal-mag-response}
\end{table*}


\begin{table*}[tb!]
\hspace{-0.4in}
\begin{tabular}{c|ccc|ccc|ccc}
\toprule
True Color & 
$\langle g-r\rangle $ & 
$\widetilde{g-r}$ & 
$\sigma_{g-r}$ & 
$\langle r-i\rangle $ & 
$\widetilde{r-i}$ & 
$\sigma_{r-i}$ & 
$\langle i-z\rangle $ & 
$\widetilde{i-z}$ &
$\sigma_{i-z}$ \\
  & (mag) & (mag) & (mag) & (mag) & (mag) & (mag) & (mag) & (mag)  & (mag) \\
  \midrule
-0.2 & -0.006 & -0.003 & 0.082 & -0.006 & -0.003 & 0.111 & 0.000  & -0.003 & 0.156 \\
-0.1 & -0.004 & -0.002 & 0.098 & -0.007 & -0.003 & 0.102 & -0.002 & -0.002 & 0.114 \\
0.0  & -0.003 & -0.002 & 0.092 & -0.004 & -0.002 & 0.074 & -0.002 & -0.001 & 0.091 \\
0.1  & -0.004 & -0.003 & 0.09  & -0.004 & -0.002 & 0.078 & -0.002 & -0.001 & 0.11  \\
0.2  & -0.002 & -0.002 & 0.074 & -0.003 & -0.002 & 0.09  & -0.002 & -0.001 & 0.111 \\
0.3  & -0.001 & -0.002 & 0.077 & -0.002 & -0.002 & 0.097 & -0.002 & -0.001 & 0.101 \\
0.4  & -0.001 & -0.001 & 0.085 & -0.001 & -0.002 & 0.096 & -0.002 & -0.001 & 0.092 \\
0.5  & 0.000  & -0.001 & 0.09  & 0.000  & -0.001 & 0.094 & -0.001 & -0.001 & 0.087 \\
0.6  & 0.000  & -0.001 & 0.103 & 0.001  & -0.001 & 0.091 & 0.000  & -0.001 & 0.083 \\
0.7  & -0.001 & -0.001 & 0.109 & 0.001  & -0.001 & 0.088 & 0.001  & -0.001 & 0.078 \\
0.8  & -0.002 & -0.001 & 0.113 & 0.002  & -0.001 & 0.092 & 0.001  & 0.000  & 0.075 \\
0.9  & -0.003 & -0.001 & 0.126 & 0.002  & -0.001 & 0.097 & 0.001  & 0.000  & 0.081 \\
1.0  & -0.006 & -0.001 & 0.131 & 0.002  & -0.001 & 0.101 & 0.004  & 0.001  & 0.084 \\
1.1  & -0.010 & -0.002 & 0.142 & 0.003  & -0.001 & 0.106 & 0.003  & 0.001  & 0.078 \\
1.2  & -0.017 & -0.003 & 0.154 & 0.002  & -0.001 & 0.112 & 0.020  & 0.001  & 0.073 \\
1.3  & -0.021 & -0.003 & 0.155 & 0.002  & 0.000  & 0.116 & -0.024 & 0.000  & 0.177 \\
1.4  & -0.027 & -0.004 & 0.17  & 0.000  & 0.001  & 0.123 & 0.006  & -0.003 & 0.119 \\
1.5  & -0.044 & -0.01  & 0.208 & 0.000  & 0.000  & 0.129 & -0.008 & -0.008 & 0.007 \\
1.6  & -0.061 & -0.017 & 0.24  & 0.000  & 0.000  & 0.137 & --    & --    & --   \\
1.7  & -0.076 & -0.026 & 0.265 & -0.004 & -0.001 & 0.138 & --    & --    & --  \\ \bottomrule
\end{tabular}
\caption{The mean ($\langle \Delta\rangle $), median ($\widetilde{\Delta}$), and standard deviation ($\sigma$) of the \balrog{} $g-r$, $r-i$, and $i-z$ color responses binned in injection color for the \protect{\starsample} sample. The quoted colors correspond to the left bin edge. Simple Gaussian statistics do not fully capture the complexity of the responses -- see Figure \ref{fig:stars-color-response}.}
\label{tab:star-color-response}
\end{table*}

\begin{table*}[tb!]
\hspace{-0.4in}
\begin{tabular}{c|ccc|ccc|ccc}
\toprule
True Color & 
$\langle g-r\rangle $ & 
$\widetilde{g-r}$ & 
$\sigma_{g-r}$ & 
$\langle r-i\rangle $ & 
$\widetilde{r-i}$ & 
$\sigma_{r-i}$ & 
$\langle i-z\rangle $ & 
$\widetilde{i-z}$ &
$\sigma_{i-z}$ \\
  & (mag) & (mag) & (mag) & (mag) & (mag) & (mag) & (mag) & (mag)  & (mag) \\
  \midrule
-0.2 & 0.081  & 0.053  & 0.211 & 0.079  & 0.043  & 0.216 & 0.092  & 0.047  & 0.239 \\
-0.1 & 0.047  & 0.030  & 0.192 & 0.053  & 0.032  & 0.201 & 0.062  & 0.030  & 0.213 \\
0.0  & 0.026  & 0.016  & 0.182 & 0.028  & 0.013  & 0.182 & 0.030  & 0.009  & 0.177 \\
0.1  & 0.012  & 0.006  & 0.179 & 0.011  & 0.002  & 0.155 & 0.019  & 0.004  & 0.163 \\
0.2  & 0.002  & -0.002 & 0.178 & 0.004  & 0.000  & 0.140 & 0.011  & 0.001  & 0.145 \\
0.3  & -0.009 & -0.006 & 0.169 & 0.001  & -0.001 & 0.140 & 0.007  & 0.000  & 0.134 \\
0.4  & -0.015 & -0.008 & 0.161 & -0.003 & -0.001 & 0.139 & 0.004  & 0.000  & 0.141 \\
0.5  & -0.019 & -0.009 & 0.158 & -0.007 & -0.003 & 0.140 & 0.001  & -0.001 & 0.160 \\
0.6  & -0.024 & -0.010 & 0.157 & -0.012 & -0.005 & 0.146 & -0.004 & -0.003 & 0.161 \\
0.7  & -0.028 & -0.011 & 0.158 & -0.015 & -0.007 & 0.147 & -0.009 & -0.005 & 0.159 \\
0.8  & -0.031 & -0.011 & 0.159 & -0.018 & -0.007 & 0.146 & -0.012 & -0.007 & 0.161 \\
0.9  & -0.036 & -0.011 & 0.162 & -0.022 & -0.008 & 0.152 & -0.016 & -0.009 & 0.171 \\
1.0  & -0.041 & -0.011 & 0.167 & -0.026 & -0.010 & 0.161 & -0.019 & -0.011 & 0.176 \\
1.1  & -0.046 & -0.011 & 0.173 & -0.029 & -0.012 & 0.170 & -0.031 & -0.016 & 0.193 \\
1.2  & -0.051 & -0.010 & 0.184 & -0.035 & -0.013 & 0.178 & -0.053 & -0.024 & 0.210 \\
1.3  & -0.059 & -0.011 & 0.194 & -0.071 & -0.030 & 0.221 & -0.049 & -0.024 & 0.215 \\
1.4  & -0.069 & -0.013 & 0.210 & -0.149 & -0.091 & 0.276 & -0.054 & -0.018 & 0.223 \\
1.5  & -0.074 & -0.015 & 0.222 & -0.171 & -0.105 & 0.288 & -0.076 & -0.028 & 0.236 \\
1.6  & -0.070 & -0.016 & 0.224 & -0.183 & -0.112 & 0.300 & -0.075 & -0.015 & 0.220 \\
1.7  & -0.066 & -0.016 & 0.224 & -0.206 & -0.126 & 0.314 & -0.050 & -0.007 & 0.240 \\
1.8  & -0.096 & -0.028 & 0.265 & -0.206 & -0.127 & 0.334 & -0.063 & -0.017 & 0.255 \\
1.9  & -0.193 & -0.092 & 0.358 & -0.221 & -0.112 & 0.363 & -0.061 & -0.003 & 0.220 \\ \bottomrule
\end{tabular}
\caption{The mean ($\langle \Delta\rangle $), median ($\widetilde{\Delta}$), and standard deviation ($\sigma$) of the \balrog{} $g-r$, $r-i$, and $i-z$ color responses binned in injection color for the \protect{\dfsample} sample. The quoted colors correspond to the left bin edge. Simple Gaussian statistics do not fully capture the complexity of the responses -- see Figure \ref{fig:gals-color-response}.}
\label{tab:gal-color-response}
\end{table*}

\begin{table*}[tb!]
\hspace{-0.65in}
\centering
\begin{tabular}{ccccc}
\toprule
True Mag & Star-\textgreater{}Star & Gal-\textgreater{}Star & Star-\textgreater{}Gal & Gal-\textgreater{}Gal \\
  & (TP; \%) & (FP; \%) & (FN; \%) & (TN; \%) \\
  \midrule
18.50  & 99.6 & 1.6  & 0.4  & 98.4 \\
18.75 & 99.6 & 2.9  & 0.4  & 97.1 \\
19.00  & 99.4 & 2.9  & 0.6  & 97.1 \\
19.25 & 99.3 & 2.6  & 0.7  & 97.4 \\
19.50  & 99.2 & 2.8  & 0.8  & 97.2 \\
19.75 & 99.1 & 2.3  & 0.9  & 97.7 \\
20.00  & 98.7 & 1.9  & 1.3  & 98.1 \\
20.25 & 98.6 & 1.8  & 1.4  & 98.2 \\
20.50  & 98.2 & 1.8  & 1.8  & 98.2 \\
20.75 & 97.8 & 1.9  & 2.2  & 98.1 \\
21.00  & 97.3 & 1.8  & 2.7  & 98.2 \\
21.25 & 96.7 & 1.7  & 3.3  & 98.3 \\
21.50  & 95.9 & 2.2  & 4.1  & 97.8 \\
21.75 & 95.1 & 2.0  & 4.9  & 98.0 \\
22.00  & 93.4 & 2.3  & 6.6  & 97.7 \\
22.25 & 90.8 & 3.2  & 9.2  & 96.8 \\
22.50  & 86.4 & 4.1  & 13.6 & 95.9 \\
22.75 & 79.5 & 5.2  & 20.5 & 94.8 \\
23.00  & 70.3 & 6.7  & 29.7 & 93.3 \\
23.25 & 58.2 & 8.3  & 41.8 & 91.7 \\
23.50  & 46.4 & 10.1 & 53.6 & 89.9 \\
23.75 & 37.5 & 12.4 & 62.5 & 87.6 \\
24.00  & 30.9 & 14.5 & 69.1 & 85.5 \\
24.25 & 25.9 & 15.0 & 74.1 & 85.0 \\ \bottomrule
\end{tabular}
\caption{Elements of the classification (or confusion) matrix for \protect{\balrog{}} sources binned by injection magnitude when normalized by percent, where the measured classification is determined by \protect{\code{EXTENDED\_CLASS\_SOF}$<=1$} for stars and \protect{\code{EXTENDED\_CLASS\_SOF}$>1$} for galaxies. The second through fifth columns correspond to the true positive (TP), false positive (FP), false negative (FN), and true negative (TN) rates of \balrog{} stars respectively. The very pure \protect{\starsample{}} sample is used to compute the TP and FN rates, while the nosier classifications of the DF \protect{\dfsample{}} injections are used for the rest. The quoted magnitudes correspond to the left bin edge. See Figure \ref{fig:efficiency-contamination}.}
\label{tab:confusion-matrix}
\end{table*}




\end{document}